\documentclass[11pt]{article}
\usepackage[utf8]{inputenc}
\usepackage[T1]{fontenc}
\usepackage[normalem]{ulem}
\usepackage{verbatim}

\usepackage{amsmath,amssymb,amsthm,amsfonts,mathrsfs,dsfont,braket,array,float}
\usepackage{color,graphicx,cite}
\usepackage{slashed}
\usepackage{enumitem}
\usepackage{simpler-wick}
\usepackage{epsfig}
\usepackage{mathabx}
\usepackage[mathscr]{eucal}
\usepackage{stmaryrd}
\usepackage{esint}
\usepackage{physics}
\usepackage{setspace}
\usepackage{braket}
\usepackage{float}
\usepackage{url}
\usepackage{dsfont}
\usepackage{mathtools}
\usepackage{bbold}
\usepackage[export]{adjustbox}
\usepackage{slashed}
\usepackage{tikz}
\usepackage{graphicx}
\usepackage{epstopdf}
\usepackage{subfigure}
\usepackage[labelfont=bf]{caption}
\usepackage{pgfplots}
\pgfplotsset{compat=1.18}
\usepackage{fancybox}
\usepackage{eurosym}
\usepackage{tcolorbox}
\usepackage{tensor}

\allowdisplaybreaks

\newmuskip\pFqmuskip

\newcommand*\pFq[6][8]{%
  \begingroup 
  \pFqmuskip=#1mu\relax
  \mathcode`\,=\string"8000
  \begingroup\lccode`\~=`\,
  \lowercase{\endgroup\let~}\pFqcomma
  {}_{#2}F_{#3}{\left[\genfrac..{0pt}{}{#4}{#5};#6\right]}%
  \endgroup
}
\newcommand{\pFqcomma}{\mskip\pFqmuskip}

\usepackage[margin = 2.4cm]{geometry}
\usepackage[ragged]{footmisc}
\newcommand{\e}{\textnormal{e}}

\newcommand{\clo}{\mathcal{O}}

\newcommand{\bbi}{\mathbb{1}}

\def\be{\begin{equation}}
	\def\ee{\end{equation}}

\def\XXint#1#2#3{{\setbox0=\hbox{$#1{#2#3}{\int}$}
     \vcenter{\hbox{$#2#3$}}\kern-.5\wd0}}

\numberwithin{equation}{section}

\newcommand{\mean}[1]{\overline{#1}}

\newcommand{\opmean}[1]{
\langle\hspace{-3pt}\langle #1 \rangle\hspace{-3pt}\rangle
}
\newcommand{\bigopmean}[1]{
\big\langle\hspace{-4pt}\big\langle #1 \big\rangle\hspace{-4pt}\big\rangle
}

\newcommand{\eqimg}[1]{
\begin{gathered}
\includegraphics[height=25mm]{#1} 
\end{gathered}
}

\usepackage{bm}
\newcommand{\mb}[1]{\bm{#1}}

\usepackage{mdframed}
\usepackage{xpatch}
\makeatletter
\xpatchcmd{\endmdframed}
  {\aftergroup\endmdf@trivlist\color@endgroup}
  {\endmdf@trivlist\color@endgroup\@doendpe}
  {}{}
\makeatother

\usepackage{formatting}
 \definecolor{darkgreen}{rgb}{0.01, 0.5, 0.24}
 \definecolor{bluenice}{rgb}{0.0, 0.53, 0.74}

\usepackage{shortcuts}

\renewcommand{\order}[1]{O\left(#1\right)}

\begin{document} 
\begin{titlepage}
\begin{center}
\phantom{ }
\vspace{2cm}

{\bf \Large{A principle of maximum ignorance for semiclassical gravity}}
\vskip .75cm
Jan de Boer${}^{\dagger}$, Diego Liska${}^{*}$, Boris Post${}^{\ddagger}$, Martin Sasieta${}^{\mathsection}$
\vskip 0.75cm
\small{${}^{\dagger}$ ${}^{*}$  ${}^{\ddagger}$  \textit{Institute for Theoretical Physics,
University of Amsterdam,}}
\vskip -.25cm
\small{\textit{ PO Box 94485, 1090 GL Amsterdam, The Netherlands}}
\vskip .3cm
\small{${}^{\mathsection}$ \textit{Martin Fisher School of Physics, Brandeis University}}
\vskip -.25cm
\small{\textit{Waltham, Massachusetts 02453, USA}}

\vskip 3cm

\begin{abstract}
\noindent The principle of maximum ignorance posits that the coarse-grained description of a system is maximally agnostic about its underlying microscopic structure. We briefly review this principle for random matrix theory and for the eigenstate thermalization hypothesis. We then apply this principle in holography to construct ensembles of random mixed states. This leads to an ensemble of microstates which models our microscopic ignorance, and which on average reproduces the effective semiclassical physics of a given bulk state. We call this ensemble the \emph{state-averaging ansatz}. The output of our model is a prediction for semiclassical contributions to variances and higher statistical moments over the ensemble of microstates. The statistical moments provide coarse-grained -- yet gravitationally non-perturbative -- information about the microstructure of the individual states of the ensemble. We show that these contributions exactly match the on-shell action of known wormhole configurations of the gravitational path integral. These results strengthen the view that wormholes simply parametrize the ignorance of the microstructure of a fundamental state, given a fixed semiclassical bulk description.

\end{abstract}
\end{center}

\small{\vfill \noindent ${}^{\dagger}$j.deboer@uva.nl\\
${}^{*}$d.liska@uva.nl\\
${}^{\ddagger}$b.p.post@uva.nl\\
${}^{\mathsection}$martinsasieta@brandeis.edu
}

\end{titlepage}

\setcounter{tocdepth}{2}

{
\parskip = 0\baselineskip 
  \hypersetup{linkcolor=black}
  \tableofcontents
}

\newpage

\newgeometry{left=2.4cm,right = 2.4cm, bottom = 2.4cm, top=2cm}

\section{Introduction}\label{SecI}

In recent years, it has become clear that Euclidean gravity is more than a low-energy effective field theory. The gravitational path integral knows about its UV completion, but only through the \emph{statistical} properties of its microscopic description. This has been made most precise in the case of 
Jackiw–Teitelboim (JT) gravity \cite{Saad:2019lba,Saad:2019pqd,Stanford:2020wkf} and pure AdS$_3$ gravity \cite{Belin:2020hea, Belin:2021ryy, Anous:2021caj, Chandra:2022bqq}. In the former case, the microscopic boundary Hamiltonian is modeled by a random Hermitian matrix $H$, and higher moments of the partition function $\Tr \e^{-\beta H}$ are described by Euclidean wormholes in the bulk. In the latter case, the 
microscopic OPE coefficients of the boundary CFT$_2$ are modeled by random variables $C_{ijk}$, and the statistical moments of the OPE coefficients are captured by multi-boundary 
wormholes. 

In a series of papers \cite{Sasieta:2022ksu,Balasubramanian:2022gmo,Balasubramanian:2022lnw} it was furthermore shown that the connection between statistical variance and semiclassical wormholes in pure gravity is not restricted to low dimensions. By injecting a thin shell of dust particles behind the horizon, the authors showed that one can construct on-shell Euclidean wormholes by gluing a portion of the Euclidean AdS$_d$ black hole along the trajectory of the thin shell. These wormholes precisely capture the variance 
of the matrix elements of the
operator creating the thin shell of dust. 
Such operators were then used to construct an ensemble of black hole microstates with semiclassical interiors, a construction inspired by partially entangled thermal states (PETS) in the SYK model \cite{Goel:2018ubv}. 
The mutual overlaps of these states are universal 
and consistent with a finite density of states controlled by the Bekenstein-Hawking entropy. 
This construction gives another perspective on the effects produced by replica wormholes, originally introduced in the West Coast model of black hole evaporation \cite{Penington:2019kki}, in higher dimensional black holes.

Central to the statistical description of the microscopic theory of semiclassical gravity is the anticipation that the high-energy sector of the dual CFT exhibits chaotic behavior.
It is expected 
that holographic CFTs at strong coupling have an ergodic phase where the universal features of quantum chaos apply \cite{Cotler:2016fpe,Altland:2020ccq,Altland:2021rqn,Hollander}, such as eigenvalue repulsion and approximate unitary invariance in narrow high-energy microcanonical windows. One furthermore expects simple CFT operators to obey the Eigenstate Thermalization Hypothesis (ETH) \cite{Lashkari:2016vgj}, characterizing the thermal behavior of chaotic high-energy eigenstates \cite{Srednicki_1994,PhysRevA.43.2046}. It is because of these properties that we can put on a statistician's hat and only ask coarse-grained questions to the CFT.

This naturally leads to the question: what is being averaged over in the statistical description? There are various directions explored in the literature, which can be broadly grouped into three categories: averaging over Hamiltonians $H$, averaging over operators $\mathcal{O}$, and averaging over states $\rho$.
While all three approaches have different ranges of applicability, what they have in common is a \emph{principle of maximum ignorance}: the statistical ensembles that are used to model semiclassical gravity are maximally agnostic about the precise fine-grained description of the underlying quantum theory, but they are compatible with coarse-grained low-energy data. 

One can formalize the notion of maximum ignorance as a constrained optimization of an information-theoretic entropy function, such as the Shannon entropy. To see how this works in practice, we briefly describe two examples which will serve as inspiration for the main topic of this paper.

\restoregeometry

\subsection*{Example 1: Averaging over Hamiltonians}

Suppose we want
to model a chaotic Hamiltonian in a narrow energy band using a random matrix $H$ drawn from some probability distribution $\mu(H)$. The only knowledge we have about the system is 
the partition function $Z(\beta)$ in some range of temperatures. In AdS/CFT, this can be computed from  
the semiclassical gravitational path integral  
with boundary manifold $M\times S_\beta$. The question is: what is the 
distribution $\mu(H)$ that best describes the system given only 
the single-boundary input 
$Z(\beta)$?

According to the principle of maximum entropy, sometimes also called Jaynes' principle \cite{balian1968random, PhysRev.106.620}, it is the distribution which maximizes the Shannon information entropy 
with the additional constraint that the average $\expval{\Tr \e^{-\beta H}}\coloneqq \int \mathrm{d}H\mu(H)\Tr \e^{-\beta H}$ should be equal to $Z(\beta)$. That is, we look for the distribution that optimizes the entropy functional
\begin{equation}
    S[\mu] = 
    \int \dd H \;\mu(H)
    \Big[
    -\log \mu(H) +
    \int\dd\beta \;\lambda(\beta) \Big(\Tr \e^{-\beta H} - Z(\beta)\Big)
    \Big].
\end{equation}
The first term is the Shannon information entropy, 
and the Lagrange multiplier $\lambda(\beta)$ enforces the constraint on the average partition function. Solving the extremization problem $\delta S/\delta \mu =0$ leads to the probability distribution 
\begin{equation}
    \mu(H) = \frac{1}{\mathcal{N}}\, \e^{-\Tr V(H)},\\[0.5em]
\end{equation}
where $\mathcal{N}$ is a normalization such that $\int \dd H \mu(H) =1$, and the potential $V(E)$ is the Laplace transform of $\lambda(\beta)$. To complete the procedure, one has to express $\lambda(\beta)$ in terms of the input $Z(\beta)$ by solving the constraint equation $\expval{\Tr \e^{-\beta H}} = Z(\beta)$. For large matrices whose eigenvalues have compact support, the constraint equation is solved to leading order by 
\begin{equation}
\frac{1}{2} \,V'(E) =  
\mathrm{p.v.}\!\int 
\dd E'\frac{\varrho(E')}{E-E'},
\end{equation}
where the coarse-grained density of states $\varrho(E)$ is defined through $Z(\beta) = \int \dd E\,\varrho(E)\,\e^{-\beta E}$.

We have glossed over many details here, but this simple exercise already shows an important feature of the maximum entropy principle: a single-trace constraint results in a single-trace matrix model. The model's \emph{output} is then a prediction for multi-trace observables, such as the averaged spectral form factor $\mathrm{SFF}(\beta,t) =\expval{\Tr \e^{-(\beta+iT)H}\Tr \e^{-(\beta-iT)H}}$. This prediction can be tested against a wormhole computation in gravity. For example, when $\beta=0$, the microcanonical version of the SFF is matched to the double cone of \cite{Saad:2018bqo,Mahajan:2021maz}. 

Conceptually, an important point in this line of reasoning is that the wormhole quantifies statistical correlations in an ensemble $\mu(H)$ that models the true Hamiltonian. Factorization may well be restored in a complete microscopic theory, including non-gravitational objects sensitive to microscopics like D-branes \cite{Blommaert:2021fob,Johnson:2022wsr,Post:2022dfi,Altland:2022xqx} or half-wormholes \cite{Saad:2021rcu}. However, remaining agnostic about this, the role of Euclidean wormholes in the AdS/CFT dictionary \cite{Cotler:2021cqa,Cotler:2022rud} is to uncover statistical properties of the pseudo-random microscopics.

\subsection*{Example 2: Averaging over operators}

Instead of modeling the statistics of energy eigenvalues, one can also model chaotic high-energy eigenstates by probing them with a simple operator $\mathcal{O}$. To model the matrix elements $\mathcal{O}_{ij}\coloneqq \expval{E_i | \mathcal{O}|E_j}$, we again invoke a principle of maximum ignorance. 

Namely, suppose we have measured $\mathcal{O}$'s thermal one-point function $F(\beta)$ and thermal time-evolved two-point function $G(\beta,t)$ as a function of $\beta$ and $t$. These functions are input data, and are available from semiclassical gravity \cite{Maldacena:2001kr, Son2002MinkowskispaceCI, Alday:2020eua} by studying the wave equation in a black hole background. We can look for the probability distribution $\mu(\mathcal{O})$ on the space of Hermitian matrices which maximizes the Shannon entropy of $\mu$ and agrees with the input data on average. 
In other words, we optimize
\begin{equation}
S[\mu] = \int \dd \mathcal{O} \,\mu(\mathcal{O}) \Big[-\log \mu(\mathcal{O}) + C_1[\mathcal{O}]+C_2[\mathcal{O}] \Big],
\end{equation}
where the volume element on the space of Hermitian matrices is $\dd \mathcal{O} = \prod_{i<j} \dd \mathcal{O}_{ij}\dd \mathcal{O}^*_{ij} \prod_k \dd \mathcal{O}_{kk}$, and
the constraints are implemented by Lagrange multiplier functions $\lambda_1$ and $\lambda_2$:
\begin{equation}
\begin{split}
    C_1[\mathcal{O}] &= \int \dd\beta\, \lambda_1(\beta) \Big(\Tr\rho_\beta \mathcal{O} - F(\beta) \Big) 
    \\[1em]
    C_2[\mathcal{O}] &= \int \dd\beta \dd t \,\lambda_2(\beta,t) \Big( \Tr \rho_\beta^{1/2}\mathcal{O} \rho_\beta^{1/2} \mathcal{O}(t) -G(\beta,t)\Big).
\end{split}
\end{equation}
Here $\rho_\beta = \e^{-\beta H}/Z(\beta)$ is the thermal density matrix, and we have analytically continued the conventional thermal two-point function $\Tr\rho_\beta \mathcal{O}(0)\mathcal{O}(t)$ with a shift $t\to t+i\beta/2$, as is standard practice. Solving the variational equation $\delta S/\delta\mu =0$ as before, gives the optimal distribution
\begin{equation}
\begin{split}
    \mu(\mathcal{O}) &= \frac{1}{\mathcal{N}}\,\exp(-\int \dd\beta\, \lambda_1(\beta) \Tr\rho_\beta \mathcal{O}-\int \dd\beta \dd t \,\lambda_2(\beta,t)  \Tr \rho_\beta^{1/2}\mathcal{O} \rho_\beta^{1/2} \mathcal{O}(t)) \\[1em]
    &= \frac{1}{\mathcal{N}}\,\exp(-\sum_i V_1(E_i)\mathcal{O}_{ii} - \sum_{ij}V_2(\bar E_{ij},\omega_{ij})|\mathcal{O}_{ij}|^2),
    \end{split}
\end{equation}
where in the second line we have written the exponent explicitly in the energy eigenbasis. We have also defined the variables 
\begin{equation}
    \bar E_{ij} = \frac{E_i+E_j}{2} \quad \text{and}\quad \omega_{ij} = E_i-E_j.
\end{equation}
The potential $V_1(E)$ is the Laplace transform of $\lambda_1(\beta)$, while $V_2(\bar E,\omega)$ is the combined Laplace/Fourier transform of $\lambda_2(\beta,t)$. As a final step, we then express $V_1$ and $V_2$ in terms of the input data $F$ and $G$ by solving the constraint equations
\begin{equation}
    \opmean{\Tr\rho_\beta\mathcal{O}} = F(\beta) \quad \text{and}\quad \opmean{\Tr \rho_\beta^{1/2}\mathcal{O} \rho_\beta^{1/2} \mathcal{O}(t)} = G(\beta,t),
\end{equation}
where $\opmean{\bullet} = \int \dd \mathcal{O}\mu(\mathcal{O})(\bullet)$. Since the distribution is Gaussian, these equations can easily be solved. 
\newgeometry{left=2.4cm,right = 2.4cm, bottom = 2.2cm, top=2cm}
\noindent In doing so, one uses the continuum approximations
\begin{equation}
    \sum_i \to \int \dd E\varrho(E) \quad \text{and} \quad \sum_{ij} \to \int \dd \bar E\dd\omega \,\varrho\left(\bar E+\frac{\omega}{2}\right)\varrho\left(\bar E - \frac{\omega}{2}\right).
\end{equation}
The result of this exercise is the familiar form of the Eigenstate Thermalization Hypothesis \cite{PhysRevA.43.2046,Srednicki_1994,DAlessio:2015qtq}
\begin{equation}\label{eq:ETH}
\mathcal{O}_{ij} = f(E_i) \delta_{ij} + \e^{-S(\bar E_{ij})/2} g(\bar{E}_{ij},\omega_{ij})^{1/2} R_{ij}.
\end{equation}
Its meaning is that high-energy eigenstates, as probed by $\mathcal{O}$, are indistinguishable from a thermal (or microcanonical) state, up to small corrections that are exponentially suppressed in the microcanonical entropy $S(E) \coloneqq \log \varrho(E)$, and which vary erratically as a random variable $R_{ij}$. The smooth functions $f$ and $g$ are the microcanonical one- and connected two-point function, related to our input data by the inverse Laplace transforms $f(E)= \mathcal{L}^{-1}[F(\beta)]$ and $g(\bar E,\omega) = \mathcal{L}^{-1}[G(\beta,t)^{c}]$.\footnote{More details about the ETH ensemble can be found in Appendix \ref{app:ETH}.}

This ETH ensemble is Gaussian because our input was quadratic in $\mathcal{O}$. However,
non-Gaussianities are necessary to explain connected contributions to OTOCs with more operator
insertions \cite{Foini:2018sdb,Anous:2019yku,Murthy:2019fgs,Belin:2021ryy}. These non-Gaussianities can be incorporated by adding higher-point single-trace correlators as input. Maximizing the Shannon entropy of $\mu(\mathcal{O})$ while keeping this collection of higher-point correlators fixed straightforwardly leads to the \emph{generalized ETH} matrix model of \cite{Jafferis:2022uhu}.

The above derivation teaches us two important lessons. First, it emphasizes that matrix elements $\expval{E_i|\mathcal{O}|E_j}$ which obey ETH are `maximally random' in the precise sense of having the statistics of a maximum entropy ensemble. Second, the ensemble was derived using only single-trace observables. In the context of AdS/CFT, these correspond to single-boundary correlation functions. Combining maximal randomness with single-boundary low-energy input, we then get a prediction for multi-boundary quantities. An example is the statistical variance of the thermal one-point function
\begin{equation}\label{eq:1pointvariance}
\begin{split}
    \opmean{\Tr\rho_{\beta_1} \mathcal{O}\, \Tr\rho_{\beta_2} \mathcal{O} } - \opmean{\Tr\rho_{\beta_1}\mathcal{O}}\opmean{\Tr\rho_{\beta_2}\mathcal{O}} &\approx \frac{1}{Z(\beta_1)Z(\beta_2)}\int \dd E \,\e^{-(\beta_1+\beta_2) E}g(E,0) \\
    &= \vcenter{\hbox{\begin{tikzpicture}\,\node at (0,0) {\includegraphics[width=4cm, valign=c]{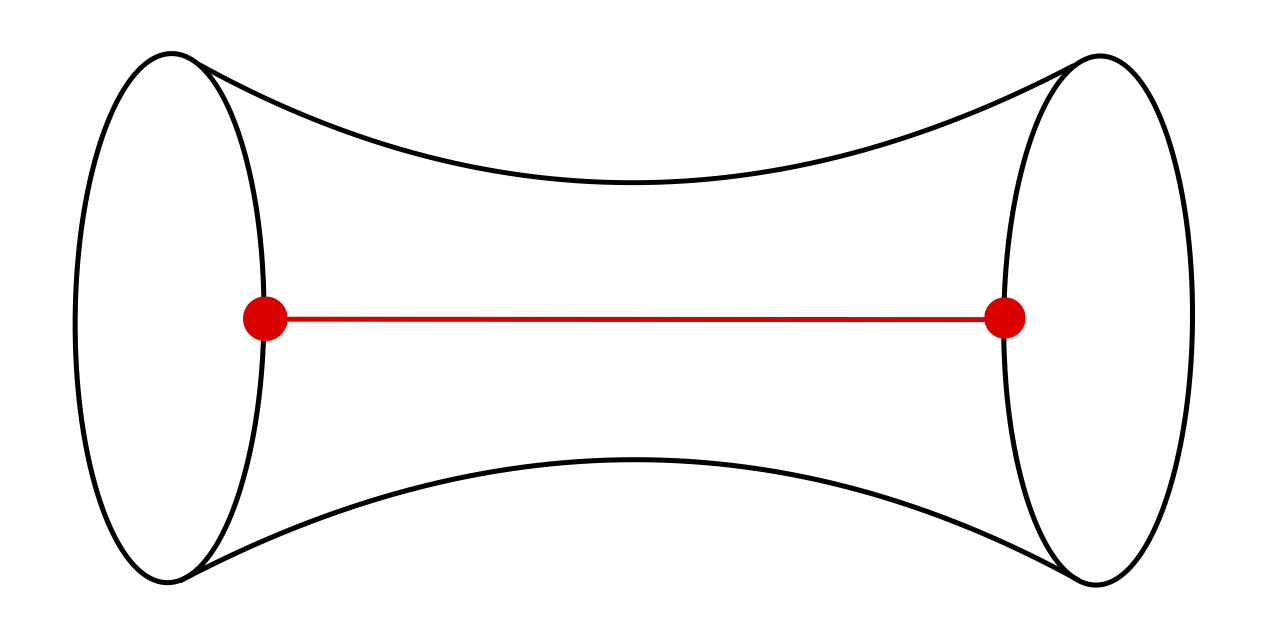}};
    \node at (-2,0){$\beta_1$};
    \node at (2,0) {$\beta_2$};
    \end{tikzpicture}}},
    \end{split}
\end{equation}
which we have evaluated in the continuum approximation $\sum_i \to \int \dd E\,\e^{S(E)}$. For this example we assumed $f(E)=0$. The right-hand side originates from a non-zero correlation $\bigopmean{\mathcal{O}_{ii} \mathcal{O}_{jj}}$ for $i=j$. 
In \cite{Sasieta:2022ksu}, this expression was shown to agree precisely with a Euclidean wormhole saddle in $D$-dimensional Einstein gravity and matter, for a specific class of heavy operator $\Op$ for which the integral \eqref{eq:1pointvariance} has a saddle point. Similar calculations have been done in JT gravity  \cite{Saad:2019pqd,Blommaert:2020seb,Jafferis:2022wez} as well as in AdS$_3$ \cite{Chandra:2022bqq,Collier:2023fwi}. This corroborates the principle of maximum ignorance as a useful guideline to understand the wormhole configurations of the gravitational path integral.

\restoregeometry

\subsection*{Example 3: Averaging over states}
Motivated by the above two examples, we set out to apply the same logic to ensembles of states. This will be the topic of the rest of the paper. Instead of an ensemble of Hamiltonians or operators, we consider a single theory with a fixed chaotic Hamiltonian $H$ with energy eigenstates $\ket{E_1},\ket{E_2},\dots,\ket{E_L}$ and spectral density $\varrho(E) = \e^{S(E)}$, as well as a fixed set of low-energy operators. For the application to holographic CFTs we will moreover be interested in the limit $L\to \infty$. We restrict ourselves to the chaotic part of the spectrum (e.g. above the black hole threshold) where we can approximate the density of states $\varrho(E)$ by a smooth function (like the Cardy density). We then consider random density matrices $\rho$ whose \emph{average} properties agree with a given semiclassical state, but whose fine-grained matrix elements $\rho_{ij} = \bra{E_i}\rho\ket{E_j}$ are treated statistically. In other words, the goal is to 

\begin{mdframed} 
construct a maximally entropic probability distribution $\mu(\rho)$ on the space of density matrices, which reproduces single-boundary low-energy input data on average.
\end{mdframed}
By `low-energy input' we mean any smooth function that is in principle available from a semiclassical gravity computation. Examples include partition functions, gravitational overlaps $\braket{\Psi}{\Psi}$ or expectation values of simple operators $\expval{\mathcal{O}}_\rho$. The resulting ensemble $\mu(\rho)$ will, of course, depend on the specific input we decide to use.

When we have found such distribution,
we define state averages by a weighted integral over the space of density matrices $\mathcal{D}$,
\begin{equation}
    \overline{(\bullet)} \coloneqq \int_{\mathcal{D}} \dd\rho\, \mu(\rho) (\bullet).
\end{equation}
For $\bullet = \rho$, this procedure gives a coarse-graining map $\rho \mapsto \bar \rho$. The mean $\bar \rho$ may be, for example, the thermal density matrix $\rho_\beta$. Holographically, this represents the fact that many microscopic configurations are indistinguishable from equilibrated black holes as probed by semiclassical observables. However, one can also compute averages of \emph{non-linear} quantities in $\rho$, such as the average von Neumann entropy $ -\overline{\Tr(\rho\log\rho)}$, moments $\overline{\Tr(\rho^k)}$, tensor products $\overline{\rho\otimes \rho}$ and multi-trace averages like 
\begin{equation}\label{multitrace_average}
    \overline{\Tr(\rho \mathcal{O})\Tr(\rho \mathcal{O})} - \Tr(\bar \rho\mathcal{O})\Tr(\bar \rho\mathcal{O}),
\end{equation}
which was dubbed the `quantum deviation' in \cite{Freivogel:2021ivu}. In general, the average tensor product state does not factorize into a product of averages, which we want to interpret holographically as due to the presence of specific Euclidean wormholes in gravity. In other words, our second goal is to
\begin{mdframed} 
match the connected contribution to multi-trace averages computed using $\mu(\rho)$ to a multi-boundary Euclidean wormhole computation in semiclassical gravity.
\end{mdframed}
Note that non-factorization of products of observables like \eqref{multitrace_average} should be contrasted to the Wilsonian notion of coarse-graining. In a local QFT, the coarse-grained ground state $\bar \rho_{\Lambda}$
at scale $E = \Lambda$ gives the same correlation functions for low-energy observables $\mathcal{O}_\Lambda$ as the ground state $\rho$ of the UV fixed
point, $\Tr_{\mathcal{H}}(\rho\hspace{0.2mm} \mathcal{O}_\Lambda) = \Tr_{\mathcal{H}_{\Lambda}}(\bar\rho_\Lambda \mathcal{O}_\Lambda)$.
 However, the variance of low-energy observables vanishes:
\begin{equation}
    \overline{\Tr_{\mathcal{H}}(\rho\hspace{0.2mm} \mathcal{O}_\Lambda)\Tr_{\mathcal{H}}(\rho\hspace{0.2mm} \mathcal{O}_\Lambda)}^{\mathrm{QFT}} = \Tr_{\mathcal{H}_{\Lambda}}(\bar\rho_\Lambda \mathcal{O}_\Lambda)\Tr_{\mathcal{H}_{\Lambda}}(\bar\rho_\Lambda \mathcal{O}_\Lambda),
\end{equation}
because integrating out short distance modes cannot generate relevant non-local correlations between different
systems.
Universality of the RG flow implies that this holds even at the non-perturbative level in any of the relevant couplings of the low-energy EFT.
On the other hand, in gravity, integrating out microscopic details induces relevant non-perturbative effects on IR observables. This is often called `UV/IR mixing'. Even while microscopic states coincide with the semiclassical state at perturbative orders in $1/S$, they differ at order $\exp(-S)$. These non-perturbative (in $G_{\text N}$) effects can be probed statistically in the semiclassical theory via the action of a spacetime instanton.

\subsection{Overview and summary of results} 

In section \ref{sec:SAA}, we explore one particular ensemble $\mu(\rho)$ that meets the goals formulated above. It is induced by the canonical purification $\rho = AA^\dagger$, which always exists by the positivity of $\rho \geq 0$. The matrix $A$ describes the expansion coefficients of the purification 
\begin{equation}
    \ket{\Psi} = \sum_{i\alpha}A_{i\alpha}\ket{E_i} \ket{E_\alpha}, \qquad \rho = \Tr_{R}\ket{\Psi}\bra{\Psi}
\end{equation}
living in the doubled Hilbert space $\mathcal{H}_L\otimes \mathcal{H}_R$. To get an ensemble of random mixed states on $\mathcal{H}_L$, we want to sample the coefficients $A_{i\alpha}$ from some probability distribution $\mu(A)$. This induces a probability distribution on the space of density matrices \cite{Zyczkowski_2001} as 
$
    \mu(\rho) = \int \dd A \,\mu(A) \,\delta(\rho - AA^\dagger) ,
$
where we have introduced a matrix delta function and the integral is over the space of complex random matrices with fixed norm $\Tr AA^\dagger = 1$.  An advantage of working with an ensemble of $A$'s is that the positivity constraint on $\rho$ is automatic.

Next, we pick a state $\ket{\Psi_{\mathrm{sc}}}$ in the doubled Hilbert space for which a semiclassical bulk dual exists, and use this semiclassical state as our reference state. Examples to have in mind are the thermofield double $\ket{\mathrm{TFD}_\beta}$, dual to the eternal two-sided black hole \cite{Maldacena:2001kr}, or an excitation of $\ket{\mathrm{TFD}_\beta}$ by a simple low-energy operator.
Given such a semiclassical state $\ket{\Psi_{\mathrm{sc}}}$ in the boundary CFT, we evolve it in Euclidean time with the left and right Hamiltonians $H_L = H \otimes \mathds{1}$ and $H_R = \mathds{1}\otimes H$, and then compute its norm. Pictorially,
\begin{equation}\label{eq:norm_input}
    \mathbf{Z}(\tau_1,\tau_2) \coloneqq \bra{\Psi_{\mathrm{sc}}}\e^{-\tau_1 H_L}\e^{ -\tau_2 H_R}\ket{\Psi_{\mathrm{sc}}}= \vcenter{\hbox{\begin{tikzpicture}\,\node at (0,0) {\includegraphics[width=4.8cm, valign=c]{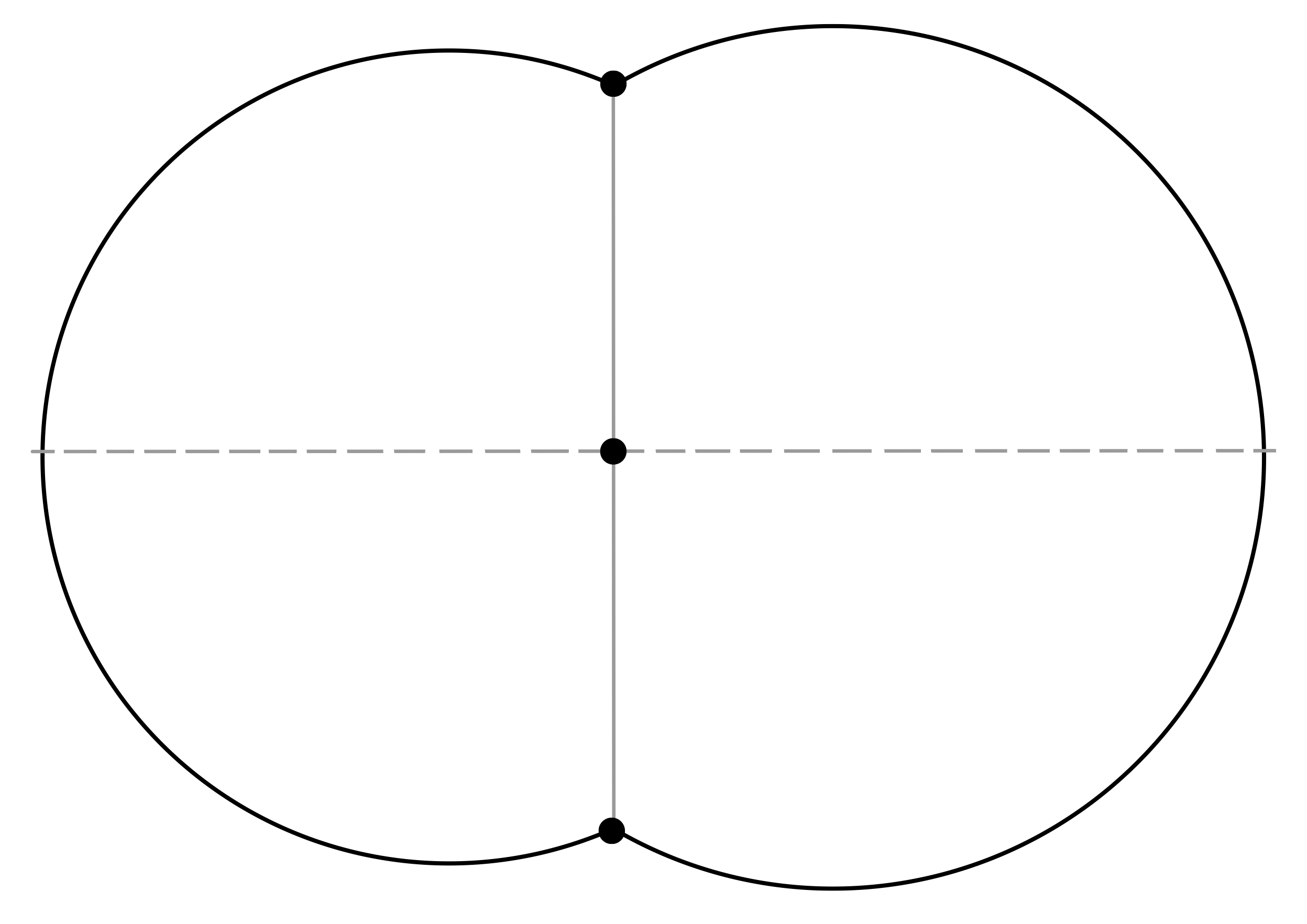}};
    \node at (-2.7,0){$\tau_1$};
    \node at (2.7,0) {$\tau_2$};
    \end{tikzpicture}}}.
\end{equation}
This produces a smooth function $\mathbf{Z}(\tau_1,\tau_2)$ which we will use as our input from gravity. Namely, the overlap \eqref{eq:norm_input} is computed holographically by a single-boundary Euclidean gravitational path integral filling in the bulk.
If we pick the reference state to be $\ket{\mathrm{TFD}_\beta}$, the right-hand side is simply the partition function $Z(\beta+\tau_1+\tau_2)/Z(\beta)$. However, for now, we will keep $\mathbf{Z}(\tau_1,\tau_2)$ generic input data, and explore the ensemble of states resulting from it.

The general idea of this construction is that $\mu(A)$ defines an ensemble of microscopic states that are indistinguishable from $\ket{\Psi_{\mathrm{sc}}}$  up to $\e^{-S}$ accuracy, but which contain non-perturbative `noise' encoded in the coefficients $A_{i\alpha} = \braket{E_i,E_\alpha}{\Psi}$. Given a random state $\ket{\Psi}$ in this ensemble, the overlap $\bra{\Psi}\e^{-\tau_1 H_L}\e^{-\tau_2 H_R}\ket{\Psi}$ is sensitive to the non-perturbative pattern of correlations encoded in these coefficients. However, we demand that the average overlap agrees with the semiclassical input:
\begin{equation}\label{eq:overlap_input}
    \int \dd A \,\mu(A) \,\bra{\Psi}\e^{-\tau_1 H_L}\e^{-\tau_2 H_R}\ket{\Psi}\, \stackrel{!}{=}\, \mathbf{Z}(\tau_1,\tau_2).
\end{equation}
We then find the maximally ignorant distribution that satisfies this constraint. That is, we perform a constrained optimization of the Shannon entropy of $\mu(A)$ and demand that the average overlap is fixed to \eqref{eq:norm_input} using a Lagrange multiplier $\lambda(\tau_1,\tau_2)$. This results in the maximum ignorance ensemble
\begin{equation}\label{eq:state_ensemble_intro}
\begin{split}
    \mu(A) &= \frac{1}{\mathcal{N}}\,\exp(-\int \dd\tau_1\dd\tau_2\, \lambda(\tau_1,\tau_2)  \braket{\Psi_{\tau_1,\tau_2}}{\Psi_{\tau_1,\tau_2}}) \\[1em]
    &= \frac{1}{\mathcal{N}}\,\exp(-\frac{1}{2}\sum_{i,\alpha} \sigma(E_i,E_\alpha)^{-1} |A_{i\alpha}|^2).
\end{split}
\end{equation}
In the second line, we wrote the potential in the energy eigenbasis $\ket{E_i}\ket{E_\alpha}$ of the doubled Hilbert space, and we defined the function $\sigma(E_1, E_2)^{-1}$ to be the double Laplace transform of $\lambda(\tau_1,\tau_2)$. As before, $\mathcal{N}$ is a normalization, while $\sigma(E_1,E_2)$ is fixed in terms of the gravitational input $\mathbf{Z}(\tau_1,\tau_2)$ by solving the constraint equation \eqref{eq:overlap_input}.

Now, our goal is to study the ensemble of mixed states induced by $\mu(A)$. For ease of exposition, we treat the coefficients $A_{i\alpha}$ in the ensemble \eqref{eq:state_ensemble_intro} as independent Gaussian random variables with variance $\sigma(E_i,E_\alpha)$.\footnote{However, the normalization condition $\Tr AA^\dagger =1$ induces small statistical correlations at subleading order in $\e^{-S}$. These correlations will be addressed in section \ref{sec:SAA_2}. For now, we only impose the average normalization condition $\overline{\Tr AA^\dagger} =1$, which constrains the variance function to satisfy $\sum_{i\alpha}\sigma(E_i,E_\alpha)=1.$} It is then easily seen that the mean $\bar \rho$ of this ensemble is diagonal in the energy eigenbasis, 
\begin{equation}\label{eq:mean_intro}
    \bar \rho_{ij} = \delta_{ij} \sum_{\alpha} \sigma(E_i,E_\alpha) \coloneqq \delta_{ij}\,\bar \rho(E_i).
\end{equation}
So the mean density matrix is an equilibrium state.\footnote{In section \ref{sec:saa24} we discuss ensembles with a time-dependent mean $\bar \rho(t)$. These can arise if we probe the reference state with local operators that do not commute with the Hamiltonian, providing them as additional input.} Technically, this arises thanks to the left-right $U(1)^L\times U(1)^L$ invariance of the potential \eqref{eq:state_ensemble_intro}. While the volume element $\dd A$ itself has a full left-right unitary invariance $A \to U_L A U_R$, the measure $\mu(A)\dd A $ breaks this symmetry to those unitaries which satisfy $[U_L,H_L] = [U_R,H_R] = 0$. In an infinitely narrow energy band, the function $\sigma(E_i,E_\alpha)$ becomes a constant and the full $U(L)\times U(L)$ invariance is restored. However, in general the ensemble depends non-trivially on the band structure through the energy dependent variance, similar to what happens in ETH.

To study the statistics of the induced ensemble, we define the fluctuation $\delta\rho = \rho - \bar \rho$. The moments of its matrix elements $\delta\rho_{ij} = \bra{E_i}\delta\rho \ket{E_j}$ are found by doing the Wick contractions in $A$. For example,
\begin{equation}\label{eq:second_moment}
\begin{split}
    \overline{\delta \rho_{ij}\delta\rho_{kl}} &= \sum_{\alpha,\beta} \wick{\c1 A_{i\alpha} \c2 A^*_{j\alpha} \c2 A_{k\beta} \c1 A^*_{l\beta}}\\[1em]  &= \delta_{il}\delta_{jk} \sum_{\alpha} \sigma(E_i,E_\alpha)\sigma(E_j,E_\alpha).
\end{split}  
\end{equation}
Even without plugging in an explicit expression for $\mathbf{Z}(\tau_1,\tau_2)$ or $\sigma(E_1,E_2)$, one can already note some interesting properties of the induced ensemble $\mu(\rho)$. First, the index structure implied by \eqref{eq:second_moment} is such that the second moment $\overline{\rho_{ij}\rho_{kl}}$ receives a connected contribution  ($\delta_{il}\delta_{jk}$) that is swapped with respect to the disconnected contribution ($\delta_{ij}\delta_{kl}$). This same phenomenon was noted in the context of the Page curve \cite{Marolf:2021ghr} and is standard whenever small microcanonical averages are Haar-typical \cite{Pollack:2020gfa}. 

Moreover, higher moments receive connected contributions from cyclic contractions of the indices:
\begin{equation}
\begin{split}
    \overline{\delta \rho_{ij} \delta \rho_{kl} \delta \rho_{mn}}^{\,\mathrm{conn}.} &= \sum_{\alpha,\beta,\gamma} \wick{\c1 A_{i\alpha} \c2 A^*_{j\alpha} \c2 A_{k\beta} \c3 A^*_{l\beta} \c3 A_{m\gamma} \c1 A^*_{n\gamma}} \\[1em]
    &= \delta_{in}\delta_{jk}\delta_{lm} \sum_\alpha \prod_{\bullet = i,j,l} \sigma(E_{\bullet},E_\alpha).
\end{split}
\end{equation}
So, even though the $\mu(A)$ ensemble \eqref{eq:state_ensemble_intro} is Gaussian, the induced ensemble $\mu(\rho)$ receives non-Gaussian corrections to higher moments of $\delta \rho$. These non-Gaussianities have precisely the same index structure as the generalized ETH ensemble of \cite{Foini:2018sdb} for the statistics of $\mathcal{O}_{ij}$.

In fact, the cumulants of $\delta\rho_{ij}$ satisfy a very similar hierarchy in terms of their scaling with $\e^{-S}$ as in generalized ETH. A simple counting argument in a microcanonical window around energy $E$ shows that the average trace condition implies $\sigma(E_1,E_2) = O\left(\e^{-2S(E)}\right)$, so that the $n$th moment has order
\begin{equation}\label{eq:non_gauss_intro}
    \overline{\delta \rho_{i_1i_2}\delta \rho_{i_2i_3}\cdots \delta \rho_{i_ni_1}}^{\,\mathrm{conn}.} = {\LaTeXunderbrace{\sum_\alpha }_{O\left(\e^S\right)} \LaTeXunderbrace{\prod_{\bullet = i_1,\dots,i_n} \sigma(E_{\bullet},E_\alpha)}_{O\left(\e^{-2n S}\right)}} = O\left(\e^{-(2n-1)S}\right).
\end{equation}
In other words, each higher cumulant of $\delta \rho$ is further exponentially suppressed in the system's microcanonical entropy.
We will now make this scaling precise, by positing an \emph{ansatz} for the variance function $\sigma(E_i,E_\alpha)$. 

\subsection*{The state-averaging ansatz}

In sections \ref{sec:SAA_2} and \ref{sec:saa23}, we examine specific examples and incorporate some assumptions about the gravitational input $\mathbf{Z}(\tau_1,\tau_2)$ and the allowed semiclassical states $\ket{\Psi_{\mathrm{sc}}}$. We consider reference states $\ket{\Psi_{\mathrm{sc}}}$ with a mean energy $E_* = (\expval{H_L}_{\mathrm{sc}}+\expval{H_R}_{\mathrm{sc}})/2$ above the black hole threshold and a corresponding inverse temperature $\beta \coloneqq S'(E_*)$. This is part of the input data, as $E_*$ is equal to minus the symmetrized first derivative of $\mathbf{Z}(\tau_1,\tau_2)$. We also assume that the state is in equilibrium $\expval{H_L}_{\mathrm{sc}}= \expval{H_R}_{\mathrm{sc}}$ and that the energy variance is small, $\expval{(H_L-H_R)^2}_{\mathrm{sc}} = \delta^2 \ll E_*^2$. The latter is a condition on the second derivative of the overlap $\mathbf{Z}(\tau_1,\tau_2)$. 

Under these assumptions, we then posit the following ansatz for the variance:
\begin{equation}\label{eq:variance_ansatz}
    \sigma(E_i,E_\alpha) = \frac{\e^{-\beta \bar E_{i\alpha}}}{Z(\beta)}\, \e^{-S(\bar E_{i\alpha})} j_0(\bar E_{i\alpha},\omega_{i\alpha};\beta),
\end{equation}
where $\bar E_{i\alpha} = (E_i+E_\alpha)/2$ and $\omega_{i\alpha} = E_i-E_\alpha$. The first piece is the usual Boltzmann factor, evaluated at the average energy. The function $j_0$ is a smooth function of its arguments, peaked at $\omega = 0$ with a characteristic width $\delta$. It is moreover weighted by a factor of the inverse density of states.

In a microcanonical window of size $L = \e^{S(E)}$ where $\bar E_{i\alpha} \approx E$ for all $i,\alpha$, the right-hand side of \eqref{eq:variance_ansatz} becomes $\e^{-2S(E)}$ as expected
. However, the ansatz \eqref{eq:variance_ansatz} postulates a non-trivial energy-dependence outside the microcanonical energy band. The most important difference with the fully unitary invariant average (e.g. \cite{Page:1993df,Page:1993wv,Pollack:2020gfa,Kudler-Flam:2021rpr}) is that the variance remains finite in the $L\to \infty$ limit. This feature makes the ansatz suited for e.g. CFTs on the cylinder. 

In section \ref{sec:SAA_2}, we analyze this ansatz for various choices of reference state. In the case that $\bar\rho$ is approximately thermal, the function $j_0$ is an envelope function constraining $E_i$ to be close to $E_\alpha$. Plugging the ansatz into \eqref{eq:mean_intro} and \eqref{eq:second_moment} we derive the following expression for the matrix elements of the reduced density matrix, which we dub the \emph{state-averaging ansatz}:
\vspace{2mm}
\begin{mdframed}
\begin{equation}\label{eq:state_averaging_ansatz}
    \expval{E_i|\hspace{0.1mm}\rho\hspace{0.1mm}|E_j} = \delta_{ij} \,\bar\rho(E_i) + \frac{\e^{-\beta \bar E_{ij}}}{Z(\beta)}\e^{-S(\bar E_{ij})/2}j_2(\bar E_{ij},\omega_{ij};\beta)^{1/2} R_{ij}.\hspace{0.5cm}
\end{equation}
\end{mdframed}
On the left-hand side is what we want to model, namely the matrix elements of $\rho$ in energy eigenstates. On the right-hand side the structure is very similar to ETH: it has a smooth diagonal piece and small off-diagonal noise. Indeed, one can view \eqref{eq:state_averaging_ansatz} as a `canonical' version of ETH, suited for modeling the microscopic density matrix of a large black hole. 

The mean $\bar\rho$ is close to $\rho_\beta$ (with small non-random corrections of order $1/S$), while $R_{ij}$ is an approximately Gaussian random variable with mean zero and variance one. The exponentially suppressed non-Gaussianities are determined by \eqref{eq:non_gauss_intro}. The function $j_2(\bar E,\omega;\beta)$ is a convolution integral of the function $j_0$ and varies smoothly with $\bar E_{ij} = (E_i+E_j)/2$ and $\omega_{ij} = E_i-E_j$. In section \ref{sec:saa23} we relate $j_2$ to the average relative entropy $\overline{S(\rho||\bar\rho)} $ between a member of the ensemble and the mean. It thus controls how well a typical state $\rho$ can be distinguished from the thermal reference state. 

\subsection*{Generalization with conformal symmetry} 

In the remainder of section \ref{sec:SAA}, we present various generalizations of the above construction. The main refinement is to distinguish between conformal primaries and descendant states. Namely, the primary sector of a holographic CFT above the black hole threshold is believed to be chaotic, whereas the descendants are fully fixed by conformal symmetry and should not be treated statistically. In section \ref{sec:conformalSAA} we present a modification of the state-averaging ansatz specifically for 2d CFTs, 
\begin{equation}
    \bra{ h, N,\bar h, \bar N} \delta\rho \ket{h', N', \bar h', \bar N'} = f_{N,N'}f_{\bar N,\bar N'}
   \e^{-\beta(|N|+|\bar N|)} \bra{ h,\bar h} \delta\rho \ket{ h',\bar h'},
\end{equation}
where the matrix elements of $\delta \rho$ between primary states $\ket{h,\bar h}$ are similar to \eqref{eq:state_averaging_ansatz} but descendants are non-random. In particular, the functions $f_{N,N'}$ (where $N,N'$ label the level of the descendant) are fully fixed by conformal symmetry. This modification will prove to be important when we study wormholes in AdS$_3$ gravity.

\subsection*{Matching the statistics of $\rho$ to semiclassical wormholes} 

Next, in sections \ref{sec:wormholes} and \ref{sec:higher_d}, we relate statistical variances computed using the state-averaging ansatz \eqref{eq:state_averaging_ansatz} to semiclassical wormholes in gravity. Specifically, in section \ref{sec:wormholes} we focus on wormholes in pure AdS$_3$ gravity with conical defects, which were matched to an ensemble of OPE data in \cite{Chandra:2022bqq}. We discuss the interplay between state averaging and operator averaging, which provides an interpretation of the AdS$_3$ wormholes in \cite{Chandra:2022bqq} as arising from averaging over states in a single CFT$_2$, instead of an average over OPE coefficients in an ensemble of theories.\footnote{This is in the same spirit as \cite{Chandra:2023rhx}. We discuss the differences with their construction in section \ref{sec:wormholes}.} 

This interplay, in its most basic form, is illustrated by the following example. Suppose we want to capture the statistical variance of a thermal one-point function of a simple operator $\mathcal{O}$, either by using ETH for its matrix elements, or by using the state-averaging ansatz centered at the thermal state $\bar \rho = \rho_\beta$. Then, the factorized contribution will give the square $\langle \mathcal{O}\rangle_\beta^2$ in both cases, while the connected contributions give different predictions: 
\begin{align}\label{state_average_wh}
   \overline{\Tr (\rho \mathcal{O})\Tr(\rho \mathcal{O})}^{\,\mathrm{conn.}} &= \sum_{ij}\overline{|\delta\rho_{ij}|^2}\,\mathcal{O}_{ij}\mathcal{O}_{ji} \qquad \hspace{1cm}\mathrm{(state \, averaging)} \\ \label{operator_average_wh}
    \opmean{\Tr(\rho_{\beta} \mathcal{O}) \Tr(\rho_{\beta} \mathcal{O}) }^{\,\mathrm{conn.}} &= \sum_{i} (\rho_{\beta})_{ii}^2 \opmean{\mathcal{O}_{ii}\mathcal{O}_{ii}}^{\,\mathrm{conn.}}\qquad \mathrm{(operator \, averaging).}  
\end{align}
We wrote the right-hand side in the energy eigenbasis to highlight the different index structures. The first line (state average) arises from the index contraction \eqref{eq:second_moment}, resulting in a double sum. The second line (ETH) arises from the contraction $\overline{R_{ii}R_{jj}}$ which sets $i=j$, resulting in a single sum. 

Despite this apparent difference, the two computations agree in the thermodynamic limit, if we plug in the state-averaging ansatz. In the continuum approximation, the double sum is replaced by $\int \dd \bar E \dd \omega\,\e^{2S(\bar E)}$, and if we assume that $B_{\mathcal{O}}(\bar E_{ij},\omega_{ij}) \coloneqq \mathcal{O}_{ij}\mathcal{O}_{ji}$ varies smoothly with $\bar E$ and $\omega$ then we obtain\footnote{Note that this amounts to a smearing of nearby energy levels in $B_\mathcal{O} =\mathcal{O}_{ij}\mathcal{O}_{ji}$, in the same way that one replaces the discrete sum over states in $\varrho(E)$ by a smooth function. In particular, we are not statistically averaging over the matrix elements $\mathcal{O}_{ij}$ in this case, so the connected `wormhole' contribution only comes from the statistics of $\delta \rho_{ij}$.}
\begin{equation}\label{eq:state_avg_wh}
    \overline{\Tr (\rho \mathcal{O})\Tr(\rho \mathcal{O})}^{\,\mathrm{conn.}} \simeq \frac{1}{Z(\beta)^2}\int \dd \bar E\dd \omega \,\e^{S(\bar E)}\e^{-2\beta \bar E} j_2(\bar E ,\omega;\beta) B_{\mathcal{O}}(\bar E,\omega).
\end{equation}
The inverse density of states $\e^{-S(\bar E)}$ appearing in the state-averaging ansatz has canceled one of the factors of $\e^{S(\bar E)}$ from the double sum. Moreover, since the function $j_2$ is peaked at $\omega = 0$, we can do a saddle-point approximation of the $\omega$-integral, which becomes exact when $\delta \to 0$. If $j_2$ depends weakly on $\bar E$ and is normalized to $1$ at $\omega=0$, the state average \eqref{eq:state_avg_wh} approximately agrees with the operator average. In fact, if one takes as the semiclassical state the thermofield double, $\ket{\Psi_{\mathrm{sc}}} = \ket{\mathrm{TFD_\beta}}$, the two predictions \eqref{state_average_wh} and \eqref{operator_average_wh} match exactly (see section \ref{sec:SAA}).

As a simple illustration, we evaluate the right-hand side of \eqref{eq:state_avg_wh} in the case of JT gravity with matter \cite{Jafferis:2022wez} in the semiclassical approximation, for which $\varrho(E) = \sinh(2\pi\sqrt{E})$. In the boundary theory, take an operator with large scaling dimension $\Delta$, for which the function $B_{\mathcal{O}}$ takes the following form \cite{Mertens:2017mtv}
\begin{equation}\label{eq:operator_JT}
    B_{\mathcal{O}}(\bar E,\omega) = \frac{\Gamma\left(\Delta \pm i\sqrt{\bar E + \frac{\omega}{2}}\pm i\sqrt{\bar E - \frac{\omega}{2}}\right)}{2^{2\Delta-1}\,\Gamma(2\Delta)}.
\end{equation}
If we plug this into \eqref{eq:state_avg_wh} and evaluate the integral on the saddle point $\omega=0$, it matches the wormhole constructed in \cite{Stanford:2020wkf},
\begin{equation}\label{eq:JT_wormhole}
    Z_{\mathrm{WH}} = \int_{0}^\infty \dd \ell\,\e^{-\Delta \ell} \e^{-2\beta E(\ell)},
\end{equation}
where $2\beta E(\ell)$ is the action of a `constrained instanton' connecting two boundaries of length $\beta$, where $\ell$ is a fixed length between the two boundaries. The factor $\e^{-\Delta \ell}$ arises from the geodesic approximation of the particle going from one boundary to the other, and $E(\ell)$ can be determined exactly in JT gravity, see \cite{Stanford:2020wkf}. We want to 
view this result as an illustration that wormholes can also correctly capture the variance of $\expval{\mathcal{O}}_\rho$ in an ensemble of states, rather than operators.

In section \ref{sec:punct-tor-worm} we generalize the above computation to the punctured torus wormhole in AdS$_3$, with a conical defect connecting the two torus boundaries. For the function $B_{\mathcal{O}}$ we now take the universal 2-point function of OPE coefficients  derived in \cite{Collier:2019weq}. Instead of averaging over OPE coefficients, we use the state-averaging ansatz with conformal symmetry and find a match with the punctured torus wormhole amplitude,
\begin{equation}
\overline{\Tr(\rho\mathcal{O})\Tr(\rho\mathcal{O})}^{\mathrm{\,conn.}} = \eqimg{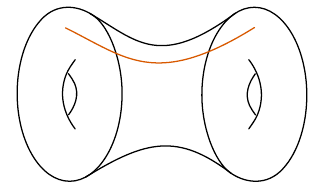}.
\end{equation}
Again, the basic reason is due to the compatibility of \eqref{state_average_wh} and \eqref{operator_average_wh}, now generalized to CFT$_2$.

This same basic mechanism can be extended to more general observables which are not thermal one-point functions, such as the genus-two partition function (section \ref{sec:genus-two-worm}), the torus two-point function or the sphere four- and six-point functions (section \ref{sec:gen-obs}). Namely, we use the plumbing construction to write these observables as a trace in a tensor product Hilbert space
\cite{Belin:2021ibv, Cho:2017oxl},
\begin{equation}
 \langle \cdots \rangle \quad \longrightarrow \quad   \Tr_{\mathcal{H}^{\otimes \mathsf{k}}}(\rho_{\mathsf{k}}\hspace{0.1mm} \mathbb{O}). 
\end{equation}
The operator $\mathbb{O}$ encodes the relevant OPE coefficients in its matrix elements.\footnote{E.g., for the genus-two partition function we have $\mathsf{k}=3$ and $\bra{ \mb h, \mb N} \mathbb{O} \ket{\mb h', \mb N'} = C_{123} C_{1'2'3'} \;\sigma(\mb h, \mb N)\sigma(\mb h', \mb N')$. The bold notation is vector notation $\mb h = (h_1,h_2,h_3)$, and $\sigma(\mb h, \mb N)$ is a smooth function fully determined by Virasoro symmetry.}
We then use a generalization of the state-averaging ansatz for density matrices $\rho_{\mathsf{k}}$ in tensor product Hilbert spaces, centered at the thermal product state $\overline{ \rho_{\mathsf{k}}} = \rho_{\beta_1}\otimes \cdots \otimes \rho_{\beta_{\mathsf{k}}}$, where $\beta_1,\dots,\beta_{\mathsf{k}}$ encode the moduli of the CFT observable. In this way, we are able to match state averages to the operator averages computed using the OPE ensemble of \cite{Chandra:2022bqq}, and reinterpret the wormholes constructed there in terms of an ensemble of random states. 

\subsection*{Non-Gaussianities and multi-boundary wormholes}

Besides matching the statistics of the state-averaging ansatz to known wormholes in the literature, we also naturally obtain predictions for new semiclassical wormholes that should capture the non-Gaussianities in $\mu(\rho)$. Following \eqref{eq:non_gauss_intro}, the state-averaging ansatz predicts connected correlations of the form 
\begin{equation}
\label{statecorre}
\mean{\delta\rho_{i_1i_2}\delta\rho_{i_2i_3}\dots\delta\rho_{i_k i_1}} = \frac{\e^{-k \beta \bar E}}{Z(\beta)^{k}} \e^{-(k-1)S(\bar E_k)} j_k(\bar E_k,\omega;\beta). 
\end{equation}
where we defined $\bar E_k = \frac{1}{k}(E_1+\dots+E_k)$. The functions $j_{k}$ are smooth functions of $\bar E_k$ and the energy differences $\omega$. These moments play a role in multi-boundary wormholes contributing to 
\begin{equation}
    \overline{\expval{\mathcal{O}}_\rho\cdots \expval{\mathcal{O}}_\rho}^{\,\text{conn.}}.
\end{equation}
Interestingly, the cubic cumulant 
predicts a three-boundary wormhole in AdS$_3$, given in \eqref{eq:manybworm}, which is not predicted by the Gaussian ensemble of OPE coefficients of \cite{Chandra:2022bqq}. Similarly, the quartic cumulant predicts a four-boundary wormhole contributing to $\overline{\expval{\mathcal{O}}^4}$ which differs from the quartic moment of \cite{Belin:2023efa} when $\mathcal{O}$ is a light operator. We give a tentative bulk interpretation to these higher cumulants in a follow-up paper \cite{deBoer:2024kat}, in which we explore the interplay between modular invariance and typicality.

\subsection*{More general semiclassical states}

In section \ref{sec:higher_d}, we generalize the state-averaging ansatz to describe more general semiclassical states of the black hole. We focus on two-sided black hole states with long semiclassical Einstein-Rosen bridges supported by heavy matter. The bulk state that we aim to model is the semiclassical dual to one of the two holographic CFTs in this setup. 

We first show that the principle of maximum ignorance leads to an ensemble of microstates captured by the state-averaging anstaz \eqref{eq:state_averaging_ansatz}. In this case, there is an important modification. Namely, the smooth function $j_2(\bar E_{ij},\omega_{ij};\beta)$ now scales with the microcanonical entropy. This scaling represents the fact that the semiclassical state in the black hole interior has been backreacted by the heavy matter, and thus is a large perturbation of the semiclassical state of an eternal black hole.

We show that, when the Einstein-Rosen bridge is large, the coarse-graining map $\rho \rightarrow \overline{\rho}$ reproduces the semiclassical state of the exterior of the black hole. The semiclassical features of the interior are encoded in the smooth function $j_2(\bar E_{ij},\omega_{ij};\beta)$. This part of the state-averaging ansatz can be signaled semiclassically in non-linear properties of the state. For instance, we show that the average purity $\overline{\text{Tr}(\rho^2)}$ admits semiclassical corrections from $j_2(\bar E_{ij},\omega_{ij};\beta)$ which are large, given by replica wormholes of the gravitational path integral in the bulk computation of the purity. This effect can make the relative entropy $\overline{S(\rho||\bar\rho)}$ large. At the same time, we observe that variances of correlation functions $\overline{\Tr(\rho\mathcal{O})\Tr(\rho\mathcal{O})}$ over the ensemble admit semiclassical connected contributions, given by the action of multi-boundary wormhole configurations of the gravitational path integral. From this perspective, both classes of wormholes arise from the same microscopic model of the semiclassical state.

\subsection{Organization of the paper}

This paper is organized as follows. In section \ref{sec:SAA} we give further details about state-averaging and the state-averaging ansatz. We discuss the measure in the space of density matrices, give explicit examples, and discuss the interpretation of several information-theoretic quantities like the averaged von Neumann entropy. We finish this section with a discussion of state-averaging in the context of CFT$_2$.  In section \ref{sec:wormholes} we use the state-averaging ansatz to exactly reproduce several on-shell wormhole amplitudes in 3D gravity. We solve explicitly the examples of the torus wormhole and the genus-two wormhole. Then, we extend the discussion to include arbitrary correlation functions on generic Riemann surfaces.  In section \ref{sec:higher_d} we discuss the state-averaging ansatz for more general semiclassical states in $D\geq 3$. We consider a family of partially entangled thermal states (PETS) constructed from thin-shell operators in the bulk. The state-averaging interpretation of these states naturally connects thermal correlation functions to replica geometries and R\'{e}nyi entropies. We end with some conclusions in section \ref{discussion}. Additionally, we present some technical details about state-averaging and comment on off-shell configurations in the appendices.

\newpage
\section{An ensemble of density matrices}\label{sec:SAA}
In this section, we will formalize the notion of state averaging by constructing a measure $\mu(\rho)\dd\rho$ on the space of density matrices $\mathcal{D}$. If $L$ denotes the complex Hilbert space dimension, this space consists of all normalized and positive semi-definite Hermitian matrices of size $L\times L$,
\begin{equation}
    \mathcal{D} = \{\rho \in \mathrm{GL}(L,\mathbb{C}) \mid \rho = \rho^\dagger, \,\Tr\rho = 1,\,\rho\geq 0\}.
\end{equation}
This is a compact manifold with boundary, where $\partial \mathcal{D}$ consists of all pure states $\rho = \ket{\psi}\bra{\psi}$ and non-maximal rank density matrices. For an excellent introduction to the geometry of $\mathcal{D}$, see \cite{bengtsson_zyczkowski_2006}. A choice of metric on $\mathcal{D}$ determines a volume element $\dd\rho$. In this paper we will need only the Hilbert-Schmidt metric $ds^2 = \Tr \delta\rho^2$, for which $\dd\rho$ is the standard unitary invariant volume element on the space of Hermitian matrices.
We refer to Appendix \ref{app:volume_element} for a discussion of more general metrics on $\mathcal{D}$. By diagonalizing
$ \rho = U \Lambda U^\dagger$, 
the volume element similarly factorizes as 
\begin{equation}
    \dd\rho = [\dd U][\dd\Lambda],
\end{equation}
where $[\dd U]$ is the Haar invariant measure on $U(L)$ and the precise form of $[\dd\Lambda]$ is given in Appendix \ref{app:volume_element}. This decomposition shows that averaging over density matrices combines both spectral and basis uncertainty. In other words, the measure $\mu(\rho)\dd\rho$ introduces randomness both in the entanglement spectrum of $\rho$ as well as in the basis that diagonalizes it.

Given a probability distribution $\mu(\rho)$, we denote state averages by an overline
\begin{equation}
    \overline{(\bullet)} = \int_{\mathcal{D}} \dd\rho\, \mu(\rho) (\bullet).
\end{equation}
To incorporate the positivity constraint on $\rho$ and the trace condition, we can introduce a theta function $\Theta(\rho) = \prod_i \Theta(\lambda_i)$ for the eigenvalues $\lambda_i$ of $\rho$, and a delta function for its trace:
\begin{equation}
    \int_{\mathcal{D}}\dd\rho = \int \dd\rho \,\Theta(\rho)\, \delta(1-\Tr \rho). 
\end{equation}
As we motivated in the introduction, the distribution $\mu(\rho)$ should then be determined from a maximization of the Shannon entropy $ - \int \dd\rho \,\mu(\rho)\log \mu(\rho)$, with additional constraints. If we do not input any constraints, the maximum ignorance ensemble is easily found to be the constant distribution 
\begin{equation}\label{eq:HilbertSchmidt}
    \mu(\rho) = \frac{1}{\mathrm{vol}(\mathcal{D})},
\end{equation}
where we have normalized $\mu(\rho)$ such that $\int_{\mathcal{D}} \dd\rho \,\mu(\rho) = 1$. The ensemble \eqref{eq:HilbertSchmidt} is sometimes called the Hilbert-Schmidt ensemble \cite{Zyczkowski_2001}. It has a full unbroken $U(L)$ unitary invariance, which automatically implies that the mean $\bar \rho$ is the maximally mixed state $\bar \rho = \mathds{1}/L$. Variances around the mean are suppressed by powers of $1/L$, due to the phenomenon of measure concentration \cite{popescu2006entanglement}.

\subsection{Random purifications}\label{sec:random_purification}

A large class of measures $\mu(\rho)\dd\rho$ is induced by the canonical purification of $\rho$ in the doubled Hilbert space $\mathcal{H}_L\otimes \mathcal{H}_R$, where $\mathcal{H}\cong \mathcal{H}_L \cong \mathcal{H}_R$. We can then write 
\begin{equation}
    \rho = \Tr_R \ket{\Psi}\bra{\Psi} = AA^\dagger, 
\end{equation}
for a complex $L\times L$ matrix\footnote{We will always take the purification Hilbert space $\mathcal{H}_R$ to have the same dimension as the original $\mathcal{H} \coloneqq \mathcal{H}_L$, although the analysis can easily be extended to the case that $\mathrm{dim}(\mathcal{H}_R)\geq \mathrm{dim}(\mathcal{H}_L)$  \cite{Zyczkowski_2011}, in which case $A$ is rectangular.} $A$ with unit norm $\Tr AA^\dagger =1$, which is unique up to a left-right unitary action on $A \to U_L A U_R$. The matrix $A$ encodes the expansion coefficients of the purified state $\ket{\Psi}$ in the basis of the doubled Hilbert space. A \emph{random} purification can thus be obtained by sampling $A$ from some probability distribution $\mu(A)$ on the space of unit-norm complex matrices. 

Any choice of $\mu(A)$ induces a probability distribution $\mu(\rho)$ formally as \cite{Zyczkowski_2001}
\begin{equation}
    \mu(\rho) = \int \dd A \,\mu(A) \,\delta(\rho - AA^\dagger),
\end{equation}
where the right-hand side contains the matrix delta function.\footnote{For Hermitian matrices $X$, we define $\delta(X) =\prod_{i<j}\delta(X_{ij}) \delta(X^*_{ij})\prod_{k}\delta(X_{kk})$.}
On a practical level, this means that when computing any state average $\overline{F(\rho)}$ one simply substitutes $\rho = AA^\dagger$ and averages over $A$ with statistical weight $\mu(A)$. The volume element is given by
\begin{equation}
    \dd A = \prod_{i,\alpha}\dd A_{i\alpha}\dd A_{i\alpha}^*.
\end{equation}
The fixed norm condition on $A$ can be incorporated by the change of variables $A = (\Tr \mathsf{A} \mathsf{A}^\dagger)^{-1/2} \mathsf{A}$.
Equivalently, we can insert the Fourier representation of the delta function enforcing the constraint $\delta(\Tr AA^\dagger - 1)$ on each member of the ensemble. 

To see how this works in practice, it is instructive to consider a simple example. Without giving any input, the maximally ignorant $\mu(A)$ is the constant distribution, as before, and it is equivalent to \eqref{eq:HilbertSchmidt}. We can explicitly compute the mean of $\rho$ in the induced ensemble:\footnote{The factor of $\e^{-\epsilon \cdot 1}$ can be absorbed into $Z$, but we wrote it to make the Fourier transform in the second line well defined.}
\begin{equation}
\begin{split}
    \overline{\rho}_{ij} &= \frac{1}{\mathcal{N}}\int \dd A \,\e^{-\epsilon \Tr AA^\dagger}\delta(\Tr AA^\dagger-1) \,(AA^\dagger)_{ij} \\[1em]
    &= \delta_{ij} \,\frac{1}{\mathcal{N}}\int_{-\infty}^\infty \dd s\,\e^{is} \,\prod_{n,\nu} \int_{-\infty}^\infty  \dd A_{n\nu}\dd A_{n\nu}^* \,\e^{-(\epsilon +is)|A_{n\nu}|^2} \sum_{\alpha=1}^L |A_{i\alpha}|^2.
    \end{split}
\end{equation}
To go from the first to the second line, we wrote the delta function as a Fourier transform. Next we perform the $L^2$ independent Gaussian integrals, giving
\begin{equation}
 \label{eq:fourier_tr}
    \bar\rho_{ij}=\delta_{ij} \, \frac{1}{\mathcal{N}}\int_{-\infty}^\infty \dd s\,\e^{is}\, \left(\frac{2\pi}{\epsilon +is}\right)^{L^2} \frac{2\pi L }{\epsilon+is}.
\end{equation}
The sum over $\alpha$ gave another factor of $L$. Finally we use the Fourier transform
\begin{equation}\label{eq:Fourier_transform}
    \int_{-\infty}^\infty \dd s \,\e^{is} \frac{1}{(x+is)^b} = 2\pi\,\frac{\e^{-x}}{\Gamma(b)},
\end{equation}
to evaluate $\mathcal{N}$ as well as the integral \eqref{eq:fourier_tr}. This gives the ratio $ \frac{L(L^2-1)!}{(L^2)!} = \frac{1}{L}$, while the $\epsilon$-dependence drops out as it should. We conclude 
\begin{equation}
    \bar \rho_{ij} =  \frac{\delta_{ij}}{L}.
\end{equation}
So the ensemble is centered at the maximally mixed state. This could have already been anticipated by the fact that the induced measure $\mu(\rho)\dd\rho = [\dd U][\dd \Lambda]$ has a full $U(L)$ invariance in this case, which forces $\bar \rho\,\, \propto\,\, \mathds{1}$. The proportionality constant $1/L$ is then fixed by the normalization condition.

In this simple example, we can also straightforwardly compute the variance $\overline{\delta\rho_{ij}\delta\rho_{kl}}$ around the mean, where we defined $\rho = \bar \rho + \delta \rho$. A similar calculation as above now gives
\begin{equation}
\overline{\rho_{ij}^{}\rho_{kl}^{}} = \frac{1}{\mathcal{N}}\int_{-\infty}^\infty \dd s\,\e^{is} \,\prod_{n,\nu} \int_{-\infty}^\infty  \dd A_{n\nu}\dd A_{n\nu}^* \,\e^{-(\epsilon +is)|A_{n\nu}|^2} \sum_{\alpha,\beta} A_{i\alpha}A^*_{j\alpha}A_{k\beta}A^*_{l\beta}.
\end{equation}
Again, we do the Gaussian integrals, which give two distinct Wick contractions. The first sets $\delta_{ij}\delta_{kl}$ while the second gives the cross-contraction $\delta_{\alpha\beta}\delta_{il}\delta_{jk}$. Then we perform the Fourier transform using the formula \eqref{eq:Fourier_transform}. The result is a sum of a disconnected piece and a connected contribution with only a single sum over $\alpha$. The connected term evaluates to the ratio $L \cdot \frac{(L^2-1)!}{(L^2+1)!}$, and so we find
\begin{equation}\label{eq:variance_HS}
\overline{\delta\rho_{ij}\delta \rho_{kl}}  = \frac{\delta_{il} \delta_{jk}}{L(L^2+1)} -\frac{\delta_{ij}\delta_{kl}}{L^2(L^2+1)}.
\end{equation}
Notice that for large $L$, the typical fluctuation $\delta \rho_{ij}$ scales as $L^{-3 /2}$, and is therefore suppressed by a factor $L^{-1/2}$ compared to the mean. This is a manifestation of the phenomenon of concentration of measure \cite{popescu2006entanglement,Hayden_2006}: in the limit $L\to \infty$, the ensemble localizes to the mean with fluctuations that are suppressed in powers of $1/L$.\footnote{The scaling $\delta \rho \sim L^{-3/2}$ should be contrasted to the case of ensembles of \emph{pure} states, in which the variance is typically of the order $L^{-2}$. This is a simple consequence of the fact that the averaged purity $\overline{\Tr \rho^2}$, which contains information about the variance, is 1 for pure states but $1/L$ for maximally mixed states. }

A useful way to quantify how far a typical state $\rho$ is from the mean $\bar\rho$ is via the relative entropy 
\begin{equation}
    S(\rho || \bar \rho \hspace{0.2mm}) = \Tr \rho \log \rho - \Tr \rho \log \bar \rho.
\end{equation}
It is a measure of the distinguishability between $\rho$ and $\bar \rho$, as formalized by the Quantum Stein Lemma \cite{inbook, Ogawa:1999gv} in quantum hypothesis testing, which states that the optimal asymptotic rate of error in distinguishing a state $\rho$ from the `true' state $\sigma$ after $N$ measurements is given by
$\exp\left[-N S(\rho || \sigma)\right]$. 

Positivity of the relative entropy implies that the averaged von Neumann entropy is always smaller than or equal to the von Neumann entropy of the averaged state,
\begin{equation}\label{eq:relentr}
    \overline{S(\rho || \bar{\rho}\hspace{0.2mm})} = S(\bar{\rho}) - \overline{S(\rho)} \geq 0.
\end{equation}
This fact also underlies the Page curve \cite{Page:1993wv}.  
In an ensemble of black hole microstates (where $A$ is a rectangular matrix), the entropy of the coarse-grained state $\bar\rho$ gives the area term, which monotonically grows with $L$, while $\overline{S(\rho)}$ first grows and then comes back down.

We can compute the averaged relative entropy in the unitary invariant ensemble discussed above at large $L$, by expanding $S(\rho || \bar{\rho}\hspace{0.2mm})$  perturbatively around the mean. In Appendix \ref{sec:stateperturbations} we show that
\begin{equation}\label{eq:rel_ent_expansion}
    S(\rho || \bar \rho\hspace{0.2mm}) = \frac{1}{2}\sum_{ij}\frac{\log \lambda_i-\log\lambda_j}{\lambda_i-\lambda_j}|\delta \rho_{ij}|^2 \,+\, O(\delta\rho^3),
\end{equation}
where $\lambda_i$, $i=1,\dots,L$ denote the eigenvalues of $\bar \rho$. Notice that the constant and linear terms vanish, so that the relative entropy is approximately quadratic for small $\delta\rho$. We can now use the expression derived in \eqref{eq:variance_HS} for the variance, as well as $\lambda_i = 1/L$. The prefactor simplifies to $\frac{\log \lambda_i-\log\lambda_j}{\lambda_i-\lambda_j} \to L$, and the double sum gives a factor $L^2$. Hence the averaged relative entropy in the Hilbert-Schmidt ensemble is
\begin{equation}
    \overline{ S(\rho || \bar \rho\hspace{0.2mm})} =\log L - \overline{S(\rho)} \approx \half \,\frac{L^2-1}{L^2+1}.  
\end{equation}
For large $L$, we see that the average relative entropy is approximately $\half$. The corrections can be obtained either by computing higher-order terms in the expansion \eqref{eq:rel_ent_expansion}, or by using the replica trick \cite{PhysRevLett.72.1148}. In either case, the subleading corrections can be shown to scale as $O\left(\frac{\log L}{L}\right)$.

\subsection*{Main takeaway}
The positivity constraint $\rho\geq 0$ allowed us to write $\rho = AA^\dagger$ and study the ensemble induced by $\mu(A)$.  Without inputting any constraints, the maximally ignorant ensemble is the fully unitary invariant Hilbert-Schmidt ensemble. It is centered on the maximally mixed state, with a typical fluctuation of the order $L^{-3/2}$. Moreover, the average relative entropy is an order one number. 

This ensemble is maximally agnostic about the basis in which the states are diagonalized. This is because we have not given the ensemble any input that depends on the Hamiltonian. In the next section, we will see that if we do add such input, the unitary invariance gets broken. In other words, the ensemble becomes `less uncertain' about the preferred basis by adding single-trace input data. Besides basis uncertainty, state averaging also introduces spectral uncertainty in the eigenvalues of $\rho$. This is in principle different from averaging over the spectrum of $H$.  However, the entanglement spectrum may depend on the smooth band structure of the Hamiltonian, as we will now show.

\subsection{Semiclassical states and average overlaps}\label{sec:SAA_2}

In order to connect the above discussion to gravity, specifically in the context of AdS/CFT, we have to supply the ensemble of states in the boundary theory with additional low-energy gravitational bulk input. As motivated in the introduction, a very general type of input to consider are gravitational overlaps between states with a gravitational bulk dual. In this subsection we will analyze the maximum-ignorance ensemble that arises from these simple input data in more detail. 

As in the previous subsection, consider a generic density matrix $\rho$ and its canonical purification $\ket{\Psi}$ in the doubled Hilbert space. We want to model $\ket{\Psi}$ on some reference state, which we take to be a semiclassical state $\ket{\Psi_{\mathrm{sc}}}$ dual to a fixed geometry in the bulk. 
For such a semiclassical state, we can use the holographic dictionary to compute its norm after some Euclidean time evolution with $H_L = H \otimes \mathds{1}$ and $H_R = \mathds{1}\otimes H$.  This defines a smooth function
\begin{equation}\label{eq:grav_input}
    \mathbf{Z}(\tau_1,\tau_2) \coloneqq \bra{\Psi_{\mathrm{sc}}}\e^{-\tau_1 H_L}\e^{ -\tau_2 H_R}\ket{\Psi_{\mathrm{sc}}},
\end{equation}
which we will use as input from gravity. Namely, we find the maximally ignorant distribution $\mu(A)$ compatible with the constraint that the \emph{average overlap} $\overline{\bra{\Psi}\e^{-\tau_1 H_L}\e^{ -\tau_2 H_R}\ket{\Psi}}$ is equal to $\mathbf{Z}(\tau_1,\tau_2)$. We can write this constraint in the energy eigenbasis of the doubled Hilbert space as
\begin{equation}\label{eq:constr_eqn}
    \sum_{i,\alpha}\e^{-\tau_1 E_i}\e^{-\tau_2 E_\alpha} \overline{|A_{i\alpha}|^2} \,\stackrel{!}{=}\, \mathbf{Z}(\tau_1,\tau_2),
\end{equation}
where $A_{i\alpha} = \braket{E_i,E_\alpha}{\Psi}$. From this expression we see that we are constraining the variance of the ensemble of random purifications. Indeed, optimizing the Shannon entropy $-\int \dd A \,\mu(A)\log\mu(A)$ with the above constraint results in the optimal distribution 
\begin{equation}\label{eq:state_ensemble}
\begin{split}
    \mu(A) 
    &= \frac{1}{\mathcal{N}}\exp(-\frac{1}{2}\sum_{i,\alpha} \sigma(E_i,E_\alpha)^{-1}\, |A_{i\alpha}|^2).
\end{split}
\end{equation}
This is a product of normal distributions with mean zero and energy-dependent variance $\sigma(E_i,E_\alpha)$,
where the functional form of $\sigma(E_i,E_\alpha)$ is fixed in terms of $\mathbf{Z}(\tau_1,\tau_2)$ by solving the constraint equation \eqref{eq:constr_eqn}. If the ensemble were exactly Gaussian, we would simply get 
\begin{equation}\label{eq:Gaussian_prediction}
    \overline{|A_{i\alpha}|^2} = \sigma(E_i,E_\alpha) \qquad \mathrm{(Gaussian)}.
\end{equation}
However, although the distribution $\mu(A)$ is Gaussian, the ensemble receives small non-Gaussian corrections due to the fixed trace condition $\delta(\Tr AA^\dagger -1)$. As in the previous section, we deal with this condition using the Fourier representation of the delta function.
In Appendix \ref{app:normalization}, we show that this corrects the variance to
\begin{equation}\label{eq:corrected_var}
    \overline{|A_{i\alpha}|^2} = \sigma(E_i,E_\alpha) \left(1+\sum_{n=1}^\infty c_n\, \sigma(E_i,E_\alpha)^n \right),
\end{equation}
where we defined the coefficients
\begin{equation}
c_n =  \frac{\int_{-\infty}^\infty \dd s\,\e^{-is} d(s) (is)^n}{\int_{-\infty}^\infty \dd s\,\e^{-is} d(s)}, \qquad d(s) = \prod_{j\beta}(\sigma^{-1}_{j\beta}-2is)^{-1}.
\end{equation}
 These corrections introduce statistical correlations between different $A_{i\alpha}$ and $A_{j\beta}$, because the function $d(s)$ depends on all $\{\sigma_{i\alpha} \coloneqq \sigma(E_i,E_\alpha)\}$.
However, we can solve for $\sigma$ in terms of $\mathbf{Z}$ order by order in $\e^{-S}$. The trace condition already implies that $\sigma_{i\alpha}$ is order $\e^{-2S}$, so the leading order solution is the Gaussian variance \eqref{eq:Gaussian_prediction}. Each further term in the series \eqref{eq:corrected_var} is then suppressed in powers of $\e^{-2S}$, as long as the $c_n$ remain bounded.

We conclude that the normalization condition only modifies the variance at subleading order. This was expected because it is only a single constraint in an exponentially large Hilbert space. Therefore, we will simplify our analysis from now on by only demanding the weaker constraint that the trace $\Tr(\rho)$ is 1 on average. The fact that this only modifies the ensemble at subleading orders in $\e^{-S}$ is then the assertion that $\Tr(\rho) $ is self-averaging. 

\subsection*{Some examples of $\ket{\Psi_{\mathrm{sc}}}$}

In section \ref{sec:random_purification} we studied the unitary invariant Hilbert-Schmidt ensemble, which has a uniform variance $\sigma(E_i,E_\alpha) = L^{-2}$. Now we give three examples of semiclassical states, and compute the resulting form of the variance function $\sigma(E_i,E_\alpha)$ from their overlaps. The three examples are illustrated in Figure \ref{fig:overlaps}.
\begin{figure}
\centering
\begin{tikzpicture}\,\node at (0,0) {\includegraphics[width=\textwidth, valign=c]{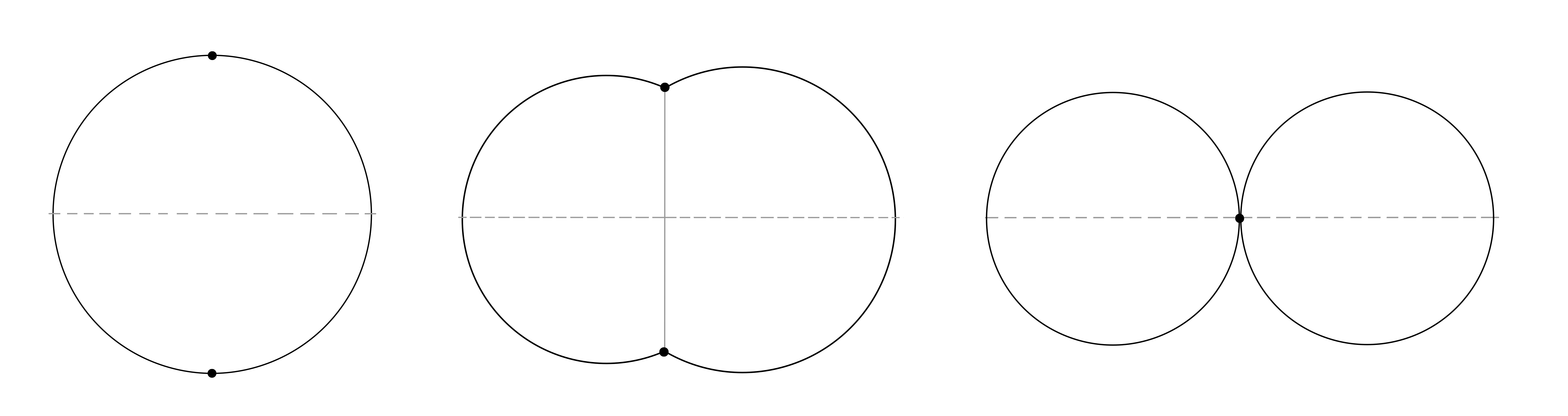}};
    \node at (-6.1,2){$\ket{\mathrm{TFD}}$};
    \node at (-1.2,2) {$\ket{\mathrm{PETS}}$};
    \node at (4.8,2) {$\ket{\Psi_L}\otimes \ket{\Psi_R}$};    \end{tikzpicture}
\caption{Three examples of gravitational overlaps of semiclassical states. Left: the thermofield double. Middle: a partially entangled thermal state. Right: a product state. The time-symmetric slice is indicated with a dashed line. }
\label{fig:overlaps}
\end{figure}
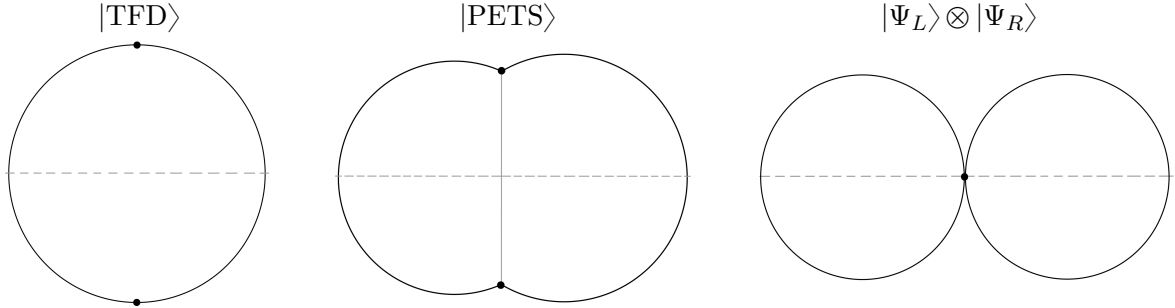
\begin{enumerate}[label=\bfseries\roman*)]
    \item \textbf{Thermofield double.} Suppose we choose as our reference state the TFD at inverse temperature $\beta$, 
    \begin{equation}
        \ket{\mathrm{TFD}_\beta} = \frac{1}{\sqrt{Z(\beta)}}\sum_{n} \e^{-\frac{\beta}{2} E_n} \ket{E_n}\otimes \ket{E_{\bar n}}. 
    \end{equation}
    Its gravitational dual is the two-sided eternal black hole \cite{Maldacena:2001kr}. It has a diagonal entanglement pattern in which the energy levels are paired and weighted by a Boltzmann factor, and the overlap \eqref{eq:grav_input} is given by the single-sided partition function at inverse temperature $\beta+\tau_1+\tau_2$:
    \begin{equation}
        \mathbf{Z}(\tau_1,\tau_2) = \frac{Z(\beta+\tau_1+\tau_2)}{Z(\beta)}.
    \end{equation}
    To solve for $\sigma(E_i,E_\alpha)$, we work in the Gaussian approximation \eqref{eq:Gaussian_prediction} and evaluate the constraint equation \eqref{eq:constr_eqn} in the continuum limit. After changing variables to $\bar E_{i\alpha} = (E_i+E_\alpha)/2$ and $\omega_{i\alpha} = E_i-E_\alpha$, 
    we can write the constraint equation as
    \begin{equation}\label{eq:constr_eqn2}
        \int \dd \bar E \dd \omega \,\e^{2S(\bar E)}\e^{-(\tau_1+\tau_2)\bar E} \e^{-\frac{1}{2}(\tau_1-\tau_2)\omega} \sigma(\bar E,\omega) \stackrel{!}{=} \frac{Z(\beta+\tau_1+\tau_2)}{Z(\beta)}.
    \end{equation}
     The right-hand side of \eqref{eq:constr_eqn2} does not depend on $\tau_1-\tau_2$, because the thermofield double is invariant under the modular evolution with $H_L-H_R$. The left-hand side must therefore be localized at $\omega = 0$, hence
    \begin{equation}\label{eq:TFD_variance}
        \sigma(\bar E,\omega)=  \frac{\e^{-\beta \bar E}}{Z(\beta)} \, \e^{-S(\bar E)}\delta(\omega).
    \end{equation}
    More precisely, in going to the continuum limit we have smeared the energy levels over bins of size $\delta \sim 1/S$, so we should really think of the delta function in \eqref{eq:TFD_variance} as the limit of a Gaussian of width $\delta$,
    \begin{equation}
    \label{eq:229}
        \sigma_\delta (\bar E,\omega)=  \frac{\e^{-\beta \bar E}}{Z(\beta)} \, \e^{-S(\bar E)}\,\frac{\e^{-\omega^2/2\delta^2}}{\sqrt{2\pi} \delta}.
    \end{equation}
    This corresponds to the same bulk geometry as the TFD perturbatively in $G_{\text N}$.

\item \textbf{Partially entangled thermal states.}
Another class of states that will be important in this paper are partially entangled thermal states (PETS), introduced in the context of the SYK model in \cite{Goel:2018ubv},  
\begin{equation}\label{eq:PETS_state_2}
    \ket{\mathrm{PETS}_{\beta_{L},\beta_{R}}}= \frac{1}{\sqrt{\mathcal{N}}}\sum_{i,\alpha}\,\e^{-\half\beta_{L}E_i}\,\e^{-\half\beta_{R}E_\alpha}\mathcal{O}_{i\alpha} \ket{E_i}\otimes \ket{E_\alpha}.
\end{equation}
These states are labeled by left and right Euclidean time variables $\beta_{L,R}$ as well as an operator $\mathcal{O}$ that prepares the state. Thanks to the diagonal entanglement in the TFD, one can think of the PETS as the Euclidean time evolution of an operator $\mathds{1}\otimes \mathcal{O}$ acting on the purifying side of the thermofield double. These states were generalized to 2d CFTs in \cite{Chandra:2023dgq}, and they were used in higher-dimensional models of black hole microstates in \cite{Sasieta:2022ksu, Balasubramanian:2022gmo,Antonini:2023hdh}.

For this class of states, we can also solve the constraint equation by going to the continuum limit,
\begin{equation}
    \int \dd \bar E \dd \omega \,\e^{2S(\bar E)}\e^{-(\tau_1+\tau_2)\bar E} \e^{-\frac{1}{2}(\tau_1-\tau_2)\omega} \sigma(\bar E,\omega) \,\stackrel{!}{=} \,\frac{1}{\mathcal{N}}\expval{\mathcal{O}^\dagger(0)\mathcal{O}(\tau)}_{\beta}.
\end{equation}
Here we rewrote the overlap \eqref{eq:grav_input} of the PETS as a Euclidean thermal two-point function in the single-sided Hilbert space, at an inverse temperature and Euclidean time given by
\begin{equation}
    \beta = \beta_{L}+ \tau_1+\beta_{R} +\tau_2, \qquad \tau = \frac{1}{2}(\beta_{L}+\tau_1-\beta_{R}-\tau_2).
\end{equation}
As before, we can solve for $\sigma(\bar E,\omega)$ using the inverse Laplace transform, which yields
\begin{equation}
    \sigma(\bar E,\omega) = \frac{1}{\mathcal{N}}\,\e^{-(\beta_{L}+\beta_{R})\bar E } \,\e^{-S(\bar E)} g_{\mathcal{O}}(\bar E,\omega;\beta_{L,R}),
\end{equation}
where $g_{\mathcal{O}}$ is related to the microcanonical two-point function of $\mathcal{O}$. If $\mathcal{O}$ is a simple operator (in the sense of ETH), the function $g_{\mathcal{O}}$ should fall off sufficiently fast as $\omega \to \infty$. We can therefore think of $g_{\mathcal{O}}$ as a `window function', determining how much the variance is spread in $E_i-E_\alpha$.

If the dependence of $g_{\mathcal{O}}$ on the mean energy $\bar E$ is of order one, we can interpret the bulk dual of the state \eqref{eq:PETS_state_2} as a small perturbation on the black hole background. However, for heavy enough operators the function $g_{\mathcal{O}}$ itself may be exponentially enhanced to $O(\e^{S(\bar E)})$, in which case one should take into account the gravitational backreaction of the bulk matter field created by $\mathcal{O}$. An example of this kind will be studied in section \ref{sec:higher_d}, where the operator $\mathcal{O}$ creates a thin shell of dust in the bulk \cite{Sasieta:2022ksu}.

\item \textbf{Product states.} As a third example, suppose that our reference state is separable
\begin{equation}\label{eq:product_input}
    \ket{\Psi_{\mathrm{sc}}} = \ket{\Psi_L}\otimes \ket{\Psi_R},
\end{equation}
so that there is no entanglement between $L$ and $R$. Then the gravitational overlap factorizes,
\begin{equation}
    \mathbf{Z}(\tau_1,\tau_2) = Z_L(\tau_1)Z_R(\tau_2),
\end{equation}
which is illustrated in Figure \ref{fig:overlaps}. If we plug this into the constraint equation \eqref{eq:constr_eqn}, we infer that the variance function also factorizes,
\begin{equation}
    \sigma(E_i,E_\alpha)= \sigma_L(E_i)\sigma_R(E_\alpha). 
\end{equation}
The functions $\sigma_{L,R}$ are related by inverse Laplace transforms to $Z_{L,R}$, and satisfy $\sum_i \sigma_L(E_i) = \sum_\alpha \sigma_R(E_\alpha) = 1$ by the normalization condition. This means that the ensemble can be written as
\begin{equation}\label{eq:product_ensemble}
    \mu(A) = \frac{1}{\mathcal{N}}\,\e^{-\half \Tr (\Sigma_L A \,\Sigma_R A^\dagger)},
\end{equation}
where $\Sigma_{L}$ is a diagonal matrix in the energy eigenbasis of $H_{L}$ with eigenvalues $\sigma_{L}(E_i)^{-1}$, and $\Sigma_R$ is diagonal in the energy eigenbasis of $H_R$ with eigenvalues $\sigma_{R}(E_\alpha)^{-1}$. This ensemble is also known as the correlated Wishart ensemble \cite{Marcenko_1967}. 

An important feature of this example is that even though the input \eqref{eq:product_input} is a product state, a typical member of the ensemble is not separable in general. Namely, the mean reduced density matrix becomes
\begin{equation}
    \bar \rho_{ij} = \delta_{ij}\sum_\alpha \overline{|A_{i\alpha}|^2} = \delta_{ij}\,\sigma_L(E_i),
\end{equation}
so that the von Neumann entropy of the average state $S(\bar \rho)$ is non-zero.

\end{enumerate}

\newgeometry{left=2.4cm,right = 2.4cm, bottom = 2cm, top=2cm}
\subsection*{Relation to holographic coarse-graining}
At this point it is interesting to compare the maximum ignorance principle to other notions of coarse-graining in the context of holography.  These are based on (among others) the diagonal projection \cite{Chandra:2022fwi,diubaldo2023ads3rmt2}, random quantum channels \cite{BRUZDA2009320,Kukulski_2021}, or complexity coarse-graining \cite{Engelhardt:2017aux,Engelhardt:2018kcs,akers2022black}. In the latter approach, one maximizes the entanglement entropy of $\rho$ compatible with its expectation values of simple operators, in an analogous fashion to how the canonical thermal state is derived as the state with maximal von Neumann entropy subject to a constraint on the average energy.

We want to emphasize that our construction is conceptually different: instead of maximizing a notion of quantum information entropy of $\rho$ with low-energy constraints, we maximized the classical Shannon entropy of \emph{a measure on the space of $\rho$'s}, and obtained as an output the coarse-graining map 
\begin{equation}\label{eq:coarse_graining_map}
    \rho  \mapsto \bar \rho = \int \dd A \,\mu(A) AA^\dagger,
\end{equation}
which sends a fine-grained microscopic state to the mean of the induced ensemble. This interplay between random matrix theory and quantum information theory moreover allows to compute averages of more general (multi-boundary) quantities encoded in higher statistical moments $\overline{\rho\otimes \dots \otimes \rho}$.  

These higher moments give us more information about the fine-grained microscopics than $\bar\rho$ can give. As a physical example, inspired by \cite{Engelhardt_2022}, suppose we want to distinguish between (1) a bulk geometry consisting of two disconnected black holes, and (2) an eternal black hole with an Einstein-Rosen bridge. In case (1), the gravitational input is a product of thermal partition functions,
\begin{equation}\label{eq:product_thermal}
    \mathbf{Z}(\tau_1,\tau_2) = \frac{Z(\beta+\tau_1)}{Z(\beta)}\,\frac{Z(\beta+\tau_2)}{Z(\beta)},
\end{equation}
and the maximally ignorant ensemble is given by \eqref{eq:product_ensemble}. We can then solve for $\Sigma_L$ and $\Sigma_R$, giving
\begin{equation}
    \Sigma_L = Z(\beta)\,\e^{\beta H_L} \qquad \mathrm{and}\qquad \Sigma_R = Z(\beta)\,\e^{\beta H_R}.
\end{equation}
Hence the mean of the induced ensemble is thermal, $\bar \rho = \rho_{\beta}$, and so its von Neumann entropy $S(\bar \rho)$ is simply given by the thermal entropy at inverse temperature $\beta$. In case (2), the semiclassical state is the thermofield double $\ket{\mathrm{TFD}_\beta}$, and by plugging in \eqref{eq:TFD_variance} we also find that
$
    \bar \rho = \rho_\beta.
$

In other words, on the level of the coarse-grained state the two cases (1) and (2) are indistinguishable and give the same prediction for the entanglement entropy $S(\bar\rho) = S(\beta)$. However, they can be distinguished by computing the average relative entropy \eqref{eq:relentr} in both cases, either via the replica trick \cite{Lashkari_2014} or by expanding perturbatively in $\delta\rho$ as \eqref{eq:rel_ent_expansion}. Indeed, a simple computation shows that the two cases give different state averages, to leading order in $\e^{-S}$:
\begin{equation}
    \mathrm{case\,(1):}\quad \overline{S(\rho || \rho_\beta)} = \half \, \e^{3S(\beta)/4}, \qquad  \mathrm{case\,(2):}\quad \overline{S(\rho || \rho_\beta)} = \half. 
\end{equation}
More generally, any measure that depends on the second moment of $\delta\rho$ can distinguish a typical state of ensemble (1)  from a typical state of ensemble (2), e.g. by running the SWAP test \cite{Marolf:2021ghr,Balasubramanian_2022}.

\restoregeometry
\subsection*{Relation to quantum chaos}

Another motivation to use the overlap $\mathbf{Z}(\tau_1,\tau_2)$ as input data to model random states comes from the field of quantum many-body chaos. 
In fact, the overlap \eqref{eq:grav_input} is related to the \emph{return amplitude}\footnote{$R(t_1,t_2)$ is also often called the `Loschmidt echo', `fidelity' or `autocorrelation function' depending on the context.} by the analytic continuation $\tau_{1,2} = it_{1,2}$ \cite{Cardy_2014, Cardy:2016lei},
\begin{equation}
    R(t_1,t_2) \coloneqq |\bra{\Psi}\e^{-it_1 H_L}\e^{-it_2 H_R}\ket{\Psi}|.
\end{equation}
The return amplitude is a well-known probe of quantum chaos. For typical states in chaotic theories, this function decays exponentially from its initial value of one to a plateau where it starts oscillating erratically. For finite systems, after a sufficiently long time, the amplitude develops patterns of `revivals' where the initial state is (partially) recovered. For systems with an infinite but discrete spectrum, the amplitude eventually collapses to an asymptotic value where it intertwines with complex Poincaré recurrences. 

In holographic CFTs, such patterns of revivals are expected to appear only at late times thanks to the chaotic nature of the energy spectrum at high energies. For example, in large-$c$ 2d CFTs with a large gap and in a finite spatial volume $L$, the timescale at which revivals appear in certain boundary states is estimated to be $O(Lc)$ \cite{Cardy_2014}.
Nonetheless, even in holographic CFTs there are atypical states for which quantum revivals in the return amplitude appear already at $O(1)$ times. This was demonstrated in the case of Virasoro coherent states in the context of 2d CFTs \cite{deBoer:2016bov,deBoer:2023lrd,Liska:2022vrd,Caputa:2022zsr}, where so-called `scarred-states' exist that fail to thermalize and do not correspond to black holes in the bulk despite having energies above $c/12$. These observations give a criterium directly in the boundary CFT for our semiclassical input. Loosely speaking, we want to choose states $\ket{\Psi_{\mathrm{sc}}}$ with return amplitudes that are smoothly decaying for sufficiently long times. 

The return amplitude was also recently studied in the context of chaotic energy eigenstates \cite{shi2023local}. In this setup, an eigenstate of an interacting Hamiltonian is reduced to a subsystem and a bath with a much larger volume.\footnote{For large volume of the bath, subsystem ETH \cite{Dymarsky:2016ntg} is the statement that the reduced density matrix of an energy eigenstate is exponentially close in trace distance to the thermal density matrix.} This is to be contrasted to our setup, where the original Hilbert space and the purification Hilbert space are of the same size. 
Nonetheless, the statistical methods used in \cite{shi2023local} as well as \cite{Murthy:2019qvb} are very similar to ours. Namely, both references expand energy eigenstates of the interacting Hamiltonian in terms of eigenstates of the non-interacting left and right Hamiltonians, with coefficients that are treated statistically. The variance of these coefficients is then modulated by an envelope function, similar to the variance function $\sigma(E_1,E_2)$ that we introduced above. Moreover, the fluctuations of the typical reduced density matrix away from thermality are found to scale as $\e^{-S(E)/2}$. This suggests that we can extend the general framework of the maximum ignorance principle to encompass the results of \cite{shi2023local,Murthy:2019qvb}. 
 
\subsection{The state-averaging ansatz}
\label{sec:saa23}

Having derived a maximum ignorance ensemble for the purification matrix $A$, and having computed the variance function $\sigma(E_1,E_2)$ for three classes of gravitational overlaps, we go on to study the \emph{induced} ensemble of mixed states $\rho = AA^\dagger$. In particular, we will be interested in computing the variance $\overline{\delta\rho_{ij}\delta\rho_{kl}}$ and higher moments, which we will show to remain finite in the limit $L\to\infty$. 

For the variance function we take the following ansatz, which encompasses the main types of gravitational input that will be relevant in the rest of this paper:
\begin{equation}\label{eq:variance_SAA}
    \sigma(E_1,E_2) = \frac{\e^{-\beta \bar E }}{Z(\beta)} \,\e^{-S(\bar E)}j_0(\bar E,\omega;\beta),
\end{equation}
where $\bar E = (E_1+E_2)/2$ and $\omega = E_1-E_2$.
The function $j_0(\bar E,\omega;\beta)$ is determined by the input $\mathbf{Z}(\tau_1,\tau_2)$ through solving the constraint equation \eqref{eq:constr_eqn}, which can be rewritten in terms of $\bar E$ and $\omega$ in the continuum approximation as
\begin{equation}\label{eq:constr_eqn3}
   \mathbf{Z}(\tau_1,\tau_2) = \frac{1}{Z(\beta)} \int \dd \bar E\dd\omega \,\e^{S(\bar E)-(\beta+\tau_+)\bar E}\e^{-\tau_- \omega}j_0(\bar E,\omega;\beta), 
\end{equation}
where we defined $\tau_+ = \tau_1+\tau_2$ and $\tau_- = (\tau_1-\tau_2)/2$. We also make the following physically motivated assumptions about the window function $j_0$ that appears in the ansatz.

We will assume that the function $j_0(\bar E,\omega;\beta)$ 
\begin{enumerate}[label=\roman*)]
\item \label{assumption1}  is even in $\omega$. This  implies that $\int \dd\omega\,j_0(\bar E,\omega;\beta)\,\omega = 0$, which is equivalent to $-\partial_{\tau_-}\mathbf{Z}\mid_{\tau_i=0} = 0$. In other words, the input state is in equilibrium 
\begin{equation}
    E_* \coloneqq \bra{\Psi_{\mathrm{sc}}}H_L\ket{\Psi_{\mathrm{sc}}}  = \bra{\Psi_{\mathrm{sc}}}H_R\ket{\Psi_{\mathrm{sc}}}.
\end{equation}
Note that this does not imply that $\mathbf{Z}$ is necessarily independent of $\tau_-$. In fact, the characteristic width of $j_0$ in $\omega$ is determined by the second derivative of $\mathbf{Z}$ with respect to $\tau_-$:
\begin{equation}
    \partial_{\tau_-}^2 \mathbf{Z}(\tau_1,\tau_2)\vert_{\tau_i=0} = \frac{1}{Z(\beta)}\int \dd \bar E \,\e^{S(\bar E)} \e^{-\beta \bar E} \int \dd\omega\, j_0(\bar E,\omega) \,\omega^2 \eqqcolon \delta^2.
\end{equation}
    \item \label{assumption2}  is peaked at $\omega = 0$, and falls off at least as fast as $\e^{-C |\omega|}$ for some $C>0$ as $\omega \to \infty$. This assumption means that $\delta$ is small enough, with the precise relation between $\delta$ and $C$ determined by the functional form of $j_0$. This ensures that the integral over $\omega$ converges.
\item \label{assumption3} depends weakly on $\bar E$ and $\beta$. In other words, it affects the saddle point $E_*$ of the $\bar E$-integral in \eqref{eq:constr_eqn3} only to subleading order in $1/S$. This statement can be rephrased in terms of $\mathbf{Z}$, by demanding that $-\partial_{\tau_+}\mathbf{Z}(\tau_1,\tau_2)$ evaluated at $\tau_+ =0$ is approximately the thermal energy $E_\beta = S'(E_*)$. This means the average energy of the semiclassical state $E_*$ determines $\beta$ (and vice versa) via the standard thermodynamic relation. 
\end{enumerate}

In section \ref{sec:higher_d} we will discuss an ensemble where $j_0$ fails to satisfy property \ref{assumption3} as $j_0$ becomes itself of the order of $\e^{S(\bar E)}$. Moreover, we have encountered one example, when the input is a product of thermal partition functions \eqref{eq:product_thermal}, for which the function $j_0$ fails to satisfy property \ref{assumption2}. However, both the TFD and the PETS  discussed in the previous section are of the form \eqref{eq:variance_SAA}. 

\subsection*{Mean density matrix}
Using the above ansatz, we first compute the mean of the induced ensemble $\mu(\rho)$. We have 
\begin{equation}
    \bra{E_i}\bar \rho\ket{E_j} = \sum_\alpha \,\wick{\c1 A_{i\alpha}\c1 A^*_{j\alpha}} = \delta_{ij} \sum_\alpha \sigma(E_i,E_\alpha).
\end{equation}
Going to the continuum limit and plugging in the ansatz for $\sigma$,  we can write this as 
\begin{equation}\label{eq:mean_d}
\begin{split}
    \bar \rho_{ii} &= \int \dd E_2 \,\e^{S(E_2)} \e^{-S(\bar E_{i2})} \frac{\e^{-\beta \bar E_{i2}}}{Z(\beta)}j_0(\bar E_{i2},\omega_{i2};\beta)\\[1em]
    &=\frac{\e^{-\beta E_i}}{Z(\beta)}\int \dd E_2\, \e^{-\frac{1}{2}(\beta_2-\beta) \omega_{i2}}\,j_0(\bar E_{i2},\omega_{i2};\beta),
\end{split}
\end{equation}
where $\beta_2 = S'(E_2)$. In the second line we expanded $S(\bar E_{i2})$ around $E_2$, neglecting terms of order $S''(E_2)$ in the thermodynamic limit. 
The integral over $E_2$ defines a smooth function $j_1$. For this integral to converge we need that $j_0$ decays at least as fast as $\e^{-\frac{1}{2}(\beta_2-\beta)|\omega|}$ for large $E_2$. Hence
\begin{equation}
   \bar \rho_{ii} =  \frac{\e^{-\beta E_i}}{Z(\beta)} j_1(E_i;\beta).
\end{equation}
 Although the function $j_1$ depends weakly on $E_i$ and $\beta$ by assumption \ref{assumption3}, the presence of a non-trivial profile for $j_1$ is in principle detectable at finite order in  $1/S$. 

\subsection*{Variance of the ensemble}
Now, let us study the variance of $\rho$ around the mean $\bar \rho$. Defining the variable $\delta\rho = \rho - \bar \rho$, 
\begin{equation}
\begin{split}
    \overline{\delta \rho_{ij}\delta\rho_{kl}} &= \sum_{\alpha,\beta} \wick{\c1 A_{i\alpha} \c2 A^*_{j\alpha} \c2 A_{k\beta} \c1 A^*_{l\beta}}= \delta_{il}\delta_{jk} \sum_{\alpha} \sigma(E_i,E_\alpha)\sigma(E_j,E_\alpha).
\end{split}  
\end{equation}
Here we evaluated the quartic moment of $A$ in the Gaussian approximation to \eqref{eq:state_ensemble}. Note that the index structure is swapped compared to the disconnected contribution $\bar\rho_{ij}\bar\rho_{kl}$. Moreover, the sum over $\alpha$ is a convolution of the variance function $\sigma$. 

Since $\sigma$ is peaked at $E_i = E_\alpha$, the convolved function will also be peaked at $E_i = E_j$. To make this precise, we fill in the ansatz for $\sigma$ and go to the continuum limit. Expanding the entropy we find
\begin{equation}
    S(\bar E_{i2}) + S(\bar E_{j2}) \approx S(\bar E_{ij})+S(E_2) + \order{\omega^2},
\end{equation}
where we neglect terms of order $S''(E_2)$ as these terms vanish in the thermodynamic limit. From this expression, we find that the variance is given by
\begin{equation}\label{eq:varianz}
    \begin{split}
        \overline{|\delta\rho_{ij}|^2} 
    &= \frac{\e^{-2\beta \bar E_{ij}}}{Z(\beta)^2} \e^{-S(\bar E_{ij})}\int \dd E_2 \, \e^{\beta \omega_{ij2}} \,j_0(\bar E_{i2},\omega_{i2};\beta) j_0(\bar E_{j2},\omega_{j2};\beta) \\[1em]
    &\coloneqq \frac{\e^{-2\beta \bar E_{ij}}}{Z(\beta)^2} \e^{-S(\bar E_{ij})} j_2(\bar E_{ij},\omega_{ij};\beta).
    \end{split}
\end{equation}
In the first line, we defined $\omega_{ij2} = \bar E_{ij} - E_2$. The function $j_2$ is peaked at $\omega_{ij}=0$, and has an envelope in $\omega$ that is determined by the integral over $E_2$. 

Combining \eqref{eq:mean_d} and \eqref{eq:varianz}, we can write an expression for the high-energy matrix elements of $\rho= \bar \rho + \delta\rho$ which takes a form very similar to ETH
\begin{equation}
\label{eq:saa-sec2}
    \rho_{ij} = \delta_{ij}\,\frac{\e^{-\beta E_i}}{Z(\beta)} j_1(E_i;\beta) + \frac{\e^{-\beta \bar E_{ij}}}{Z(\beta)} \e^{-S(\bar E_{ij})/2} j_2(\bar E_{ij},\omega_{ij};\beta)^{1/2}R_{ij}.
\end{equation}
Here $R_{ij}$ is a complex random variable of mean zero and unit variance. This variable is approximately Gaussian, but it receives exponentially small non-Gaussian corrections (see discussion below). We will refer to equation \eqref{eq:saa-sec2} and its generalization to higher moments as the \emph{state-averaging ansatz}.

Notice that in a microcanonical window where $E_i \approx E_j \approx E$, the suppression of the diagonal and off-diagonal terms precisely agrees with the unitary invariant ensemble of section \ref{sec:random_purification}; namely, $\bar \rho_{EE} \sim \e^{-S(E)}$ and $\delta\rho_{EE'}\sim \e^{-3S(E)/2}$. However, outside the microcanonical window the state-averaging ansatz depends non-trivially on the spectral density $\varrho(\bar E_{ij})$ evaluated at the mean energy. In particular, the variance remains finite in the $L\to \infty$ limit, so that the state-averaging ansatz can be used for theories with a (discrete) infinite spectrum, such as a CFT on the cylinder. 

\subsection*{Average relative entropy}
Since the variance \eqref{eq:varianz} is suppressed by a factor $\e^{-S(\bar E)}$ compared to the mean squared, we can perturbatively expand averages of arbitrary functionals $F[\rho]$ in a series in $\delta\rho$. As an example, we evaluate the average relative entropy for the state-averaging ansatz using the perturbative expansion \eqref{eq:rel_ent_expansion} of $S(\rho ||\bar\rho)$ to quadratic order. The quadratic coefficient is called the Fisher information metric (see Appendix \ref{app:volume_element}). For an ensemble centered at the thermal state $\bar\rho=\rho_\beta$ we obtain
\begin{equation}
    \overline{S(\rho||\rho_\beta)} \approx \half \sum_{ij}Z(\beta) \e^{\beta \bar E_{ij}}\frac{\beta\omega_{ij}/2}{\sinh(\beta \omega_{ij}/2)} \overline{|\delta\rho_{ij}|^2}.
\end{equation}
Now we can use the variance \eqref{eq:varianz} and make the continuum approximation $\sum_{ij}\to \int \dd \bar E\dd\omega \,\e^{2S(\bar E)}$. The resulting integral over $\bar E$ will then have a saddle point precisely at the thermal energy $E_\beta$. Evaluating the $\bar E$ integral on its saddle, we get the following approximation
\begin{equation}
     \overline{S(\rho || \rho_\beta)} \approx \half \int \dd \omega \frac{\beta \omega/2}{\sinh(\beta\omega/2)}j_2(E_\beta,\omega).
\end{equation}
If $j_2$ is a Gaussian envelope in $\omega$ of width $\delta$, the average tends to $\half$ in the limit where $\delta\to 0$. More generally, 
we see that the envelope function $j_2$ appearing in the state-averaging ansatz \eqref{eq:saa-sec2} has a nice interpretation as controlling the distinguishability between a typical state $\rho$ and the mean.  

\subsection*{Higher moments of the ensemble}

The random variable $R_{ij}$ appearing in the state-averaging ansatz \eqref{eq:saa-sec2} has exponentially suppressed non-Gaussian corrections. For example, the cubic cumulant is computed by a fully connected Wick contraction\footnote{As before, we are using the Gaussian approximation for the ensemble of $A$'s here. We have explicitly checked that the corrections from the fixed trace condition $\delta(\Tr AA^\dagger -1 )$ are further subleading in $\e^{-S}$.} of the form 
\begin{equation}
\begin{split}
    \overline{\delta \rho_{ij} \delta \rho_{kl} \delta \rho_{mn}}^{\,\mathrm{conn}.} &= \sum_{\alpha,\beta,\gamma} \wick{\c1 A_{i\alpha} \c2 A^*_{j\alpha} \c2 A_{k\beta} \c3 A^*_{l\beta} \c3 A_{m\gamma} \c1 A^*_{n\gamma}}. 
\end{split}
\end{equation}
This gives a cyclic pattern of contractions for $\delta\rho^3$ which is precisely of the generalized ETH form \cite{Foini:2018sdb}. If we plug in the ansatz for the variance function \eqref{eq:variance_SAA}, the resulting $n$th cumulant can be brought into the form
\begin{equation}
\label{eq:saa-non-gauss-sec2}
    \overline{\delta \rho_{i_1i_2}\delta \rho_{i_2i_3}\cdots \delta \rho_{i_ni_1}}^{\,\mathrm{conn.}} = \frac{\e^{-n\beta \bar E_n} }{Z(\beta)^n} \e^{-(n-1)S(\bar E_n)}j_n\big(\bar E_n,\{\omega_{kl}\}\big),
\end{equation}
where we defined
\begin{equation}
    \bar E_n = \frac{1}{n}(E_1+\dots + E_n),
\end{equation}
and $\{\omega_{kl}\}$ collectively denotes all energy differences. The function $j_n$ is an $n$-fold convolution of $j_0$. Hence it is strongly peaked at $\omega_{kl}=0$, with some window function determined fully by $j_0$. The entropy dependence of the $n$th cyclic non-Gaussianity is $\e^{-(2n-1)S(E)}$, in a microcanonical window around energy $E$. 

The fact that the non-Gaussianities precisely follow the structure dictated by typicality is a feature of our construction that should be contrasted to ETH. Namely, applying a principle of maximum ignorance to operator averaging (as described in the introduction), one has to input all higher statistical moments of $\mathcal{O}_{ij}$ essentially by hand, through higher-point thermal correlation functions of $\mathcal{O}$. By contrast, we have only given a single quadratic input for $A$, from which the higher cyclic non-Gaussianities follow immediately. The basic reason why this happens is that a quadratic input for $\mathcal{O}$ leads to an exactly Gaussian model, whereas a quadratic input for $A$ leads to only an \emph{approximately} Gaussian induced ensemble for $\rho = AA^\dagger$.

\subsection{Some extensions of the ansatz} \label{sec:saa24}
\subsection*{Adding extra input}

In the previous two subsections we have considered a simple class of input data, namely state overlaps after some Euclidean time evolution. However, we could supply the ensemble with additional input data, such as the expectation value of some local operator in the semiclassical state,
\begin{equation}\label{eq:extra_input}
    \bra{\Psi_{\mathrm{sc}}}\mathcal{O}^L\otimes \mathcal{O}^R\ket{\Psi_{\mathrm{sc}}}.
\end{equation}
If we now demand that the ensemble $\mu(A)$ reproduces this expectation value on average, the principle of maximum ignorance leads to the following ensemble
\begin{equation}\label{eq:extra_input_potential}
    \mu(A) = \frac{1}{\mathcal{N}
    }\exp(-\half \sum_{i\alpha} \sigma(E_i,E_\alpha)^{-1} |A_{i\alpha}|^2 + \lambda \sum_{ij\alpha\beta}\mathcal{O}^L_{ij}\,\mathcal{O}^R_{\alpha\beta}\,A_{i\alpha}A^*_{j\beta}).
\end{equation}
Here $\lambda$ is a Lagrange multiplier, which should be fixed in terms of \eqref{eq:extra_input}. We have written the potential directly in the energy eigenbasis, to highlight the fact that the ensemble no longer factorizes as a product of independent Gaussian integrals. Instead, there are now correlations between different $A_{i\alpha}$ and $A^*_{j\beta}$ for all mixed indices. 

One implication is that the mean of the ensemble is no longer strictly diagonal
\begin{equation}
    \bar\rho_{ij} = \sum_\alpha \,\overline{A_{i\alpha}A^*_{j\alpha}} \eqqcolon \bar\rho(E_i,E_j).
\end{equation}
In other words, the coarse-grained density matrix can acquire a non-trivial time-dependence $\bar \rho(t)$ under the evolution with $H_L$. Moreover, higher moments of the ensemble can also get non-Gaussian corrections
\begin{equation}
    \overline{\delta \rho_{ii}\delta \rho_{jj}} =  \sum_{\alpha,\beta} \overline{A_{i\alpha}A_{i\alpha}^*A_{j\beta}A_{j\beta}^*}^{\,\mathrm{conn.}}\qquad i\neq j,
\end{equation}
which are not of cyclic `generalized ETH' form. However, if $\mathcal{O}^{L,R}$ are simple operators, one expects their matrix elements to be approximately diagonal in the energy eigenbasis,
\begin{equation}
    \mathcal{O}^L_{ij} \approx \delta_{ij} f_L(E_i),
\end{equation}
and similarly for $\mathcal{O}^R$. Here, we have dropped the terms that are of $O(\e^{-S})$. The diagonal contribution simply gives a shift of the variance function $\sigma^{-1}_{i\alpha}$ by the smooth function $-\frac{\lambda}{2}f_L(E_i)f_R(E_\alpha)$. The only non-Gaussianities appearing in \eqref{eq:extra_input_potential} then come from the exponentially small off-diagonal components of $\mathcal{O}^{L,R}$. For the second and higher moments of $\delta\rho$, these subleading corrections compete with the subleading corrections coming from the exact normalization condition \eqref{eq:corrected_var}. Which of these subleading corrections dominates should be established on a case-to-case basis. 

\subsection*{Tensor products}
\label{sec:tensor-prod}

In section \ref{sec:wormholes}, we will also need a generalization of the state-averaging ansatz to tensor product Hilbert spaces. Suppose $\rho$ is a mixed state on $\mathcal{H}_1\otimes \mathcal{H}_2$, which we purify on the doubled Hilbert space
\begin{equation}
    \mathcal{H}_{1L}\otimes \mathcal{H}_{1R}\otimes \mathcal{H}_{2L} \otimes \mathcal{H}_{2R}.
\end{equation}
Also suppose the Hamiltonian is a sum of non-interacting terms on each copy,
\begin{equation}
    H_{1L} = H\otimes \mathds{1} \otimes \mathds{1} \otimes \mathds{1}, \quad  H_{1R} =  \mathds{1} \otimes H\otimes \mathds{1} \otimes \mathds{1}, \qquad \mathrm{etc.}
\end{equation}
We use indices $i_1,\alpha_1$ to label the energy eigenstates of $H_{1L}$ and $H_{1R}$, and $i_2,\alpha_2$ to label $H_{2L}$ and $H_{2R}$. Then a random pure state on the total Hilbert space is of the form
\begin{equation}
    \ket{\Psi} = \sum_{i_1,i_2,\alpha_1,\alpha_2} A_{i_1i_2\alpha_1\alpha_2} \ket{E_{i_1}}\otimes \ket{E_{\alpha_1}}\otimes \ket{E_{i_2}}\otimes \ket{E_{\alpha_2}}.
\end{equation}
The coefficients are now complex tensors, which we sample from a probability distribution $\mu(A)$. Taking a partial trace over the purifying Hilbert spaces gives a random mixed state 
\begin{equation}
    \rho = AA^\dagger
\end{equation}
which can be expanded in the basis $\ket{E_{i_1},E_{i_2}}\bra{E_{j_1},E_{j_2}}$. Generalizing the construction of section \ref{sec:SAA_2}, we give as input the gravitational overlap of a semiclassical state in the doubled tensor product Hilbert space, after Euclidean time evolution by the four independent boundary Hamiltonians
\begin{equation}
    \mathbf{Z}(\tau_{1L},\dots,\tau_{2R}) \coloneqq \bra{\Psi_{\mathrm{sc}}}\e^{-\tau_{1L}H_{1L}-\tau_{1R}H_{1R}}\e^{-\tau_{2L}H_{2L}-\tau_{2R}H_{2R}}\ket{\Psi_{\mathrm{sc}}}.
\end{equation}
The simplest example of such an input is a product of TFD states in each $\mathcal{H}_{iL}\otimes\mathcal{H}_{iR}$, in which case $\mathbf{Z}$ factorizes into a product of thermal partition functions. Another, more general example is the state defined on the time-symmetric slice of the multi-BTZ black holes analyzed in \cite{Balasubramanian_2014}, for which the overlap is computed holographically by a higher genus handlebody in the bulk AdS$_3$. 

Given such a $\mathbf{Z}$, we then find the distribution $\mu(A)$ that maximizes the Shannon entropy with the constraint that the average overlap agrees with the gravitational saddle-point
\begin{equation}
    \int \dd A\,\mu(A) \bra{\Psi}\e^{-\tau_{1L}H_{1L}-\tau_{1R}H_{1R}}\e^{-\tau_{2L}H_{2L}-\tau_{2R}H_{2R}}\ket{\Psi} \,\stackrel{!}{=}\, \mathbf{Z}(\tau_{1L},\dots,\tau_{2R}).
\end{equation}
This leads to the maximally ignorant tensor model
\begin{equation}
    \mu(A) = \frac{1}{\mathcal{N}}\exp(-\frac{1}{2}\sum_{\boldsymbol{i},\boldsymbol{\alpha}}\sigma(E_{\boldsymbol{i}},E_{\boldsymbol{\alpha}})^{-1}|A_{\boldsymbol{i}\boldsymbol{\alpha}}|^2),
\end{equation}
where used the multi-index notation $\boldsymbol{i} = (i_1,i_2)$, $\boldsymbol{\alpha} = (\alpha_1,\alpha_2)$. The volume element is simply the product $\dd A = \prod_{\boldsymbol{i},\boldsymbol{\alpha}}dA_{\boldsymbol{i}\boldsymbol{\alpha}}dA_{\boldsymbol{i}\boldsymbol{\alpha}}^*$. As before, we see that the ensemble is approximately Gaussian, and the normalization condition $\braket{\Psi}{\Psi} =1$ gives small non-Gaussianities. Again we can approximate the coefficients $A_{\boldsymbol{i}\boldsymbol{\alpha}}$ as independent Gaussian random variables, up to subleading corrections in $\e^{-S}$. All of the above remarks straightforwardly extend to the case of $\mathsf{k}$-fold tensor products $\mathcal{H}^{\otimes \mathsf{k}}$.

Although the above construction may seem like an uninteresting exercise in bookkeeping, there are at least two interesting new features. These are that
\begin{enumerate}[label=\roman*)]
    \item the mean of the ensemble can now depend on multiple parameters. For example, in sections \ref{sec:genus-two-worm} and \ref{sec:gen-obs} we will take the mean to be the product of thermal states
    \begin{equation}
    \bar\rho =\rho_{\beta_1}\otimes \dots\otimes \rho_{\beta_\mathsf{k}}.
\end{equation}
These `inverse temperatures' $\beta_1,\dots,\beta_{\mathsf{k}}$ are part of the input data provided by $\ket{\Psi_{\mathrm{sc}}}$. For example, in the case of CFT$_2$, these parameters are related to the moduli of the Riemann surface \cite{Balasubramanian_2014} that defines the boundary manifold of the Euclidean geometry dual to $\ket{\Psi_{\mathrm{sc}}}$.
\item the induced ensemble for $\rho$ has a pairwise selection rule for the higher moments of the fluctuation
\begin{equation}
    \overline{\delta \rho_{\boldsymbol{i}\boldsymbol{j}} \delta \rho_{\boldsymbol{k}\boldsymbol{l}}} = \delta_{\boldsymbol{i}\boldsymbol{l}}\, \delta_{\boldsymbol{j}\boldsymbol{k}}\, \overline{| \delta \rho_{\boldsymbol{i}\boldsymbol{j}} |^2},
\end{equation}
where the Kronecker delta's are defined element-wise in the multi-index notation, and we defined 
\begin{equation}
   \delta \rho_{\boldsymbol{i}\boldsymbol{j}} \coloneqq \bra{E_{i_1},\dots,E_{i_{\mathsf{k}}}} \delta \rho \ket{E_{j_1},\dots,E_{j_{\mathsf{k}}}}.
\end{equation}
In other words, the index contractions are pairwise Gaussian in the product Hilbert space.
This selection rule will be used in section \ref{sec:genus-two-worm} and \ref{sec:gen-obs} to study general observables in AdS$_3$ gravity. 
\end{enumerate}

\subsection{State-averaging with conformal symmetry}
\label{sec:conformalSAA}

As a final useful generalization of the state-averaging ansatz, we extend the above framework to two-dimensional conformal field theories. Recall that the dynamical information of a 2d CFT is a list of primary operators with conformal weights together with a list of OPE coefficients. Because of conformal symmetry, the Hilbert space of any CFT organizes into primary states $\ket{h,\bar h}$ and descendant states, the latter being given by states of the form $L_{-M}\bar L_{-N} \ket{h,\bar h}$, where $M = {m_1,\cdots,m_j}$ stands for an arbitrary ordered partition of $m=|M|$, with $m_j\leq \cdots \leq m_2\leq m_1$. We call these states a level $(m,n)$ descendant of $\ket{h,\bar h}$.

 The level $m$ descendant states of a given primary state\footnote{For simplicity, we will temporarily disregard the antiholomorphic component of each primary state. Also, we will only be considering non-degenerate representations of the Virasoro algebra.}
$\ket{h}$ have the same energy ($h + m$). This leads to a degeneracy that corresponds to the number of partitions of the integer $m$; because of this degeneracy, we cannot specify a unique a basis of eigenstates. Despite this, in every CFT$_2$ there is a natural basis of eigenstates given by the states $L_{-M} \ket{h}$. This basis is not orthonormal but can be transformed into an orthonormal basis using the inverse of the real Gram matrix $B_{h,m}$:
\begin{equation}
\label{eq:cftbasis}
    \ket{h,M} \coloneqq \sum_{\underset{|N|=m}{N,}}\left[B_{h,m}^{-\frac{1}{2}}\right]_{NM} L_{-N}\ket{h}, \qquad \left[B_{h,m}\right]_{MN} = \bra{h}L_{M}L_{-N}\ket{h},
\end{equation}
Now, it is expected that for generic irrational CFTs the \emph{primary} states exhibit the hallmark features of chaotic systems, while descendants are fully fixed by conformal symmetry and should not be averaged over.\footnote{See also  \cite{Lashkari:2016vgj,Dymarsky:2019etq,Besken:2019bsu,Datta:2019jeo} for a discussion of the validity of ETH for descendant eigenstates, as well as \cite{DiUbaldo:2023qli} for a manifestly conformally invariant and modular invariant approach to random matrix averaging in the spectrum of a CFT$_2$.} Hence we present a modified state-averaging ansatz where primary matrix elements display statistical characteristics while descendant matrix elements are entirely determined by conformal symmetry. 

For primary states, the ansatz assumes a very similar form as in the previous sections,
\begin{align}
\label{eq:19s4}
    \bra{h}\rho\ket{h'} &= \delta_{h,h'}\bar\rho(h)+ \frac{\e^{-\beta\left[ \frac{(h+h')}{2}- \frac{c}{24}\right]}}{Z(\beta)} \e^{-S_0\left(\frac{h+h'}{2}\right)/2}j(h,h';\beta)^{1/2}\, R_{hh'},
\end{align}
Here, $R$ is a random variable with unit variance and, just as in equation \eqref{eq:saa-sec2}, $j$ is a smooth function highly peaked at $\omega = 0$, with at least $j(\beta; h,h+\omega) \sim \e^{-\beta |\omega|}$ for $|\omega| \gg 1$. The microcanonical entropy of primary states $S_0(E)$ given by the Cardy formula
\begin{equation}
    S_0(h) \coloneqq \log \varrho_0(h) \approx 2\pi \sqrt{\frac{c}{6}\left(h-\frac{c}{24}\right)}.
\end{equation}
Next, for descendant states, we consider an ansatz that is fully determined by the primary spectrum:
\begin{equation}
\label{eq:19s42}
     \bra{h,M}\delta\rho\ket{h',N} = f_{M,N}(h, h';\beta) \bra{h}\delta\rho\ket{h'},
\end{equation}
with $f_{M,N}(h,h';\beta)$ being a smooth function of the variables $h, h'$. Our motivation for this ansatz stems from an analogous expression that is valid for three-point correlation functions between a primary operator $\clo$ and two descendant states
\begin{equation}
\label{threepf}
    \bra{h}L_{M} \clo(z) L_{-N}\ket{h'} =  \bra{h}\clo(1)\ket{h'} \nu(h,M;h_\clo;h',N|z),
\end{equation}
where $\nu(h,M;h_\clo;h',N|z)$ is a smooth function of $h,\bar h$ and $z$, fully determined by Virasoro symmetry in terms of the weights of the primary operators.\footnote{The function $\nu$ will be important in section \ref{sec:wormholes}. For a review on the properties of $\nu$ we  we refer the reader to \cite{Cho:2017oxl}, Appendix A.} In the section \ref{sec:wormholes}, we will find that in order to match CFT observables to gravitational amplitudes, we must take the function $f$ to be of the form 
\begin{equation}
\label{eq:btzCF}
    f_{M,N}(h,h';\beta) =  \e^{-\beta n}\delta_{M,N} + \order{h-h'},
\end{equation}
with $n = |N|$, where we have allowed for the possibility of corrections that vanish as $h=h'$. An important feature of \eqref{eq:btzCF} is that the function is invariant under changes in the eigenbasis of descendant states, as the function is proportional to the identity matrix when $h = h'$. 

There are multiple ways in which we could extend the ansatz \eqref{eq:19s4} to include the holomorphic and anti-holomorphic sectors of the CFT. The question is whether or not the ansatz should take into account the spin quantization condition $\bar h - h \in \mathbb{Z}$. As it turns out, the on-shell gravitational amplitudes discussed in section \ref{sec:wormholes} are not sensitive to this condition. For the purposes of this paper, we can regard the holomorphic and antiholomorphic sectors of the CFT independently, and use a factorized ansatz of the form 
\begin{align}
\label{eq:saa-conf}
    & \bra{h,\bar h} \delta\rho \ket{h', \bar h'} = \e^{-\beta\left[\frac{\Delta+\Delta'}{2} - \frac{c}{12}\right]} \,
   \e^{-\frac{1}{2}S_0\left(\frac{h+h'}{2}\right)-\frac{1}{2}S_0\left(\frac{\bar h+\bar h'}{2}\right)} 
   j(h,h',\bar h,\bar h')^{1/2}\,
   R_{h,h'} \bar R_{\bar h,\bar h'} \\[1.5em] \label{eq:saa-conf2}
  &  \bra{ h, N,\bar h, \bar N} \delta\rho \ket{h', N', \bar h', \bar N'} = f_{N,N'}f_{\bar N,\bar N'}
   \;
   \bra{ h,\bar h} \delta\rho \ket{ h',\bar h'}.
\end{align}
Here $R$ and $\bar R$ are a set of approximately uncorrelated random variables with unit variance and $j$ only has a narrow support around $\omega = h-h' = 0$ and $\bar \omega = \bar h - \bar h' =0$. 

In Appendix \ref{sec:CJ}, we discuss an off-shell amplitude which depends more sensitively on the details of the ensemble of states. For this off-shell amplitude, the correlations between $R$ and $\bar R$ are relevant to leading order. However, for all the on-shell wormholes that will be discussed in the next section, the uncorrelated ansatz \eqref{eq:saa-conf}, \eqref{eq:saa-conf2} gives a good approximation to the gravitational result.

\section{Wormholes from random states}
\label{sec:wormholes}

So far, we have developed the technology to accommodate state averaging in infinite-dimensional Hilbert spaces. We have also generalized our model to incorporate conformal symmetry. In this section, we test the output of our ensembles in the context of 3D gravity. We will focus on the ensemble of microstates of a holographic 2d CFT that models a small perturbation around a bulk thermal state. The resulting ensemble of microstates has the form of the state-averaging ansatz given in equations \eqref{eq:saa-conf} and \eqref{eq:saa-conf2}.

The state-averaging ansatz outputs connected configurations to statistical moments over the ensemble of microstates. We will match these contributions to the on-shell gravitational actions of known wormhole configurations in 3D gravity, constructed in \cite{Chandra:2022bqq}. In that paper the wormholes were interpreted as arising from an average over approximate CFTs. We will show that the state-averaging approach also matches the bulk wormhole configurations for a single instance of a CFT.

Our discussion begins with two important examples: the punctured torus wormhole in section \ref{sec:punct-tor-worm} and the genus-two wormhole in section \ref{sec:genus-two-worm}. After these examples, we extend our results to correlation functions on general Riemann surfaces in section \ref{sec:gen-obs}. We end with a comparison between the operator- and the state-averaging approach to 3D gravity in section \ref{sec:comparison}.

\subsection{Punctured torus wormhole}
\label{sec:punct-tor-worm}

We start by considering the torus wormhole punctured by a primary operator $\clo(x)$ at the boundary. Conformal symmetry allows us to place this operator at any point in the plane, so we will restrict ourselves to the case where $x=1$. Our proposal is that we can model the gravitational amplitude using an ensemble of \emph{mixed states} centered at the thermal state $\mean{\rho} = \rho_\beta$, 
\begin{equation}
\label{eq:torusworm}
    \eqimg{torusonepoint.pdf}
    \;\stackrel{?}{=}\; \mean{ \Tr\rho \clo(1)\Tr\rho\clo(1)}-\mean{\Tr\rho\clo(1)}\;\mean{\Tr\rho\clo(1)}. 
\end{equation}
In this approach, it is the ensemble of states that produces the correlation between the two asymptotic boundaries; there is no ensemble over CFTs or the operators $\clo$.

To evaluate \eqref{eq:torusworm}, we can use the ansatz presented in \eqref{eq:19s4} and \eqref{eq:19s42}. Since this ansatz takes conformal symmetry into account, it naturally organizes the sum into primary and descendant sectors:\footnote{In this section, we will be ignoring the anitholomorphic counterpart of this amplitude, as the final answer trivially factorizes into holomorphic and antiholomorphic components.}
\begin{equation}
\label{eq:semiWHam}
    \mean{\Tr \delta\rho \clo(1) \Tr \delta\rho\clo(1)} = \sum_{h,h'} \mean{|\bra{h}\delta\rho\ket{h'}|^2} \sum_{M_1,\dots,N_2} f_{M_1,N_1}f_{M_2,N_2} \bra{h',N_1}\clo\ket{h,M_1}\bra{h,N_2}\clo\ket{h',M_2}.
\end{equation}
The Virasoro conformal blocks corresponding to the torus one-point function are defined as,\footnote{For convenience, we have included the Boltzmann factor $\e^{-\beta(h-\frac{c}{24})}$ in the definition of the blocks.}
\begin{equation}
    \mathcal{F}_\clo (h;\beta) \coloneqq \e^{-\beta\left(h-\frac{c}{24}\right)}\sum_m \e^{-\beta m} \sum_{|M|=m} \frac{\bra{h,M}\clo(1)\ket{h,M}}{\bra{h}\clo(1)\ket{h}},
\end{equation}
with $n=|N|$. To recover the conformal blocks in \eqref{eq:semiWHam}, we must take $f$ to be of the form\footnote{This form can also be obtained from the overlap $\mathbf{Z}(\tau_1,\tau_2)$ defined in \eqref{eq:grav_input} in section \ref{sec:SAA}, as conformal blocks also appear in this amplitude. Here we are neglecting terms of the order of the difference $h-h'$, because they will not be relevant for the analysis. }
\begin{equation}
    f_{M,N}(h,h')= \e^{-\beta n} \delta_{M,N} + \order{h -  h'}.
\end{equation}
This leads to the following expression for the amplitude of the torus wormhole 
\begin{equation}
\begin{split}
 \sum_{h,h'} &\mean{|\bra{h}\delta\rho\ket{h'}|^2}\;  |\bra{h'}\clo\ket{h}|^2 \\ &\times \left(\sum_{N_1,N_2} \e^{-\beta (n_1+n_2)}\frac{\bra{h',N_1}\clo\ket{h,N_1}\bra{h,N_2}\clo\ket{h',N_2}}{\bra{h'}\clo\ket{h}\bra{h}\clo\ket{h'}} + \order{h-h'}\right).
\end{split}
\end{equation}
For primary operators, the three-point function $\bra{h',N_1}\clo\ket{h,M_1}/\bra{h'}\clo\ket{h}$ is a smooth function of the conformal weights $h$ and $h'$, meaning that we can consider the Taylor series of this amplitude around $h - h' = 0$:
\begin{equation}
    \sum_{N_1} \e^{-\beta n_1} \frac{\bra{h',N_1}\clo\ket{h,N_1}}{\bra{h'}\clo\ket{h}} = \e^{\beta\left(h-\frac{c}{24}\right)}\mathcal{F}_\clo (h;\beta) + \order{h-h'}.
\end{equation}
Plugging  this expression into \eqref{eq:semiWHam} leads to the amplitude
\begin{equation}
\label{eq:intwh}
   \mean{\Tr \delta\rho \clo(1) \Tr \delta\rho\clo(1)} 
   =
   \sum_{\clo_h,\clo_h'}  
    \mean{|\bra{h}\delta\rho\ket{h'}|^2}\;
    C_{\clo_h \clo \clo_h'} C_{\clo_h' \clo \clo_h} 
    \;
    \e^{2\beta \left(h-\frac{c}{24}\right)}
    \Big[
    \mathcal{F}_{\clo}(h;\beta)^2 + \order{h-h'}
    \Big].
\end{equation}
Here, there is a slight abuse of notation because 
the only meaningful quantity for the OPE coefficients is the product $C_{\clo_h \clo \clo_h'}C_{\bar \clo_{\bar h} \bar \clo  \bar\clo_{\bar h'}} = \bra{h',\bar h'}\clo\ket{h,\bar h}$, since for individual OPE coefficients there is no holomorphic factorization. This problem is no longer present once we include the anthiholomorphic sector in our analysis. 

The next step of the computation is to approximate the sum in \eqref{eq:intwh} by an integral. The key insight is that the function $\mean{|\bra{h}\delta\rho\ket{h'}|^2}$ is smooth. This feature allow us to average over the exponentially many OPE coefficients that lie within the microcanonical windows around $h$ and $h'$. The microcanonical average of the square $C_{\clo_h \clo \clo_h'}C_{\clo_h' \clo \clo_h}$ is a known universal result that can be derived from crossing symmetry and modular invariance \cite{Collier:2019weq}, 
\begin{equation}
\label{opemean}
\bigopmean{C_{\clo_{h_1}\clo_{h_2}\clo_{h_3}}C_{\clo_{h_3}\clo_{h_2}\clo_{h_1}} } = C_0(h_1, h_2,h_3) \coloneqq\frac{\Gamma_b(2Q)\Gamma_b(\frac{Q}{2}\pm i P_1 \pm i P_2 \pm i P_3)}{\sqrt{2}\Gamma_b(Q)^3\prod_{k=1}^3\Gamma_b(Q\pm 2i P_k)}.
\end{equation}
where the brackets $\opmean{\bullet}$ denote operator averaging. We would like to remark that this is an average over operators within a microcanonical window in a single instance of the CFT. There is no ensemble averaging over different theories in this framework.

Formula \eqref{opemean} is written in terms of the Liouville variables $b, Q$ and $P$ which are related to the conformal weights and the central charge of the CFT via 
\begin{equation}
    c = 1+ 6 Q^2, \quad Q = b + b^{-1}, \quad \text{and} \quad h = \frac{c-1}{24} + P^2.
\end{equation}
The function $\Gamma_b(x)$ is called the Barnes double Gamma function. For our purposes, it suffices to know that $\Gamma$ is a smooth meromorphic function in $x$ and continuous in $b$. Again, in \eqref{opemean} there is a slight abuse of notation, since the full equation also includes an antiholomorphic counterpart.

Using the microcanical average of the OPE coefficients and  substituting for the moments of $\delta\rho$ yields the result
\begin{equation}
\label{eq:wmampl}
\begin{split}
\mean{\Tr \delta \rho \clo(1)\Tr \delta \rho \clo(1)}
&\approx \int \dd {\sf h} \dd \omega \; \e^{S_0({\sf h}+\omega/2)+S_0({\sf h}-\omega/2)-S_0({\sf h})} j({\sf h},\omega;\beta) \\ &\phantom{\int \dd {\sf h} \dd \omega----}\times C_0\left({\sf h}+\frac{\omega}{2},h_\clo,{\sf h}-\frac{\omega}{2}\right) \left[\mathcal{F}_\clo({\sf h};\beta)^2 + \order{\omega}\right].
\end{split}
\end{equation}
Here, $\omega$ corresponds to the energy difference $\omega = h-h'$, and we have denoted the mean weight using ${\sf h} = (h+h')/2$ to avoid any confusion with the traditional (anti-)holomorphic notation $h,\bar h$. The approximate sign in this equation corresponds to the usual thermodynamic approximation $\sum_h \rightarrow \int \dd h\; \e^{S_0(h)}$.  

The ensemble of states we are considering is the one constructed from the input data of the thermofield double -- the canonical purification of the thermal state. In section \ref{sec:SAA_2}, we derived a profile for the function  $j(\sf{h},\omega;\beta)$ associated to this ensemble, see equation \eqref{eq:229}. This profile reads 
    \begin{equation}
        j(\sf{h},\omega;\beta) = \frac{\e^{-\omega^2/2\delta^2}}{\sqrt{2\pi} \delta},
    \end{equation}
where the smearing window $\delta$ is of the size $1/S_0$. The main takeaway of this computation is that the function $j(\sf{h},\omega;\beta)$ is a function highly peaked at $\omega = 0$. Taking advantage of this fact, we can simplify equation \eqref{eq:wmampl} even further. In such a narrow window, the functions $C_0$ and $\mathcal{F}_{\clo} + \order{\omega}$ can be treated as constant functions with respect to $\omega$.
This allows us to compute the integral over $\omega$ by pulling out the factors of $C_0$ and $\mathcal{F}_{\clo} + \order{\omega}$ outside of the integral and evaluating them at $\omega = 0$.\footnote{ For chaotic CFTs, the expectation is that the support of $j$ can be taken to be as narrow as $\e^{-S_0(h)}$, leaving only parametrically many states to perform the coarse-graining over the OPE coefficients. However, we do not need such a narrow (non-perturbative) profile for $j$ as the functions $C_0$ and $\mathcal{F}_{\clo} + \order{\omega}$ are smooth and almost constant functions inside a window whose width is perturbatively small in $h$.}

To fully perform the $\omega$ integral, we also need to expand Cardy's formula to second order, 
\begin{equation}
    S_0\left({\sf h} + \frac{\omega}{2}\right)+S_0\left({\sf h} - \frac{\omega}{2}\right) = 2 S_0({\sf h}) - 6\pi \omega^2 \left(\frac{24{\sf h}}{c} -1\right)^{-\frac{3}{2}} + \order{\omega^4}.
\end{equation}
These considerations lead to the following amplitude for the torus wormhole from state averaging 
\begin{equation}
\label{eq:torusworm2}
\eqimg{torusonepoint.pdf}
    =\int \dd {\sf h}\, \e^{S_0({\sf h})}j'({\sf h};\beta) C_0({\sf h},h_{\clo}, {\sf h})\mathcal{F}_{\clo}({\sf h};\beta)^2,
\end{equation}
 Here, we have taken out the functions $C_0$ and $\mathcal{F} + \order{\omega}$ from the $\omega$ integral, and absorbed the remaining piece into a new function $j'$,
\begin{equation}
j'(\mathsf{h};\beta) \coloneqq
\int\dd\omega \; \e^{-6\pi \omega^2 \left(\frac{24h}{c}-1\right)^{-\frac{3}{2}}} j(\mathsf{h},\omega;\beta) = 1 + O(\delta).
\end{equation}
As $\delta$ approaches zero in the thermodynamic limit, the state-averaging answer \eqref{eq:torusworm2} precisely matches the gravitational computation of \cite{Chandra:2022bqq}, as well as the computation using the Virasoro TQFT \cite{Eberhardt:2023mrq}.
\subsection*{Saddle-point analysis}
For \eqref{eq:torusworm2} to be a good approximation of the original amplitude \eqref{eq:torusworm}, it must be that the integral is dominated by states above the black hole threshold ${\sf h}>\frac{c}{24}$. Otherwise we cannot justify the use of the Cardy formula or expression \eqref{opemean} for the mean value of OPE coefficients. 
This is not evident from \eqref{eq:torusworm2} alone because $C_0 \sim \e^{-S_0({\sf h})}$ cancels the only entropy factor in \eqref{eq:torusworm2}. However, we can show that this amplitude is dominated by high-energy states as long as the weight of the external operator $h_{\clo}$ is larger than $\beta c /12$. To further clarify this point, we finish this section with a discussion of the integrand in \eqref{eq:torusworm2} as ${\sf h}$ approaches infinity.

For this analysis,  
we need to use more refined versions of the Cardy formula, the torus one-point blocks, and $C_0$. Expressions for these formulas to leading order in the large $h$ limit are given by (see \cite{Collier:2019weq,Hadasz:2009db})
\begin{align}
\label{eq:form1}
    S_0(h) 
    &=  
    \sqrt{\frac{c}{6} h}
    - \frac{1}{2}
    \log h 
    +\order{1}
    , \\ 
    \log C_0(h_{\clo}, h, h) 
    &= -\sqrt{\frac{c}{6} h}
    + h_{\clo}  
    \log h +\order{1},\label{eq:form2}\\
    \mathcal{F}_{\clo}(h_{\text{m}};\beta) &= \e^{-\beta\left(h-\frac{c}{24}\right)}\left[\frac{\e^{-\frac{\beta}{24}}}{\eta(\frac{i\beta}{2\pi})} + \order{h^{-1}}\right],
    \label{eq:form3}
\end{align}
where $\eta$ corresponds to the Dedekind eta function. Using these formulas to evaluate the wormhole amplitude leads to the integral
\begin{equation}
\label{eq:wmsaddle}
   \int\dd h \;h^{h_{\clo}-\frac{1}{2}}\e^{-2\beta h}.
\end{equation}
Here, we have neglected an overall multiplicative factor. To justify the $h$ expansion, we need to check that this integral is dominated by large values of $h$. Equation \eqref{eq:wmsaddle} has a saddle point at $h^{\star} =\frac{1}{2\beta} (h_{\clo}-\frac{1}{2})$. Interestingly, the saddle point is not controlled by an entropy factor, but by the smooth function $h^{h_\clo}$. We may think of this function as an `effective' entropy $S_{\tt eff}(h) \sim h_\clo \log(h)$. The validity of the saddle point relies on the $h_\clo$ being large, and at least larger than $\beta c/12$\footnote{In principle, one could study this amplitude when the saddle point is not above the black hole threshold. In this regime, the amplitude is dominated by enigmatic states that are at the edge of the integration domain. }. This is compatible with the idea that the operator $\clo$ creates a conical defect that propagates through the bulk. Note that for the operator $\clo$ to create a conical defect but not a black hole, we should also require that $\beta c/12$ is less than $c/24$, this can always be achieved as long as the inverse temperature is small enough. Finally, the spread of the saddle, to leading order in $h_\clo$, is of the order of $\Delta h = \sqrt{h_\clo}/(2\beta)$. Since $\Delta h / h^{\star} \sim \frac{1}{\sqrt{h_\clo}}$, the wormhole amplitude is fully dominated by states that are above the black hole threshold.

\subsection{Genus-two wormhole and product Hilbert spaces}
\label{sec:genus-two-worm}

As a second application of the state-averaging ansatz, we consider the genus-two wormhole, which connects two genus-two Riemann surfaces through the bulk. Microscopically, this amplitude contributes to the variance of the genus-two partition function.

In the CFT, one can construct the genus-two partition function as two spheres glued together by three cylinders of lengths $\beta_1, \beta_2$, and $\beta_3$. Such a construction leads to a partition function of the form 
\begin{equation}
\label{rawgen2}
      Z_{g=2}(\beta_1,\beta_2,\beta_3) = \sum_{ijk} |\bra{i} \clo_j \ket{k}|^2 \e^{-\beta_1 E_i-\beta_2 E_j -\beta_3 E_k},
\end{equation}
where the indices $i, j$ and $k$ run over a complete set of states \cite{Belin:2021ibv}. This construction is known as the plumbing construction, which we describe in more detail in section \ref{sec:gen-obs}. 

Formula \eqref{rawgen2} sums over the full spectrum of the CFT. However, we need to separate the contribution of primary states from their descendants. We will be using the basis of states given in \eqref{eq:cftbasis}, and the generalization of the three-point function $\nu$ given in \eqref{threepf} for correlations between any three Virasoro descendants:\footnote{As in the previous section, we will be ignoring the antiholomorphic counterpart of each expression and focus on the holomorphic part only. The notation $\clo^{(N_2)}_2(z)$ in \eqref{genthrep} corresponds to the descendant field associated to the state $L_{-N_2}\ket{h_{\clo}}$ via the state-operator correspondence.} 
\begin{equation}
\label{genthrep}
    \bra{h_1} L_{N_1} \clo_2^{(N_2)}(z) L_{-N_3}\ket{h_3} = C_{123} \;\nu(h_1, N_1; h_2, N_2; h_3,N_3 | z).
\end{equation}
We can then rewrite the genus-two partition function as:
\begin{multline}
\label{eq:g2pf}
  \!\! Z_{g=2}(\mb \beta) = 
  \hspace{-3mm}
   \sum_{\clo_1,\clo_2,\clo_3} 
   \hspace{-2mm}
   |C_{123}|^2\; 
   \e^{-\mb \beta \cdot \mb h}
   \sum_{\mb n= (0,0,0)}^{\infty}  \e^{-\mb \beta \cdot \mb n}\\\times
   \sum_{\substack{|N_i| = |N'_i| = n_i}} 
    \left[B_{h_1,n_1}^{-1}\right]_{N_1N'_1}\!
    \left[B_{h_2,n_2}^{-1}\right]_{N_2N'_2}\!
    \left[B_{h_3,n_3}^{-1}\right]_{N_3N'_3}
    \times\nu(\mb h,\mb N|1)
    \nu(\mb h, \mb N|1).
\end{multline}
Here we are using bold vector notation to write the triplets $\mb{\beta} = (\beta_1,\beta_2,\beta_3)$, $\mb{h} = (h_1,h_2,h_3)$, $\mb{n} = \big(|N_1|,|N_2|,|N_3|\big)$ and $\mb{N} = (N_1,N_2,N_3)$ labeling the temperature, conformal weight and level of each copy of the Hilbert space. Since we will only consider the function $\nu(\mb h, \mb N|z)$ at $z =1$, we will often omit this label. 

We find it useful to write \eqref{eq:g2pf} directly in terms of the orthonormal basis \eqref{eq:cftbasis}. To do this, we first  define the normalized Virasoro three-point function $\mathcal{C}$ as 
\begin{equation}
    \mathcal{C}(\mb h, \mb M|z) \coloneqq 
    \sum_{N_1,N_2,N_3} 
    \left[B_{h_1,n_1}^{-\frac{1}{2}}\right]_{M_1N_1}
    \left[B_{h_2,n_2}^{-\frac{1}{2}}\right]_{M_2N_2}
    \left[B_{h_3,n_3}^{-\frac{1}{2}}\right]_{M_3N_3}
    \nu(\mb h, \mb N|z).
\end{equation}
Since this basis absorbs the Gram matrices, the genus-two conformal block takes a somewhat simpler form
\begin{equation}
    \mathcal{F}_{g=2}(\beta_i;h_i) \coloneqq \sum_{\mb m = (0,0,0)}^{\infty} \e^{-\mb \beta \cdot (\mb h+ \mb m)} \sum_{|M_i| = m_i}\mathcal{C}(\mb h, \mb M)\,\mathcal{C}(\mb h, \mb M).
\end{equation}
Here we have omitted the label $z=1$ in $\mathcal{C}$.

Building the conformal blocks using the plumbing construction as above fixes a choice of conformal frame. Conformal blocks constructed in different conformal frames differ from each other by a factor given by the conformal anomaly. In the case of the torus one-point function, this difference is accounted for the Casimir energy of the cylinder. For higher genus Riemann surfaces, the plumbing frame fixes the conformal anomaly in such a way that the blocks are $c$-independent functions to leading order at very small temperatures -- i.e. the Casimir energy is zero because the fixture is done on the plane instead of the cylinder. We discuss this construction in more details in section \ref{sec:gen-obs} where it will play an important role in understanding more general wormhole amplitudes. 

In the plumbing frame, the genus-two partition function has a natural interpretation in terms of a thermal correlation in a tripled Hilbert space $\mathcal{H}^{\otimes 3}$:
\begin{equation}
    Z_{g=2}(\beta_1,\beta_2,\beta_3) = \Tr \mathbb{O} \,\e^{-\beta_1 H_1 -\beta_2 H_2-\beta_3 H_3},
\end{equation}
with $H_i =  L_0$ acting on the $i^{\text{th}}$ copy of the Hilbert space and $\mathbb{O}$ is an operator defined via its matrix elements:
\begin{equation}
    \bra{ \mb h, \mb N} \mathbb{O} \ket{\mb h', \mb N'} = C_{123} C_{1'2'3'} \;\mathcal{C}(\mb h, \mb N)\,\mathcal{C}(\mb h', \mb N').
\end{equation}

Having written the genus-two partition function as a thermal correlation function, we can try to evaluate its corresponding wormhole amplitude using an ensemble of states centered at $\mean{\rho} = \rho_\beta\otimes\rho_\beta\otimes\rho_\beta$. 
Using the arguments discussed in section \ref{sec:tensor-prod}, the state-averaging ansatz yields\footnote{Note that we have dropped the normalization factors of $Z(\beta_n)^{-2}$.}
\begin{equation}
\label{eq-gen-two-ten}
    \mean{ \delta \rho_{\mb i \mb j} \delta\rho_{\mb j \mb i} } =  
    j(\bar{\mb{E}},\mb \omega;\mb\beta)\prod_{n=1}^{3} \e^{-2 \beta_n \bar E_n}
    \e^{-S(\bar E_n)},
\end{equation}
where $\mb i, \mb j$ are multi-indices, $\mb i = (i_1,i_2,i_3)$,
$\mb\omega = \big(E_{i_1}-E_{j_1},E_{i_2} - E_{j_2},E_{i_3} - E_{j_3}\big)$ are the energy differences, and  $\bar{\mb{E}} = (\bar E_1,\bar E_2,\bar E_3)$ are the mean energies of each component of the Hilbert space. As with the state-averaging ansatz, we require that the smooth function $j$ is highly peaked at $\mb \omega  = 0$.

The generalization of this ansatz to one that respects conformal symmetry is straightforward. We will require that the ansatz is valid only for primary states $\ket{\mb h}$ and fix the matrix elements of descendants states to be
\begin{equation}
    \bra{\mb h, \mb M} \delta\rho \ket{\mb h',\mb N'} = \left( \e^{-\mb \beta \cdot \mb n} \delta_{\mb N,\mb N'}  +\order{\mb h - \mb h'}\right)\bra{\mb h} \delta\rho \ket{\mb h'}.
\end{equation}
By following the same steps as in the torus one-point function, this ansatz leads to the following wormhole amplitude
\begin{align}
    \mean{\Tr (\delta\rho\mathbb{O}) \Tr(\delta\rho\mathbb{O}^{\dagger}) } 
        &= \sum_{\mb h, \mb h'} C_{123}^2 C_{1'2'3'}^2 
        \overline{|\bra{\mb h'}\delta\rho\ket{\mb h}|^2} \\ \notag
        &\phantom{.}\hspace{1cm}\times\left[\sum_{\mb M,\mb M'} 
        \e^{-\mb \beta \cdot (\mb m + \mb m')} \mathcal{C}(\mb h, \mb M)\mathcal{C}(\mb h', \mb M)\mathcal{C}(\mb h, \mb M')\mathcal{C}(\mb h', \mb M')+\order{\mb h - \mb h'}\right] \\[1em] 
        &\approx \int \prod_{i=1}^3\dd {\sf h}_i\;\e^{S_0({\sf h}_i)} j'(\mb{{\sf h}};\mb\beta) C_0({\sf h}_1,{\sf h}_2,{\sf h}_3)^2 \mathcal{F}_{g=2}(\mb{{\sf h}};\mb{\beta})^2
\end{align}
where in the last line we are using the notation ${\sf h}_i = (h_i+h_i')/2$. 
To go from the second to the third line, we used the continuum approximation of the sum and assumed that $j(\mb \omega, {\sf h};\mb\beta)$ only has support within a narrow window around  $\mb \omega = 0$. 
Note that this condition is necessary to recover the genus-two conformal blocks. The function $j'(\mb{h};\mb\beta)$ is a redefinition of $j$ that results from performing the $\mb{\omega}$ integral. 
Taking $j'$ to be constant reproduces the gravitational computation of \cite{Chandra:2022bqq}:
\begin{align}
\label{eq:g2worm}
        \mean{\Tr\mathbb{O} \delta\rho\Tr\mathbb{O}^{\dagger} \delta\rho}  = \eqimg{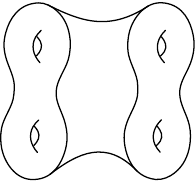}. 
\end{align}

\subsection{General observables in 3D gravity}
\label{sec:gen-obs}

Building on the previous examples, we now discuss an approach that extends to a more general set of observables: correlation functions of Virasoro primaries on general Riemann surfaces. The key insight of this section is that any such observable can be constructed using a \emph{pair-of-pants decomposition}, which reduces the observable into a set of three-punctured spheres sewn together by cylinders. Such a way of building observables in the CFT is known as the plumbing construction.

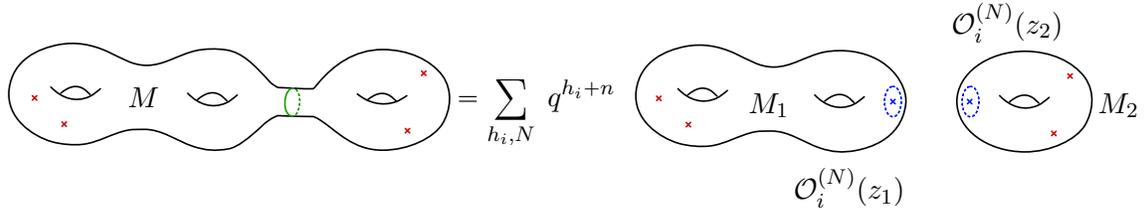
\begin{figure}
    \centering
    \begin{tikzpicture}
    
    \node at (0,0) [midway] {\includegraphics{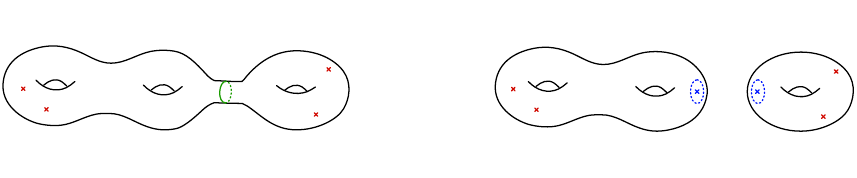}};
    
    \node at (1,-0.2) [left] {$ = \displaystyle \sum_{h_i,N} \; q^{h_i+n}$};

    \node at (3.1,-1.2) [right] {$\displaystyle \clo^{(N)}_i(z_1)$};
    \node at (5.2,1.) [right] {$\displaystyle \clo^{(N)}_i(z_2)$};

    \node at (-5.75,0) [right] {$M$};
    \node at (2.5,-0.1) [right] {$M_1$};
    \node at (7.15,-0.1) [right] {$M_2$};
    
    \end{tikzpicture}
    \caption{The correlation function on $M$ can be expressed in terms of simpler correlation functions by cutting open the path integral and inserting a complete set of boundary states. This process is repeated until the observable can be expressed as a product of sphere three-point functions.}
    \label{fig:Amplitude}
\end{figure}

In the plumbing construction, one cuts open the path integral of the CFT defined on a manifold $M$ by inserting a resolution of the identity along a circle. This allows us to express complicated correlations functions in terms of simpler building blocks. Figure \ref{fig:Amplitude} shows this relation pictorially when the manifold $M$ is divided into two manifolds $M_1$ and $M_2$ with coordinates $z_1$ and $z_2$, respectively. The gluing is performed by doing the identification $z_1 = q z_2$ along the circle where the path integral is cut open. We can also understand this relationship algebraically. If the sewing happens at the points $z_1 = 0$ and $z_2=\infty$, then the following relation applies to a general correlation function 
\begin{equation}
\label{eq:plumbing1}
\begin{split}
    \langle X(z_1 = x_a) Y(z_2 = y_b) \rangle_{M} &= \sum_{i,N} \langle X(x_a) \clo_i^{(N)}(z_1 = 0) \rangle_{M_1} \langle \clo_i^{(N)}(z_1 = \infty) Y(y_b) \rangle_{M_2}. 
\end{split}
\end{equation}
Here $X(x_a)$ and $Y(y_b)$ stand for an arbitrary product of local fields inserted at the points $z_1 = x_1,x_2,\dots x_a$ and $z_2 = y_1,y_2,\dots y_b$. Note that on the left-hand side we are using the coordinates on $M_1$ to describe the fields in $X$ and the coordinates on $M_2$ to describe the fields in $Y$. On the right-hand side, we inserted a resolution of the identity $\sum \ket{h_i,N}\bra{h_i, N}$ and used the operator-state correspondence to write the states using the local fields $\clo_i^{(N)}(z_1)$ at $z_1 = 0$ and $z_1 = \infty$. It is important to keep track of the coordinate system used to define the identity operator, as the fields $\clo_i^{(N)}(z_1)$  and $\clo_i^{(N)}(z_2)$ are related to each other by a conformal transformation 
\begin{equation}
    \clo_i^{(N)}(z_1) = q^{-h_i - n} \clo_i^{(N)}(z_2).
\end{equation}
Recalling that the definition of the field $\clo_i^{(N)}(\infty)$ is given by the limit $\lim_{z_1\rightarrow\infty}z_1^{2(h_i+n)}\clo_i^{(N)}(z_1)$, yields the relation $\clo_i^{(N)}(z_1=\infty) = q^{h_i + n} \clo_i^{(N)}(z_2 = \infty)$. This relation allows us to express the correlation function $\langle \clo_i^{(N)}(z_1 = \infty) Y(y_n) \rangle_{M_2}$ purely in terms of the coordinates in $M_2$: 
\begin{equation}
\label{eq:plumbing}
\begin{split}
    \langle X(z_1 = x_a) Y(z_2 = y_b) \rangle_{M} &= \sum_{i,N} q^{h_i+n} \langle X(x_a) \clo_i^{(N)}(z_1 = 0) \rangle_{M_1} \langle \clo_i^{(N)}(z_2 = \infty) Y(y_b) \rangle_{M_2}.
\end{split}
\end{equation}
By iterating this process, we can write any correlation as a product of three-point functions sewn together by cylinders whose length is given by the sewing parameters $q_i$. The parameter $q$ and the inverse temperature $\beta$ are related to each other via the identification 
\begin{equation}
    q = \e^{2\pi i \tau}, \quad \tau = i\beta/2\pi.
\end{equation}
An illustrative example of this construction is the four-point function of Virasoro primaries on the sphere
\begin{equation}
    G_4(q) \coloneqq \langle\clo_1(\infty)\clo_2(1)\clo_3(q)\clo_4(0)\rangle.
\end{equation}
For simplicity we consider the case when $1>|q|>0$, although the final result can be analytically continued outside this region. 
Inserting a resolution of the identity between the operators $\clo_3$ and $\clo_2$ leads to the expression 
\begin{equation}
\label{eq:4point-decom}
\begin{split}
    G_4(q) &=\sum_{h_s, N} \bra{h_1}\clo_2(1)\ket{h_s,N}\bra{h_s,N}\clo_3(q)\ket{h_4} \\ 
    &= q^{-h_3 - h_4} \sum_{h_s,N} \bra{h_1}\clo_2(1)\ket{h_s,N}\bra{h_s,N}\clo_3(1)\ket{h_4}\;
    q^{h_s + n}.
\end{split}
\end{equation}
Here, we can explicitly see the gluing of the two spheres via the parameter $q$. There is also an additional factor of $q^{-h_3-h_4}$ in front of the amplitude due to the operator product expansion of $\clo_3(q)$ approaching $\clo_4(0)$. This factor can be understood from the plumbing construction as a consequence of the conformal anomaly. For this example, the anomaly arises from the change of local to global coordinates. In the plumbing construction the operators $\clo_1$ and $\clo_2$ are defined in the coordinate system $z_1$ of the first sphere, while the operators $\clo_3$ and $\clo_4$ are defined in the coordinate system  $z_2$ of the second sphere. Changing all operators to be on the same coordinate system, $z_1$ in this example, accounts for the additional factor of $q^{-h_3-h_4}$ in equation \eqref{eq:4point-decom}.

Now, we can write the four-point function as a thermal expectation value:
\begin{equation}
    G_4(q) = q^{-h_3-h_4}\Tr(\mathbb{O}_4q^{L_0}),
\end{equation}
where the operator $\mathbb{O}_4$ is defined via its matrix elements as
\begin{equation}
    \bra{h_s',N'}\mathbb{O}_4\ket{h_s,N} = C_{12s'}C_{s34}\mathcal{C}(h_1;h_2;h_s',N')\mathcal{C}(h_s,N;h_3;h_4).
\end{equation}
In general, one can always write correlation functions as thermal expectation values using the plumbing construction. 
\begin{figure}
    \centering
    \begin{tikzpicture}

    \node at (0,0) [midway] {\includegraphics{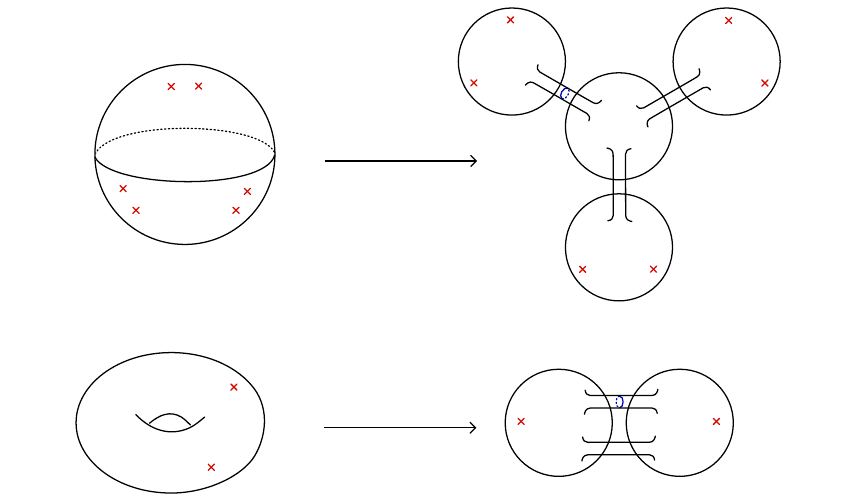}};

    \node at (-6,3.6) [right] {a)};
    \node at (-6,-1) [right] {b)};
    
    \end{tikzpicture}
    \caption{The plumbing construction of the sphere six-point function and the torus two-point function.}
    \label{fig:observables}
\end{figure}
As a second example, let us  consider the sphere six-point function depicted in figure \ref{fig:observables}. Considering the OPE of $\clo_1\clo_2$, $\clo_3\clo_4$ and $\clo_5\clo_6$, we find
\begin{multline}
\label{six-pp}
   G_6(q_i) \coloneqq \langle \clo_1(\infty)\clo_2(1)\clo_3(q_1+q_1 q_2)\clo_4(q_1)\clo_5(q_1 q_3)\clo_6(0) \rangle \\= (q_1q_2)^{-h_3-h_4}(q_1q_3)^{-h_5-h_6}\sum_{ijk} \; \lambda_{12 i}\lambda_{23j}\lambda_{56k} \lambda_{ijk} \;q_1^{h_i}q_2^{h_j}q_3^{h_k}.
\end{multline}
For clarity, we have not organized the spectrum into primary and descendant fields. The sum here is over the all operators of dimensions $h_i$, or equivalently, all the eigenstates $\ket{i}$ of $L_0$. We denote the generic three-point coefficients with a different label $\lambda_{ijk}$ as we will reserve the label $C_{ijk}$ for three-point coefficients between primary operators. 

Expression \eqref{six-pp} can be interpreted as a thermal correlator in a tripled Hilbert space
\begin{equation}
     G_6(q_i) = (q_1q_2)^{-h_3-h_4} (q_1q_3)^{-h_5-h_6} \Tr_{\mathcal{H}^{\otimes 3}}\mathbb{O}_6 q_1^{H_1}q_2^{H_2}q_3^{H_3},
\end{equation}
with $H_i = L_0$ acting on the $i^{\text{th}}$ copy of the Hilbert space, and $\mathbb{O}_6$ defined via its matrix elements:
\begin{multline}
\bra{\mb h, \mb N}
\mathbb{O}_6
\ket{\mb h', \mb N'} \coloneqq C_{12i}C_{j34}C_{k56} C_{i'j'k'}\\
   \times
   \mathcal{C}(h_1;h_2;h_i,N_i)
    \mathcal{C}(h_j,N_j;h_3;h_4)
    \mathcal{C}(h_k,N_k;h_5;h_6)
    \mathcal{C}(h_i',N_i';h_j',N_j';h_k',N_k').
\end{multline}
An interesting example is that of the torus two-point function depicted in figure \ref{fig:observables}. Although the observable itself is a thermal correlator, we can also be express it as a thermal expectation value in a doubled Hilbert space
\begin{equation}
    \Tr_{\mathcal{H}} \clo_1(x)\clo_2(1)q^{L_0} = q_1^{h_1} \Tr_{\mathcal{H}\otimes \mathcal{H}}  \mathbb{O}_{T_2}q_1^{L_0}q_2^{L_0},\quad x=\frac{1}{q_1},\quad q = q_1q_2. 
\end{equation}
The difference between the first and the second description is that, in the latter one, all the information about the position $x$ and the temperature $q = \e^{-\beta}$ are encoded in the auxiliary temperature parameters $q_1$ and $q_2$. Moreover, the operator $\mathbb{O}_{T_2}$ is not a local operator, but instead, it encodes information about the OPE coefficients in the pants decomposition:
\begin{equation}
    \bra{\mb h,\mb N}\mathbb{O}_{T_2}
    \ket{\mb h', \mb N'} \coloneqq 
    C_{ij2}C_{j'i'1} 
    \mathcal{C}(h_i,N_i;h_1;h_j,N_j)
    \mathcal{C}(h_j',N_j';h_2';h_i',N_i').
\end{equation}

\subsection*{Wormhole amplitudes from fuzzy cylinders}

The idea is to apply the state-averaging ansatz to the thermal correlator in the tensor product Hilbert space. The averaged state in this ensemble is given by the product state $\mean{\rho_{\sf k}} = \rho_{\beta_1}\otimes\cdots\otimes \rho_{\beta_{\sf k}}$.
Writing a state $\rho$ in this ensemble as $\rho = \mean{\rho_{\sf k}} + \delta\rho$, the state-averaging ansatz makes a prediction for the moments of $\delta\rho$. Taking conformal symmetry into account, the ansatz yields\footnote{
The ansatz in \eqref{eq:tensor} corresponds to the simplest generalization of the state-averaging ansatz to a tensor product Hilbert space. One could, in principle, include extra contractions if needed to accommodate for more complicated wormhole amplitudes.
}
\begin{equation}
\label{eq:tensor}
\begin{split}
 \mean{\bra{\mb{h}}\delta \rho \ket{\mb{h}'}\bra{\mb{h}'}\delta \rho \ket{\mb{h}}}
    &= 
    j(\mb{{\sf h}}, \mb \omega;\mb\beta)
    \prod_{i=1}^{\sf k}
   \e^{-2\beta_i {\sf h}_{i}}\e^{-S_0( {\sf h}_{i})},
    \\
    \bra{\mb{h},\mb{N}}\delta \rho \ket{\mb{h}',\mb{N}'}
    &=  \left(\e^{-\mb \beta \cdot \mb n} \delta_{\mb N,\mb N'}  +\order{\mb h - \mb h'}\right)
    \bra{\mb{h}}\delta \rho \ket{\mb{h}'},
\end{split}
\end{equation} 
where ${\sf h } = (h+h')/2$ denotes the mean weight. We can use this ansatz to compute the quantity
\begin{equation}
    \mean{\Tr \delta\rho \mathbb{O}\Tr \delta\rho \mathbb{O}^{\dagger}}
\end{equation}
associated to the wormhole amplitude of a generic correlation function. Since the operator $\mathbb{O}$ is not  in general self-adjoint, we have explicitly included its Hermitian conjugate $\mathbb{O}^\dagger$ in the amplitude. 

An analysis analogous to the amplitude of the the punctured torus wormhole yields the result
\begin{equation}
\label{eq:statav}
    \mean{\Tr \delta\rho \mathbb{O}\Tr \delta\rho \mathbb{O}^{\dagger}} = 
    \left(\prod_{i=1}^{\sf k}\int\dd h_i\;\e^{S_0(h_i)}\right) j'(\mb h;\mb\beta) \opmean{\mathbb{O}_{\mb{hh}'}\mathbb{O}^{*}_{\mb{h}\mb{h}'}}_{\mb h=\mb h'}\;\mathcal{F}(\mb h;\mb \beta)^2,
\end{equation}
where $\mathbb{O}_{\mb h\mb h'} = \bra{\mb h}\mathbb{O}\ket{\mb h'}$, and the conformal blocks $\mathcal{F}$ are defined as 
\begin{equation}
\label{eq:gen-block}
    \mathcal{F}(\mb h,\mb \beta) = \sum_{|N_i| = n_i\geq 0} \e^{-\mb \beta \cdot (\mb h + \mb n)} g(\mb h, \mb N,\mb h, \mb N).
\end{equation}
The smooth function $g$ is given by the overlap
\begin{equation}
     g(\mb h, \mb N,\mb h', \mb N') \coloneqq \frac{\bra{\mb h,\mb N}\mathbb{O}\ket{\mb h',\mb N'}}{\bra{\mb h}\mathbb{O}\ket{\mb h'}}.
\end{equation}
Equation \eqref{eq:gen-block} is the definition of the conformal block associated to the original observable in the plumbing frame. 

Applying equation \eqref{eq:statav} to the example of the sphere six-point  function gives the wormhole
\begin{multline}
\label{six-point-wha}
\eqimg{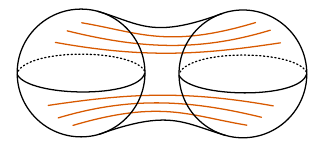} =\int \dd h_i\dd h_j\dd h_k \;\e^{S_0(h_i)+S_0(h_j)+S_0(h_k)} \;  C_0(h_1,h_2,h_i) \\[-0.7cm]\times C_0(h_3,h_4,h_j)  C_0(h_5,h_6,h_k)C_0(h_i,h_j,h_k) \;\mathcal{F}(\mb h;\mb \beta)^2,
\end{multline}
Here we have taken $j'(\mb h)$ to be a constant function. The conformal block of this amplitude is given by the formula
\begin{multline}
\label{sixpointblock}
   \mathcal{F}(\mb h;\mb \beta) =  q_1^{h_i}q_2^{h_j}q_3^{h_k}
   \sum_{\substack{|N_m|=n_m\geq0\\m=1,2,3}}q_1^{n_1}q_2^{n_2}q_3^{n_3} 
   \mathcal{C}(h_1;h_2;h_i,N_1)
    \mathcal{C}(h_j,N_2;h_3;h_4)\\\times
    \mathcal{C}(h_k,N_3;h_5;h_6)
    \mathcal{C}(h_i,N_1;h_j,N_2;h_k,N_3).
\end{multline}
Since these blocks and amplitudes are given in the plumbing frame, there are no additional factors of the form $(q_1q_2)^{-2h_3-2h_4}$.

A  similar expression holds for the case of the torus two-point function,
\begin{multline}
\!\!\eqimg{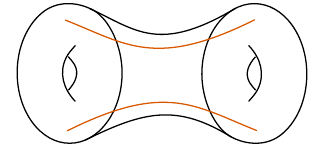} \!\!\!\!=\!\int \dd h_i\dd h_j\,\e^{S_0(h_i)+S_0(h_j)}  C_0(h_i,h_j,h_2)C_0(h_i,h_j,h_1) \mathcal{F}(\mb h;\mb \beta)^2,
\end{multline}
with a analogous expression to \eqref{sixpointblock} describing the torus two-point conformal block.

Having derived expressions for generic wormhole amplitudes from the point of view of state averaging, we end the discussion with a comparison to the operator approach to 3D gravity.

\subsection{Comparison with the operator approach}
\label{sec:comparison}

Wormholes amplitudes have also been interpreted in terms of an ensemble of CFT data constrained by the conformal bootstrap \cite{Belin:2020hea,Chandra:2022bqq}. In this interpretation, to calculate the wormhole amplitudes, one first writes correlation functions in terms of the OPE data and then one applies the operator average $\opmean{\bullet}$ we introduced in equation \eqref{opemean}. Let us consider an observable of the form 
\begin{equation}
    G(\mb \beta) = \sum C\dots C \times \mathcal{F}(\mb{\beta};\mb h).
\end{equation}
Where $C\dots C$ denotes a configuration of OPE coefficients. Using the notation of section \ref{sec:gen-obs}, we can also denote $C\dots C$ by $\mathbb{O}_{\mb h, \mb h}$. From the operator average point of view, the wormhole amplitude associated to this observable is given by the expression
\begin{equation}
\label{eq:opaverage}
   \opmean{G(\mb \beta)G(\mb \beta)^{*}}^{\text{conn.}} = \int\left(\prod_{i}\dd h_i \dd h'_i\;\e^{S_0(h_i)+S_0(h_i')}\right) 
   \opmean{\mathbb{O}_{\mb h,\mb h}\mathbb{O}^*_{\mb h',\mb h'}}^{\text{conn.}}\; \mathcal{F}(\mb \beta,\mb h)^2
\end{equation}
where by $\opmean{\bullet_1\bullet_2}^{\text{conn.}}$ we mean the connected correlation $\opmean{\bullet_1\bullet_2} -\opmean{\bullet_1}\opmean{\bullet_2}$. This time the operator average $\opmean{\bullet}$ is done deliberately. This is unlike state averaging where wormhole correlations are attributed to a lack of knowledge about the state, and the operator average $\opmean{\bullet}$ appears naturally as a coarse-graining within a small microcanonical window. 

When comparing \eqref{eq:opaverage} with the result from state averaging \eqref{eq:statav}, it appears that the two approaches produce different results. However, assuming that $C_{ijk}$ are Gaussian random variables, \eqref{eq:opaverage} reduces to \eqref{eq:statav} up to non-perturbative effects. The reason is that the leading contraction in the operator approach is the one that sets $\mb h = \mb h'$. We can see how this works explicitly with the example of the torus two-point wormhole. 

From the operator approach, the result that we need is 
\begin{equation}
\label{eq:opeavv}
\opmean{\mathbb{O}_{\mb h,\mb h}\mathbb{O}^*_{\mb h',\mb h'}}^{\text{conn.}} = \opmean{C_{ij1}C_{ij2} C_{i'j'1}C_{i'j'2}}^{\text{conn.}}  = \delta_{ii'}  \delta_{jj'} \opmean{C_{ij1}^2}\opmean{C_{ij2}^2}.
\end{equation}
Note that when transitioning from the sum over states $\sum_i$ to the integral $\int \dd h_i\,\varrho_0(h_i)$, we have to replace the discrete function $\delta_{ii'}$ with its continuous version $\e^{-S_0(h_i)}\delta(h_i-h_i')$. These eliminate half of the factors of the Cardy formula in \eqref{eq:opaverage}. State averaging requires us to integrate over the slightly different average
\begin{equation}
\opmean{\mathbb{O}_{\mb h,\mb h'}\mathbb{O}^*_{\mb h,\mb h'}}_{\mb h=\mb h'} = \opmean{C_{ij1}C_{i'j'2} C_{ij1}C_{i'j'2}}_{\mb h=\mb h'} = \opmean{C_{ij1}^2}\opmean{C_{ij2}^2},
\end{equation}
which, nonetheless, matches the operator approach. 

It is well-known that the OPE coefficients of chaotic 2d CFTs cannot be exactly Gaussian pseudo-random variables \cite{Belin:2021ryy}. Important non-Gaussian contractions are needed for the CFT to be modular invariant and crossing symmetric. These non-Gaussianities can be incorporated into the operator- and state-averaging approach to 3D gravity in very natural ways. We study in detail these corrections in a forthcoming paper \cite{deBoer:2024kat}. 

\section{Wormholes in higher dimensions}
\label{sec:higher_d}

We have seen that the state-averaging ansatz, for a small perturbation of a bulk thermal state, {\it outputs} the correct wormhole amplitudes in 3D gravity. In this section, we broaden the discussion to $D\geq 3$. We first derive a generalized version of the state-averaging ansatz \eqref{eq:state_averaging_ansatz} that serves as the maximally ignorant model of more general families of semiclassical states; these include large perturbations of thermal states. The generalization comes from the fact that the ansatz fails to satisfy property \ref{assumption3} outlined in section \ref{sec:saa23}. The envelope function $j_0(E,\omega)$ will scale with the microcanonical entropy, and the state itself will modify the microcanonical saddle point equations determining correlation functions and other quantities. This feature of the ensemble is dual to the fact that the black hole interior contains large perturbations which backreact on the semiclassical geometry.

To be specific, we consider semiclassical states of black holes with long Einstein-Rosen bridges in the black hole interior. As the reference CFT state with such a semiclassical dual, $\ket{\Psi_{\text{sc}}}$, we will consider one of the so-called `partially entangled thermal states' (PETS) \cite{Goel:2018ubv,Sasieta:2022ksu,Balasubramanian:2022gmo,Balasubramanian:2022lnw,Chandra:2023rhx,Antonini:2023hdh}. A PETS is an entangled state in the Hilbert space of two holographic CFTs, $\ket{\Psi_\Op} \in \mathcal{H}_\Le\otimes \mathcal{H}_\Ri$. Such state is prepared by a Euclidean CFT path integral on a flat cylinder, with length moduli $\frac{\tbeta_L}{2},\frac{\tbeta_R}{2}$, and an additional operator insertion $\Op$ in between the Euclidean time evolutions:

\be\label{eq:PETSdraw}
\begin{gathered}
\includegraphics[width = 5cm]{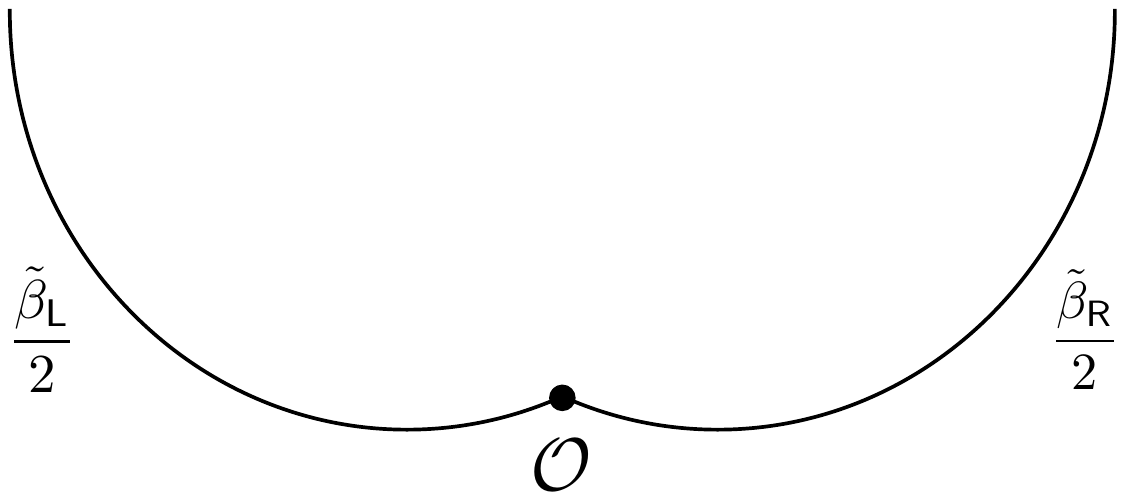} \nonumber
\end{gathered} 
\ee 
The PETS can be expanded in the energy basis of the doubled Hilbert space as
\be\label{eq:PETS}
\ket{\Psi_\Op} = \dfrac{1}{\sqrt{Z_1}}\sum_{i\alpha} \e^{-\frac{\tbeta_\Le}{2}E_i-\frac{\tbeta_\Ri}{2}E_\alpha} \Op_{i\alpha}\ket{E_i}\ket{E_\alpha}\;,
\ee 
where $Z_1$ is a normalization factor, and $\Op_{i\alpha} = \bra{E_i} \Op \ket{E_\alpha}$ are the matrix elements of the operator $\Op$ in the energy eigenbasis, which enter here as the coefficients of the two-sided state. In order for the state \eqref{eq:PETS} to admit a semiclassical bulk description, the operator $\Op$ must be chosen accordingly. In particular, the operator must itself have a semiclassical description, in terms of an operator of the low-energy gravitational EFT. This does not preclude the operator of backreacting on the spacetime -- semiclassical heavy matter can be constructed out of long wavelength quanta.

A specific family of operator which are universal and convenient to construct semiclassical wormhole solutions in $D\geq 3$ are the so-called {\it thin shell operators}. This class of operator exists in any microscopic theory describing Einstein gravity with massive matter particles in the spectrum. In AdS/CFT, the corresponding microscopic CFT operator consists of the product of $O(N^2)$ conformal primaries of low conformal dimension $1 \ll \Delta \ll N^2$, slightly smeared in order to regularize its energy, and distributed in an approximately homogeneous way along the sphere, $\Op = \prod_{i=1}^n \Op_\Delta(\Omega_i)$. The number of insertions is chosen to scale with the central charge, $n\sim N^2$. This class of operator admits a semiclassical description: it creates a homogeneous thin interface of dust particles, with total rest mass $m$, localized close to the asymptotic region of AdS space. The particles that form the shell are heavy enough to behave classically. Therefore, the heavy shell propagates coherently and backreacts on the geometry at leading order in the semiclassical $G_{\text N}\rightarrow 0$ expansion.

Given this choice of operator, for $\tbeta_\Le, \tbeta_\Ri \ll 1$ in AdS units, the semiclassical description the PETS will include a thin shell and a two-sided black hole. In \cite{Balasubramanian:2022gmo} it was shown that the preparation moduli $\tbeta_\Le, \tbeta_\Ri$ can always be chosen as a function of the rest mass of the shell $m$, in such a way that the thin shell lives in the black hole interior at the moment of time symmetry. More precisely, there will be two black holes, of physical temperatures $\beta_\Le$ and $\beta_\Ri$, solving the system of equations \cite{Sasieta:2022ksu,Balasubramanian:2022gmo}
\begin{align}
&\beta_L = \tbeta_\Le + \Delta \tau_- \;, \label{eq:spbh1}\\
&\beta_R = \tbeta_\Ri + \Delta \tau_+ \;,\hspace{2cm}\label{eq:spbh2}
\end{align}
corresponding to the ADM energies $E_\Le$ and $E_\Ri$. These equations relate the physical temperature of the black holes, to the preparation temperatures of the PETS. The functions $\Delta \tau_\pm$ represent the Euclidean time ellapsed by the thin shell in the saddle point geometry and can be found in \cite{Sasieta:2022ksu,Balasubramanian:2022gmo}. 

The two black holes share a long Einstein-Rosen bridge connecting them at the moment of time symmetry. The bridge is supported by the presence of the thin shell of rest mass $m$. In this construction, the ADM energies $E_\Le$ and $E_\Ri$ can be fixed, and one considers a one-parameter family of semiclassical states, parametrized by the rest mass $m$ of the thin shell in the black hole interior. For our purposes in what follows, we will completely fix the semiclassical state by keeping $m$ fixed.

\subsection{Ensemble from maximum ignorance}

Up to this point, the description of the PETS \eqref{eq:PETS} is exact, and there is no ensemble. The PETS is a genuine microscopic state with a bulk {\it effective} semiclassical dual corresponding to a two-sided black hole with a long Einstein-Rosen bridge supported by the thin shell. 

On the other hand, there are multiple microscopic states which share the same semiclassical description. We will now apply the principle of maximum ignorance to find this ensemble of microstates. As outlined in section \ref{sec:SAA}, the input data that we use to define the ensemble of states is the Euclidean overlap function $\mathbf{Z}(\tau_1,\tau_2) \coloneqq \bra{\Psi_\Op}\e^{-\tau_1 H_L}\e^{ -\tau_2 H_R}\ket{\Psi_\Op}$ of the PETS, which in a coarse-grained sense is uniquely determined from its semiclassical bulk description. In particular, such smooth function is computed in a bulk saddle point approximation by the following geometry (cf. \cite{Sasieta:2022ksu,Balasubramanian:2022gmo}):
\be\label{eq:PETSdraw2}
\mathbf{Z}(\tau_1,\tau_2) =\hspace{.2cm} \begin{gathered}
\includegraphics[width = 8.5cm]{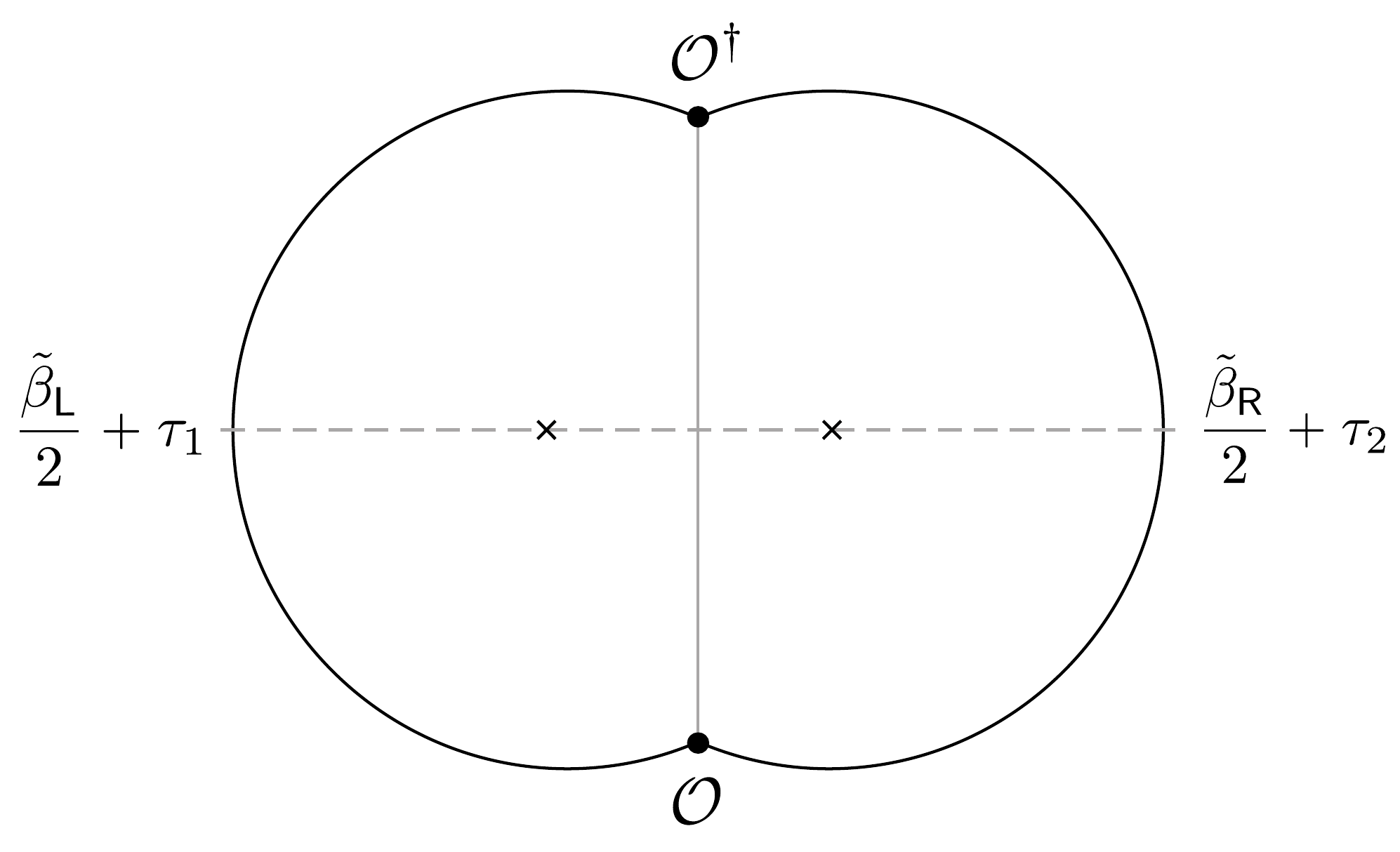} 
\end{gathered} 
\ee 
For our purposes it will be sufficient to consider $\tau_1,\tau_2\ll 1$ and model the microscopic state in the microcanonical bands associated to large black holes in AdS.

Applying the principle of maximum ignorance with this input leads to an ensemble of microstates of the form
\be\label{eq:PETSensemble}
\ket{\Psi} = \dfrac{1}{\sqrt{Z_1}}\sum_{i,\alpha} \e^{-\frac{\tbeta_\Le}{2}E_i-\frac{\tbeta_\Ri}{2}E_\alpha} A_{i\alpha}\ket{E_i}\ket{E_\alpha},
\ee 
specified by a Gaussian ensemble of the coefficients of the two-sided state
\be\label{eq:thinshellheavy} 
A_{i\alpha} = \e^{-S(\bar{E}_{i\alpha})/2}\, g(\bar{E}_{i\alpha}, \omega_{i\alpha})^{1/2} a_{i\alpha}, 
\ee 
where $a_{i\alpha}$ are independent Gaussian random variables of zero mean and unit variance. This ensemble can be recognized as the {\it generalized ETH ensemble} of microscopic operators that model the thin shell operator $\Op$ \cite{Sasieta:2022ksu}. The generalization from ETH comes from the fact that the operator $\Op$ is extensive in the entropy $S(E)$, at least for energies $E\sim m$ associated to the mass of the shell. The motivation behind this ansatz is that the operator still admits a semiclassical description and thus can still be considered as simple. When $E_i,E_j \gg m$ the ansatz reduces to the standard ETH ansatz. In the generalization, the function $g(\bar{E}_{i\alpha}, \omega_{i\alpha})$ will actually scale with the entropy. When inserted into thermal correlators, the operator $A$ will be able to modify the dominant microcanonical saddle point in the thermodynamic limit. In the semiclassical description, this phenomenon is dual to the classical backreaction that the thin shell exerts on the Euclidean geometry at leading order in $G_{\text N}\rightarrow 0$.

The smooth envelope function $g(\bar{E}_{i\alpha}, \omega_{i\alpha})$ is completely determined from the double inverse Laplace transform of $\mathbf{Z}(\tau_1,\tau_2)$. Defining the variable
\be
    f(E_i,E_\alpha)\coloneqq S(\bar{E}_{i\alpha}) - \log g(\bar{E}_{i\alpha}, \omega_{i\alpha}),
\ee
the exact match requires that
\be
f(E_i,E_\alpha) =  \alpha_- S(E_i) + \alpha_+ S(E_\alpha) + I_{s}(E_i,E_\alpha),
\ee
where $\alpha_\pm(E_i,E_\alpha)$ are $\order{1}$ coefficients, and $I_{s}(E_i,E_\alpha) = \order{m\ell}$ is a smooth function which can be found in \cite{Sasieta:2022ksu,Balasubramanian:2022gmo}. For our purposes it is enough to know that this function satisfies the property that
\be\label{eq:propertyenvfunction}
\dfrac{\partial}{\partial E_\pm} f(E_-,E_+) = \Delta\tau_\pm(E_-,E_+)\;,
\ee 
for the functions $\Delta\tau_\pm$ entering in \eqref{eq:spbh1} and \eqref{eq:spbh2}.

In this way, the microscopic thin shell operator $\Op$ is modeled as an ensemble of operators $A_{ij} 
$ which are indistinguishable at the level of the semiclassical thermal two-point function. For the PETS, this means that its semiclassical description will also be indistinguishable from the semiclassical description of the ensemble of states \eqref{eq:PETSensemble}. The ensemble of two-sided pure microstates induces an ensemble of reduced density matrices $\rho_{\Le}$ and $\rho_{\Ri}$, all of which are compatible with the bulk semiclassical description of certain region of spacetime. In particular, without loss of generality, we will focus on the reduced state to CFT$_\Le$, which from \eqref{eq:PETS} has the form
\be
\rho_\Le = \dfrac{1}{Z_1} \e^{-\frac{\tbeta_\Le}{2}H} A^\dagger \e^{-\tbeta_\Ri H} A\e^{-\frac{\tbeta_\Le}{2}H}.
\ee 
Plugging the maximally ignorant ansatz for the coefficients \eqref{eq:ETH} we get the ensemble of reduced density matrices
\begin{align}\label{app1:reducedstate}
(\rho_\Le)_{ij} &= \dfrac{1}{Z_1}\sum_{l} \e^{-\frac{\tbeta_\Le}{2}(E_i+E_j)}\e^{-\tbeta_\Ri E_l} A_{il}A^*_{jl}  \\
&= \dfrac{\e^{-\tbeta_{\Le} \bar{E}_{ij}}}{Z_1} \Big(F_1(\bar{E}_{ij}) \delta_{ij}\,   + F_2(\bar{E}_{ij},\omega_{ij}) R_{ij}\Big),
\end{align}
for the smooth functions
\begin{align}
F_1(\bar{E}_{ij}) &= \sum_l \e^{-\tbeta_\Ri E_l - f(\bar{E}_{ij},E_l)},\\
F_2(\bar{E}_{ij},\omega_{ij}) &= \sum_{l} \e^{-\tbeta_\Ri E_l - \frac{1}{2}(f(E_i,E_l)+f(E_j,E_l)) - \frac{S(E_l)}{2}}.
\end{align}

As discussed in section \ref{sec:saa23},
$R_{ij}$ are a set of independent approximate Gaussian random variables of zero mean and unit variance. Both functions $F_1(\bar{E})$ and $F_2(\bar{E},\omega)$ admit semiclassical limits in the CFT as $N\rightarrow \infty$, where the sum over $l$ is replaced by a continuous integral, and the integral is then evaluated in a saddle point approximation:
\begin{align}
F_1(\bar{E})  &
\sim \e^{S(E_*)-\tbeta_\Ri E_* - f(\bar{E},E_*)},\label{app1:saddlepoint} \\
F_2(\bar{E},\omega) &
\sim \e^{\frac{S(E_{\star})}{2}-\tbeta_\Ri E_{\star} - \frac{1}{2}\big(f(E_i,E_\star)+f(E_j,E_\star)\big)}.\label{app2:saddlepoint}
\end{align} 
From \eqref{app1:saddlepoint} and \eqref{app2:saddlepoint}, the microcanonical energies $E_*(\bar{E}_{ij})$ and $E_\star(\bar{E}_{ij},\omega_{ij})$ solve the respective saddle point equations:
\begin{align}
\beta_{E_*} &= \tbeta_\Ri + \Delta \tau_+(\bar{E}_{ij}, E_*),\label{saddlepoint1}\\
\beta_{E_\star} &= 2\tbeta_\Ri + \Delta \tau_+(E_i, E_\star) + \Delta \tau_+(E_j, E_\star),\label{saddlepoint2}
\end{align}
where we have defined the inverse temperatures $\beta_{E_*} = S'(E_*)$ and $\beta_{E_\star} = S'(E_\star)$. We have also used the relation of the functions $\Delta \tau_\pm$ with the envelope function \eqref{eq:propertyenvfunction}.

Likewise, the value of the normalization of the state in \eqref{app1:reducedstate} is given by the Euclidean two-point function 
\begin{equation}
    Z_1(\tilde \beta_L,\tilde \beta_R) =  \Tr( \e^{-\frac{\tbeta_\Le}{2}H} A^\dagger \e^{-\tbeta_\Ri H} A\e^{-\frac{\tbeta_\Le}{2}H}),
\end{equation}
which can be evaluated semiclassically. The dominant saddle point geometry is the Euclidean manifold illustrated in \eqref{eq:PETSdraw2} which serves to prepare the semiclassical state of the two-sided black hole. In this case the saddle point energies are $E_\Le$ and $E_\Ri$, which correspond to the physical ADM masses of the two black holes. The saddle point value of the normalization yields \cite{Sasieta:2022ksu}
\begin{align}\label{eq:normPETS} 
Z_1(\tilde \beta_L,\tilde \beta_R) \sim \e^{-\tbeta_{\Le} F(\beta_\Le) -\tbeta_{\Ri} F(\beta_\Ri) -I_s(E_\Le,E_\Ri) }.
\end{align} 
Here, the function $F(\beta_E)$ is defined as 
\begin{equation}
    F(\beta_E) \coloneqq E - \beta_E^{-1} S(E),
\end{equation}
and denotes the free energy of the black hole, where $S(E)$ is its Bekenstein-Hawking entropy. Plugging \eqref{app1:saddlepoint}, \eqref{app2:saddlepoint} and \eqref{eq:normPETS} back into the state \eqref{app1:reducedstate}, we find an ensemble of microscopic states of the form
\be 
\label{eq:sec4-20}
(\rho_\Le)_{ij} = \dfrac{\e^{-\beta_\Le \bar{E}_{ij}}}{Z(\beta_\Le)}\left( \delta_{ij} \e^{-w_\Le(\bar{E}_{ij})} + \e^{-v_\Le(\bar{E}_{ij},\omega_{ij})}R_{ij}\right).
\ee 

All factors in this formula need explanation. First, the envelope functions $\omega_L$ and $v_L$ are given by 
\begin{align}
w_\Le(\bar{E}) &=(\tbeta_{\Le}-\beta_{\Le}) \Delta F_\Le + \tbeta_{\Ri} \Delta F_{\Ri,*} + \Delta I_* + \alpha_*\,S(\bar{E}), \label{app1w}\\
v_\Le(\bar{E},\omega) &=(\tbeta_{\Le}-\beta_{\Le}) \Delta F_\Le + \tbeta_{\Ri} \Delta F_{\Ri,\star} + \Delta I_\star + \alpha_\star\,S(\bar{E}).\label{app1v}
\end{align}
For clarity, we have expressed these functions in terms of the differences between the free energies and the differences between the actions:
\begin{align}
    &\Delta F_\Le \coloneqq F (\beta_{\bar{E}}) - F(\beta_\Le), &\quad 
    \Delta F_{\Ri,a} \coloneqq F(\beta_{E_a}) - F(\beta_\Ri) \hspace{3.7cm} \\
    &\Delta I_* \coloneqq I_s(\bar{E},E_*) - I_s(E_\Le,E_\Ri), &\quad
    \Delta I_\star \coloneq \frac{1}{2}\Big[ I_s(E_i,E_\star) + I_s(E_j,E_\star)\Big] - I_s(E_\Le,E_\Ri).
\end{align}
where the index $a=*,\star$. Finally, the coefficients $\alpha_*$ and $\alpha_\star$ in \eqref{app1w} and \eqref{app1v} read:
\begin{align}
\alpha_* = \frac{1}{\beta_{\bar{E}}}\left(\tbeta_{\Le}-\beta_{\Le}\right) - \alpha_-(\bar{E},E_*),\quad 
\alpha_\star = \frac{1}{\beta_{\bar{E}}}\left(\tbeta_{\Le}-\beta_{\Le}\right) - \half \Big(\alpha_-(E_i,E_\star)+\alpha_-(E_j,E_\star)\Big).
\end{align}

Having found \eqref{eq:sec4-20}, the next step is to study the behaviour of the functions $w_L$ and $v_L$. We begin by considering the expansion of $w_\Le(E)$ around $E = E_{\Le}$. When $E$ approaches $E_{\Le}$, the saddle point equation \eqref{saddlepoint1} determining $E_*$ is solved by setting $E_*(E_\Le)$ to $E_\Ri$. Here, $E_\Ri$ is simply the ADM mass of the right black hole in the semiclassical state. Substituting into \eqref{app1w} we get that $w_\Le(E_{\Le}) =0$. A property of \eqref{app1w} which is straightforward to verify with careful evaluation of each of the functions is that the first derivative $w_\Le'(E)$ vanishes at $E = E_{\Le}$. Therefore, the leading term of the function is quadratic
\be 
w_\Le(\bar{E}) = \dfrac{\kappa}{2}(\bar{E} - E_\Le)^2 + \order{(\bar{E} - E_\Le)^3}.
\ee 
The coefficient $\kappa$ is some constant whose value is unimportant for our purposes in this section. The variance of the ensemble is controlled by the smooth function 
\be 
j_2(\bar{E},\omega;\beta) \coloneqq \e^{S(\bar{E})-2v_\Le(\bar{E},\omega)}.
\ee 
From the definitions above, one can verify that $j_2(\bar{E},\omega)$ is an even function of $\omega$. Note that this variance can be large, of order $\e^{S(\bar E)}$, depending on the behaviour of $v_L$. This large envelope will have important consequences in the next subsection where we study the ensemble of microstates in the limit where $m \gg E_L,E_R$.

In this way, we have derived the state-averaging ansatz \eqref{eq:state_averaging_ansatz} for the maximally ignorant ensemble of microscopic states in \eqref{eq:PETSensemble}:
\be\label{eq:PETSensemblestates}
(\rho_\Le)_{ij} = \dfrac{\e^{-\beta_\Le \bar{E}}}{Z(\beta_\Le)} \left(
\delta_{ij}\; \e^{-\frac{\kappa}{2} (\bar{E}-E_\Le)^2}   + \e^{-S(\bar{E})/2} j_2(\bar{E},\omega)^{\frac{1}{2}}R_{ij}
\right).
\ee 
In this section, we have used the simplest model of state averaging, with only the input of $\mathbf{Z}(\tau_1,\tau_2)$ which, as we noted in section \ref{sec:SAA}, results on an equilibrium average state. As developed in section \ref{sec:saa24}, time-dependence of the average state can be included with additional input, such as the two-point function of a simple operator, although we shall not include these effects here. Instead, we will now study some applications of the above result.

\subsection*{Coarse-graining and the black hole exterior}

Assume we want to compute a one-point function of some simple operator $\phi_\Le$ in the ensemble \eqref{eq:PETSensemblestates}. By simple, we mean that the operator $\phi_\Le$ belongs to the generalized free field algebra of the CFT$_\Le$ in the large-$N$ limit, and that it is completely uncorrelated with the state. From \eqref{eq:PETSensemblestates}, this will be semiclassically equivalent to the computation of the expectation value in the average state $\overline{\rho}$ obtained by keeping the diagonal part of \eqref{eq:PETSensemblestates}. In the semiclassical limit, we have
\be\label{eq:1ptfnintegral}
\text{Tr}(\overline{\rho_\Le}\phi_\Le) \approx \dfrac{1}{Z(\beta_\Le)} \int \text{d}E\; \e^{S(E)-\beta_\Le E} \,\e^{-\frac{\kappa}{2}(E - E_\Le)^2} \phi_{\Le}(E,E)\;,
\ee 
where $\phi_{\Le}(E,E)$ is the smooth diagonal envelope function of $\phi_\Le$, assuming ETH form for this operator.

In this form the integral \eqref{eq:1ptfnintegral} will admit a saddle point approximation at $E = E_\Le$, for operators $\phi_\Le$ which are light and do not modify this saddle point. That is
\be 
\text{Tr}(\overline{\rho_\Le}\phi_\Le) = \text{Tr}(\rho_{\beta_{\Le}}\phi_\Le) \;,
\ee 
in the large-$N$ limit. This equation represents the fact that the geometry on the exterior of the black hole is {\it exactly} Schwarzschild-AdS with ADM mass $E_\Le$, so simple probes cannot detect the shell in the black hole interior at the level of the classical geometry, even if the entanglement wedge of the CFT$_L$ contains the black hole interior.

Deviations from exact thermality in this state do exist at order at subleading orders in the semiclassical $G_{\text N}\rightarrow 0$ expansion. These deviations can be detected via the value of $\kappa$. The reason is that the semiclassical state of the bulk quantum fields at $t=0$ in the exterior region is not exactly thermal. This state has been prepared using a Euclidean manifold which has some additional features which affects the quantum state of the bulk fields. As we will see next, these effects can be suppressed arbitrarily by taking the volume of the interior of the black hole to be very large. Alternatively, one can wait some time until the state of the bulk field has thermalized.

\subsection{Large black hole interiors}

As mentioned above, the shell's rest mass $m$ parametrizes the semiclassical phase space of these black hole states, at fixed values of the ADM masses of the two black holes $E_\Le$ and $E_\Ri$. Many of the features of the microscopic ensemble simplify considerably in the limit $m \gg E_{\Le},E_{\Ri}$. This regime is always accessible, given that the shell lives in the black hole interior, and its rest mass does not affect the ADM masses of the two black holes. Moreover, in this regime the Einstein-Rosen bridge of the semiclassical state at $t=0$ is very large, of volume $\text{Vol} \sim m G_{\text N} $ in AdS units (see \cite{Balasubramanian:2022gmo}).

Formally, we will take the limit $m\rightarrow \infty$. Physically, in this limit, the shell's trajectories effectively pinch off in Euclidean space to the asymptotic region of each of the Euclidean manifolds. The manifolds, constructed by a cut-and-glue procedure, consequently factorize into a collection of Euclidean black hole solutions, glued in the asymptotic region along the trajectory of the thin shell. The value of the on-shell action and one-loop determinants evaluated on these manifolds factorize accordingly. In particular, the state of the quantum fields outside the $\Le$ horizon is prepared \`{a} la Hartle-Hawking by the Euclidean black hole geometry at inverse temperature $\beta_\Le$. Therefore, in the semiclassical description, the state is {\it exactly} thermal to all orders in the semiclassical $G_{\text N}\rightarrow 0$ expansion. 

This can be seen microscopically from \eqref{eq:PETSensemblestates}. In this limit, it is possible to see that $\Delta \tau_\pm \rightarrow 0$ (cf. \cite{Balasubramanian:2022gmo}). The microcanonical energies solving \eqref{saddlepoint1} and \eqref{saddlepoint2} become 
\begin{gather}
E_* = E_\Ri\; \quad \text{and}\quad 
\beta_{E_\star} = 2\beta_\Ri.
\end{gather}
From here, we can evaluate the functions in the state-averaging ansatz \eqref{eq:PETSensemble}. They become constant:
\begin{align}
w_\Le(\bar{E}) &=  0 \;, \label{app1w2}\\
v_\Le(\bar{E},\omega) &=  \beta_{\Ri}  \Big(F(2\beta_{\Ri}) - F(\beta_\Ri)\Big).
\end{align}
Hence the ensemble of microscopic states compatible with the semiclassical description yields the ensemble
\be\label{eq:ensemblelargemass}
(\rho_\Le)_{ij} = \,\delta_{ij}\overline{\rho}(E_i)  + \dfrac{\e^{-\beta_\Le \bar{E}_{ij}}}{Z(\beta_\Le)} \frac{\sqrt{Z(2\beta_\Ri)}}{Z(\beta_\Ri)} R_{ij},
\ee 
for $\overline{\rho}(E_i) = \rho_{\beta_{\Le}}(E_i)$. This ensemble looks {\it exactly} thermal for uncorrelated probes,
\be 
\text{Tr}(\overline{\rho_\Le}\phi_\Le) = \text{Tr}(\rho_{\beta_{\Le}}\phi_\Le),
\ee 
to all orders in the semiclassical $G_{\text N}\rightarrow 0$ expansion.\footnote{In some sense, one can view the large $m$ limit as a `thermalization' limit, in which the microstates compatible with the semiclassical description have become more generic in Hilbert space. Semiclassically, a proxy of the genericity is that the states develop long Einstein-Rosen bridges, as well as large python's lunches. All of the microstates $\rho_L$ compatible with the semiclassical description have thermalized, and cannot be distinguished from thermal states, unless exponentially small effects are measured on each of their matrix elements. These effects are absent in the perturbative semiclassical description of the exterior of the black hole.} Moreover, in this case it is obvious that the entanglement entropy of the average state is thermal
\be\label{eq:entropyrel}
S(\overline{\rho_L}) = S(\beta_{\Le})\;,
\ee 
at the native temperature of the left black hole. The right hand side of \eqref{eq:entropyrel} is given by the Bekenstein-Hawking entropy of the apparent horizon of the left black hole.

We reach the conclusion that the coarse-graining map $\rho_\Le\rightarrow \overline{\rho_\Le}$ effectively erases all the information that the microstate contains about the black hole interior. The apparent horizon and its entropy arise as coarse-grained notions. This matches precisely the discussions of \cite{Engelhardt:2017aux}, and in particular it is an explicit manifestation of the diagonal projection of \cite{Chandra:2022fwi}.

However, our model of the state \eqref{eq:PETSensemble} includes non-perturbative microscopic entries of the state $\rho_L$. The smooth envelope function, $j_2(\bar{E}_{ij},\omega_{ij})$, is semiclassically accessible, and contains the information about the interior of the black hole, in the case in which the entanglement wedge of the CFT$_L$ contains the interior. Note that this information is non-perturbative, but still effective, since it does not require the specification of the individual matrix elements of the microstate. This is in accord with the fact that the interior is semiclassical, and that our ansatz models many microstates which share this description. As we shall see now, different classes of wormholes contain the information about $j_2(\bar{E}_{ij},\omega_{ij})$ in the semiclassical theory.

\subsection{Wormholes in $D\geq 3$}

Given our maximally ignorant ansatz to model the state, the ensemble {\it outputs} semiclassical contributions to variances in different quantities. We will now match these with known wormhole contributions in low-energy Einstein gravity coupled to the thin shell. Moreover, we will show how multi-boundary wormholes and replica wormholes arise from the same ensemble. Explicitly, we want to check whether 
\begin{equation}
\eqimg{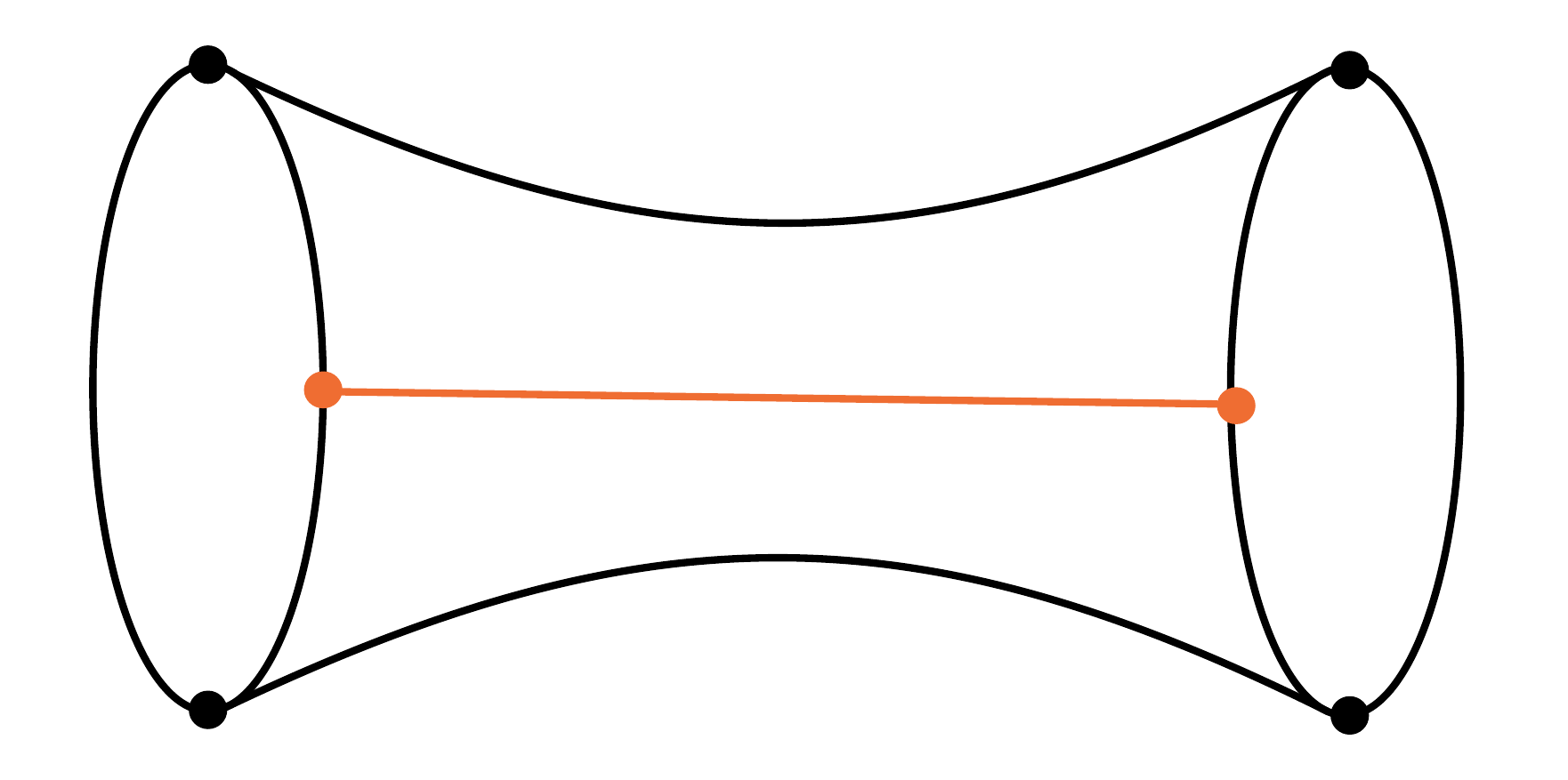}
    \;\stackrel{?}{=}\; \overline{\text{Tr}(\rho_\Le \phi_\Le)\text{Tr}(\rho_\Le \phi_\Le)}^{\text{conn.}} \;,
\end{equation}
and whether
\begin{equation}\label{eq:replicawh}
\eqimg{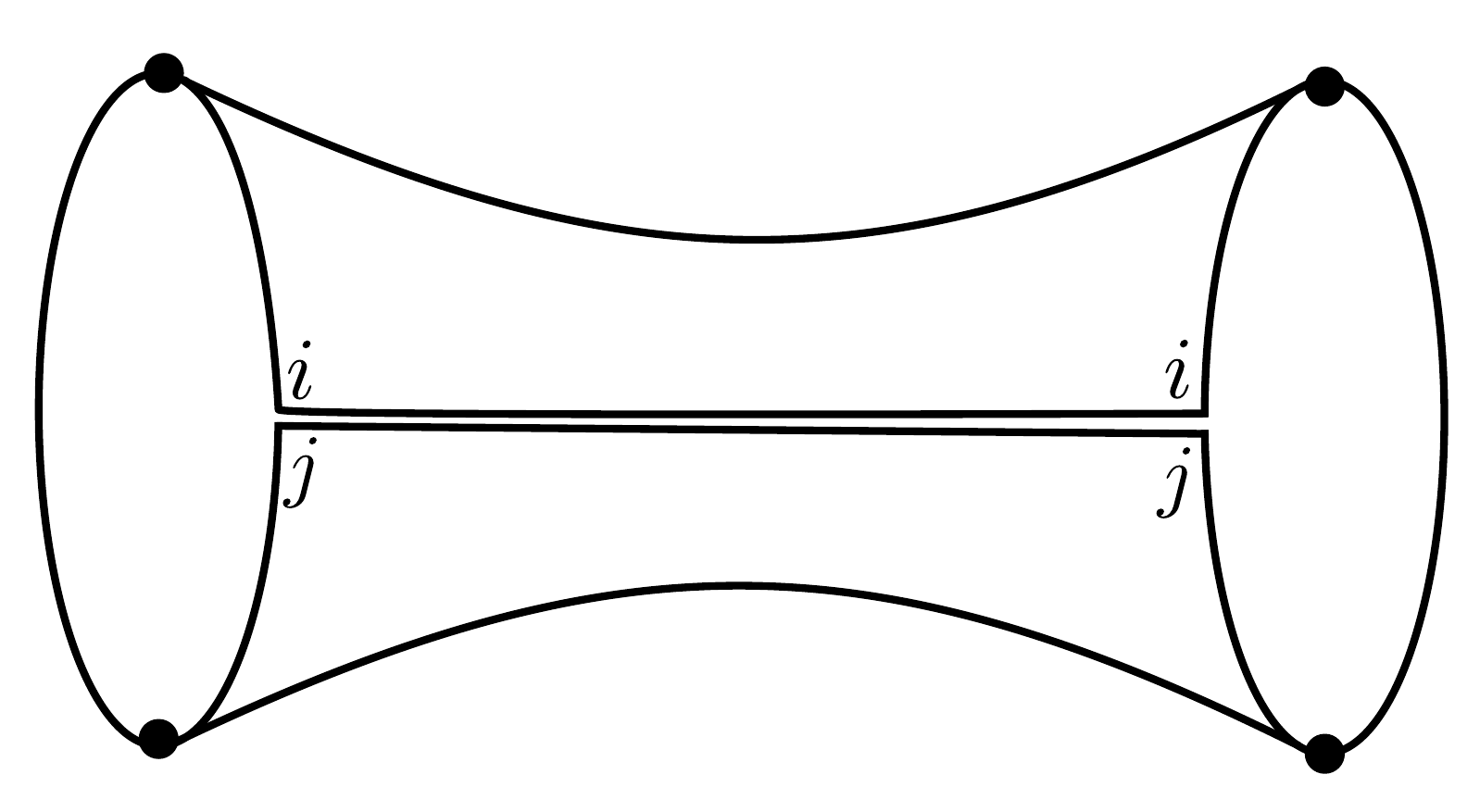}
    \;\stackrel{?}{=}\; \overline{\text{Tr}(\rho_\Le ^2)} - \text{Tr}(\overline{\rho_\Le}^2) \;.
\end{equation}
In the left-hand side of \eqref{eq:replicawh} we represent the replica wormhole contribution to the purity. Each replica prepares a copy of $(\rho_L)_{ij}$. As illustrated in the figure, the Euclidean boundary is connected, since the replicas are glued together to compute $\overline{\text{Tr}(\rho_\Le ^2)}$. The `replica wormhole' arises from a saddle point geometry in which the shell operators travel between the operator insertions on each replica. The topology of this saddle is nevertheless $D_2\times \mathbf{S}^{d-1}$, where $D_2$ is a two-dimensional disk. 

In this way, our maximal ignorant model of the semiclassical state provides a unifying flavor to the rule of summing over all possible geometries to compute moments of correlation functions and R\'{e}nyi entropies using the gravitational path integral. Both effects arise from the variance of the individual matrix elements of the microstates, which using the state-averaging ansatz \eqref{eq:PETSensemblestates}, evaluate to
\be\label{eq:variancethinshellensemble}
\overline{\delta \rho_{ij} \delta \rho_{kl}}^{\text{conn.}} = \delta_{il}\delta_{jk} \,\dfrac{\e^{-2\beta_\Le \bar{E}_{ij}}}{Z(\beta_\Le)^2} \e^{-S(\bar{E}_{ij})} j_2(\bar{E}_{ij},\omega_{ij})\;.
\ee 

\subsection*{Multi-boundary wormhole}

On the one hand, this model outputs the variance in the expectation value of a simple hermitian operator $\phi_{\Le}$. For simplicity, we will assume that the operator has a vanishing one-point function, and that it admits an ETH form $(\phi_{\Le})_{ij} = \e^{-S(\bar{E}_{ij})/2} g_{\phi_\Le}(\bar{E}_{ij},\omega_{ij})^{1/2} R^{\phi_\Le}_{ij}$. The variance thus reads
\be 
\begin{split}
\overline{\text{Tr}(\rho_\Le \phi_\Le)\text{Tr}(\rho_\Le \phi_\Le)}^{\text{conn.}} &= \sum_{ij} \dfrac{\e^{-2\beta_\Le \bar{E}_{ij}}}{Z(\beta_\Le)^2} \e^{-2S(\bar{E}_{ij})} j_2(\bar{E}_{ij},\omega_{ij}) |(\phi_L)_{ij}|^2\\ 
&\approx \dfrac{1}{Z(\beta_\Le)^2}\int \text{d}E  \text{d}\omega \e^{-2\beta_\Le E+S(E+\omega) + S(E-\omega) - S(E) - 2v_\Le(E,\omega)}  g_{\phi_\Le}(E,\omega)\;,
\end{split}
\ee
where in the last step we have taken the semiclassical limit to approximate the discrete sums by continuous integrals. We now perform the $\omega$ integrals, which we assume are peaked at $\omega =0$ due to the concavity of the entropy, as well as the form of the envelope function $g_{\phi_\Le}(E,\omega)$ and from the fact that $v_\Le(E,\omega)$ is an even function of $\omega$. From this observation we arrive to
\be\label{eq:varianceshellensemblegen}
\overline{\text{Tr}(\rho_\Le \phi_\Le)\text{Tr}(\rho_\Le \phi_\Le)}^{\text{conn.}} \approx \dfrac{1}{Z(\beta_\Le)^2}\int \text{d}E \e^{-2\beta_\Le E+S(E) - 2v_\Le(E,0)}   g_{\phi_\Le}(E,0)\;,
\ee
This integral admits a saddle point approximation at
\begin{gather}\label{eq:saddlepointeuclid1}
\beta_{E_2} = 2\beta_\Le + 2\partial_{E_2} v_\Le(E_2,0) = 2\tbeta_\Le + 2\Delta \tau_-(E_2,E_\star)\;,
\end{gather}
where in the last step we have used the identity $\partial_{E}v_\Le(E,0)= (\tbeta_{\Le}-\beta_{\Le}) + \Delta \tau_-(E,E_\star)$, which can be checked directly from the definitions above. This equation is thus coupled to the saddle point equation determining $E_\star$ \eqref{saddlepoint2}, which we inport here
\be\label{eq:saddlepointeuclid2}
\beta_{E_\star} = 2\tbeta_\Ri + 2\Delta \tau_+(E_2, E_\star)\;.
\ee 
These two coupled sadddle point equations are identical to the equations determining the geometry of the Euclidean wormhole, explicitly constructed in \cite{Sasieta:2022ksu,Balasubramanian:2022gmo}. The Euclidean wormhole can be deconstructed as a pair of Euclidean black holes glued along the trajectory of two thin shells:
\be\label{eq:euclideanwormhole}
\begin{gathered}
\includegraphics[width = 7cm]{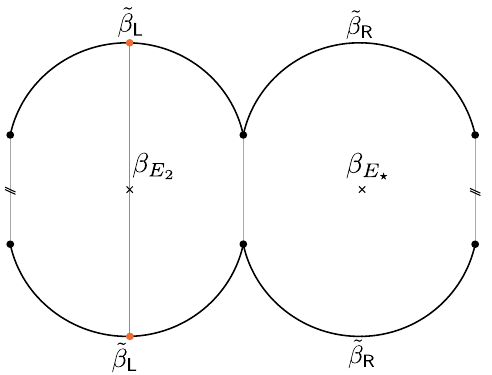} \nonumber
\end{gathered} 
\ee 
Here the gray trajectories correspond to the two thin shells, and the one on the left is identified with the one on the right. The orange trajectory corresponds to the two point function of $\phi_L$ in this geometry, in the worldline approximation. 

The Euclidean time periodicities of the black holes, $\beta_{E_2}$ and $\beta_{E_\star}$, satisfy the same saddle point equations as \eqref{eq:saddlepointeuclid1} and \eqref{eq:saddlepointeuclid2}. The factors of $2\tbeta_\Le$ and $2\tbeta_\Ri$ come from the Euclidean time on each part of the asymptotic boundary, while the functions $\Delta \tau_\pm(E_2, E_\star)$ precisely correspond to the Euclidean time elapsed by the thin shells, as mentioned before. The total periodicity of each black hole is the sum of these two terms. The saddle point value of the integral coincides with the action of the wormhole. For details on the matching of the actions, we refer the reader to  \cite{Sasieta:2022ksu,Balasubramanian:2022gmo}. The two-point function of the operator is captured by the envelope function $g_{\phi_\Le}(E_2,0)$ in \eqref{eq:varianceshellensemblegen}.

\subsection*{Replica wormhole}

On the other hand, the variance \eqref{eq:variancethinshellensemble} outputs an additional semiclassical contribution the second moment of the purity
\be\label{eq:purityconnected}
\begin{split}
\overline{\text{Tr}(\rho_\Le^2)} - \text{Tr}(\overline{\rho_\Le}^2) &=  \sum_{i,j} \dfrac{\e^{-\beta_\Le E_i -\beta_\Le E_j}}{Z(\beta_\Le)^2} \e^{-S(\bar{E}_{ij})} j_2(\bar{E}_{ij},\omega_{ij})  \\&
\approx \dfrac{1}{Z(\beta_\Le)^2}\int \text{d}E \text{d}\omega \; \e^{S(E+\omega) + S(E-\omega) -2\beta_\Le E -2v_\Le(E,\omega)}.
\end{split}
\ee 
Evaluating the integral over $\omega$ in the saddle point at $\omega =0$ yields the result
\be\label{eq:varianceshellensemblegenpurity}
\overline{\text{Tr}(\rho_\Le^2)} - \text{Tr}(\overline{\rho_\Le}^2) \approx \dfrac{1}{Z(\beta_\Le)^2}\int \text{d}E \e^{2S(E) -2\beta_\Le E -2v_\Le(E,0)}\;.
\ee
The remaining saddle point equation is
\be\label{eq:saddleptreplica1}
\beta_{E_2'} = \tbeta_\Le  + 2\partial_{E_2'} v_\Le(E_2',0) = \tbeta_\Le + \Delta \tau_-(E_2',E_\star)\;.
\ee 
where we recall that
\be\label{eq:saddleptreplica2}
\beta_{E_\star} = 2\tbeta_\Ri + 2\Delta \tau_+(E_2', E_\star)\;.
\ee 
Again, these two equations exactly match the equations that determine the geometry of the replica wormhole in \eqref{eq:replicawh}: 
\be\label{eq:replicawormholedecons}
\begin{tikzpicture}
\node at (0,0) {\includegraphics[width = 8cm]{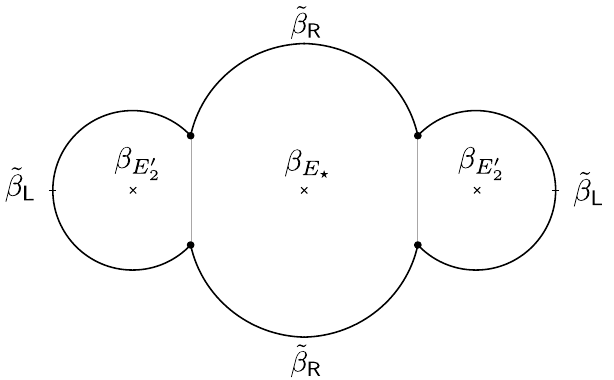}};
\node at (0,2.8){ replica I};
\node at (0,-2.9){ replica II};
\end{tikzpicture}
\nonumber
\ee 
Such wormholes were considered in \cite{Chandra:2022fwi} (see also \cite{Antonini:2023hdh} for the low-temperature versions). The solution consists of three Euclidean black holes, glued in the way illustrated on the figure by the trajectories of two thin shells. The shell's, which are represented in gray, travel between the different replicas.

The Euclidean time periodicities of each of the black holes is completely determined from the saddle point equations \eqref{eq:saddleptreplica1} and \eqref{eq:saddleptreplica2}. Namely, the total periodicities, $\beta_{E_2'}$ and $\beta_{E_\star}$, are the respective boundary periodicities, $\tbeta_\Le$ and $2\tbeta_\Ri$, with the addition of the Euclidean time ellapsed by the shell on each Euclidean black hole, given by $\Delta \tau_\mp(E_2',E_\star)$ respectively. Likewise, evaluating the saddle point action of \eqref{eq:purityconnected} yields by construction the on-shell gravitational action of the replica wormhole, with the appropriate normalization of the state given by $Z_1$.

\subsection*{Large black hole interior}

The identification of statistical variances in the maximal ignorant ensemble and bulk wormhole physics becomes more transparent in the large mass limit $m\rightarrow \infty$, where both the construction of wormholes and the ensemble averages become straightforward. In this limit, $\Delta \tau_\pm \rightarrow 0$ and $\tbeta_{\Le} = \beta_\Le, \tbeta_{\Ri} = \beta_\Ri$. Therefore, we get that $\beta_{E_2}= 2\beta_\Le$ and for the Euclidean wormhole, while $\beta_{E_2'}= \beta_\Le$ for the replica wormhole. In both cases $\beta_{E_\star} = 2\beta_\Ri$. In the saddle point approximation, the variance over the ensemble \eqref{eq:ensemblelargemass}
\be\label{eq:Euclideanwormholelargemass} 
\overline{\text{Tr}(\rho_\Le \phi_\Le)\text{Tr}( \rho_\Le \phi_\Le)}^{\text{conn.}} = \dfrac{Z(2\beta_\Ri)}{Z(\beta_\Ri)^2} \text{Tr}(\rho_{\beta_\Le} \phi_\Le \rho_{\beta_\Le}\phi_\Le)\;,
\ee 
explicitly reduces to the wormhole contribution \eqref{eq:varianceshellensemblegen}. In this limit, the gray trajectories of the thin shells in the Euclidean wormhole illustrated in \eqref{eq:euclideanwormhole} effectively pinch off. The wormhole effectively factorizes into two Euclidean black hole solutions, one of inverse temperature $2\beta_{\Ri}$, and another one of inverse temperature $2\beta_{\Le}$, where the latter contains a pair of $\phi_{\Le}$ operator insertions at antipodal points in the Euclidean time circle. The first geometry thus contributes as $Z(2\beta_{\Ri})$, while the second contributes as $\text{Tr}(\e^{-\beta_{\Le}H} \phi_\Le \e^{-\beta_{\Le}H}\phi_\Le)$. This yields the same answer as \eqref{eq:Euclideanwormholelargemass} up to the normalization factors $(Z_1)^2 = Z(\beta_\Ri)^2Z(\beta_\Le)^2$, which needs to be included to compare correlation functions on a normalized state.

Similarly, the variance in the purity is given in our ensemble \eqref{eq:ensemblelargemass} by
\be\label{eq:rwormholelargemass} 
\overline{\text{Tr}(\rho_\Le^2)} - \text{Tr}(\overline{\rho_\Le}^2) =  \dfrac{Z(2\beta_\Ri)}{Z(\beta_\Ri)^2}\;.
\ee
In this limit, the replica wormhole illustrated in \eqref{eq:replicawormholedecons} factorizes into a product of three Euclidean black hole solutions, two of which have inverse temperature $\beta_{\Le}$, and one of which has inverse temperature $2\beta_{\Ri}$. Its intrinsic action is thus $Z(2\beta_{\Ri})Z(\beta_{\Le})^2$. This yields \eqref{eq:rwormholelargemass} with the inclusion of the normalization factor $(Z_1)^2 = Z(\beta_\Ri)^2Z(\beta_\Le)^2$.

\section{Discussion}
\label{discussion}

Guided by the principle of maximum ignorance, we have identified a rather general ansatz to model semiclassical bulk states microscopically \eqref{eq:state_averaging_ansatz}. We have called this ansatz, and its generalization to higher moments \eqref{statecorre}, the  \emph{state-averaging ansatz}. Using this model, we have accurately described many semiclassical states, such as the Hartle-Hawking state in the eternal black hole, product states, and more general two-sided states with matter in the black hole interior. 

The {\it output} of our approach is a prediction of several semiclassical amplitudes that capture the statistical moments of the ensemble of states. These moments provide coarse-grained, yet non-perturbative, information about the structure of individual microstates.  
Throughout the paper, we have shown -- in different gravitational systems and in different number of dimensions -- that the statistical moments of our ensemble are semiclassical. They are captured by the on-shell gravitational action of wormhole configurations. From this point of view, wormholes simply parametrize the ignorance of the microstructure of a fundamental state given a fixed semiclassical bulk description. 
 
We now provide a list of comments, in order to structure possible avenues of future research:

\paragraph{Universality of the state-averaging ansatz.} As emphasized in section \ref{sec:SAA}, we believe that the state-averaging ansatz has a wide range of applicability, and universally describes ensembles of mixed states in systems with infinite-dimensional Hilbert spaces. In this paper, we have used this ansatz to successfully describe states prepared by a Euclidean path integral in chaotic conformal field theories: these include thermal states, product states, and partially entangled thermal states (PETS). Similar ensembles have also been used to model states in quantum chaotic systems in different contexts (see, e.g. \cite{Murthy:2019qvb,shi2023local,Dymarsky:2016ntg}), all of which share the form of the state-averaging ansatz.
An informal explanation of the universality of this ansatz is given by relation between the averaged purity and the sum of the moments $\delta \rho$
\begin{equation}
    \sum_{ij}\mean{\delta\rho_{ij}\delta\rho_{ji}} = \mean{\Tr\rho^2} - \Tr\mean{\rho}\,^2 \geq 0.
\end{equation}
In a microcanonical window with $L$ independent states, if the typical state of the ensemble is close to the maximally mixed state we expect the right-hand side to be of the order of $L^{-1}$, and thus, $\mean{\delta\rho_{ij}\delta\rho_{ji}} \sim L^{-3}$.\footnote{Note that the same argument applied to an ensemble of pure states yields an ansatz that is of the form $\mean{\delta\rho_{ij}\delta\rho_{ji}} \sim L^{-2}$.} This argument fixes the microcanonical entropy-scaling of any realistic ansatz for an ensemble of mixed-states and agrees with \eqref{eq:state_averaging_ansatz}.

\paragraph{Non-Gaussian contractions and new wormhole amplitudes.} In recent years, it has been understood that the statistical description of chaotic systems contains important non-Gaussian corrections that are needed to reproduce other signals of chaos, such as out-of-time-order correlators. The ETH ansatz incorporates these non-Gaussianities by allowing nonzero cyclic contractions of the type 
\begin{equation}
\label{foinikur}
    \opmean{\clo_{i_1i_2}\clo_{i_2i_3}\dots\clo_{i_k i_1}} \sim \e^{-(k-1) S(\bar E)}.
\end{equation}
Explicit formulas for these contractions can be found in the context of CFTs by studying the modular properties of thermal correlators. 

As explained in the paper, the state-averaging ansatz naturally incorporates a different class of non-Gaussian contractions, originating from the maximally ignorant model of the state. These contractions behave very similarly to \eqref{foinikur}, and are given by \eqref{statecorre}. For the convenience of the reader, we reproduce equation \eqref{statecorre} below
\begin{equation}
\label{statecorre2}
\mean{\delta\rho_{i_1i_2}\delta\rho_{i_2i_3}\dots\delta\rho_{i_k i_1}} = \frac{\e^{-k \beta \bar E_k}}{Z(\beta)^{k}} \e^{-(k-1)S(\bar E_k)} j_k(\bar E_k,\omega). 
\end{equation}
Equations \eqref{foinikur} and \eqref{statecorre2} are two sources of corrections that should be considered when computing wormhole amplitudes. For example, a straightforward computation from the perspective of state averaging leads to the following prediction for the multi-boundary torus wormhole with three asymptotic boundaries:
\begin{equation}
\label{eq:manybworm}
\begin{gathered}
  \includegraphics[width = 30mm]{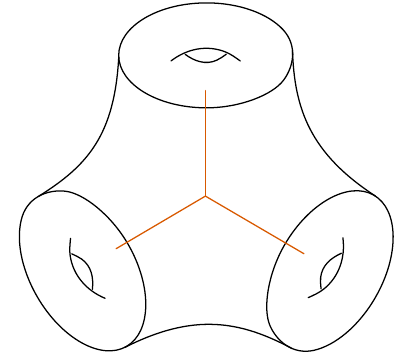}
  \end{gathered} 
 = 
  \frac{1}{Z(\beta)^3} 
  \int \dd h\; \e^{S_0(h)}
    \mathcal{F}_{\clo}(h)^3
    \opmean{C_{\clo 12}C_{\clo 23}C_{\clo 31}}.
\end{equation}
Here $\mathcal{F}_{\clo}$ corresponds to the torus one-point conformal block, and we have used \eqref{statecorre2} for $k=3$. The complete answer also includes the antiholomorphic counterpart of this expression. Using modular invariance and crossing symmetry it is possible to find an analytic expression for the microcanonical average $\opmean{C_{\clo 12}C_{\clo 23}C_{\clo 31}}$.\footnote{Which remarkably is order $\e^{-2S}$, in accordance with the generalized ETH ansatz.} In a follow-up paper \cite{deBoer:2024kat}, we elaborate on these amplitudes for $k\geq 3$. 

\paragraph{Off-shell configurations.} The logical next step in our analysis is to extract information about off-shell amplitudes from the maximally ignorant model of the state. For example, non-perturbative information like the ramp in the spectral form factor can, in principle, be extracted from the analytic continuation of the thermal two-point function 
\begin{equation}
\label{thermsf}
    \Tr \e^{-\beta_1 H} \clo \e^{-\beta_2 H} \clo = \sum_{ij} \e^{-\beta_1 E_i-\beta_2 E_j} |\bra{i}\clo \ket{j}|^2.
\end{equation}
In \cite{Saad:2019pqd}, it was shown how to extract the ramp from this correlation function in the context of JT gravity. The dual geometry that contains this information is the one with the topology of a disk with a handle in the bulk. 

The state-averaging approach to wormhole amplitudes has the added advantage that it can naturally describe spectral correlations. This paper focuses on on-shell partition functions, but we have also included some comments about off-shell configurations in Appendix \ref{sec:CJ}. The upshot of the discussion is that, to model spectral statistics, one has to include an additional contraction to the state-averaging ansatz, 
\begin{equation}
  \mean{ \delta\rho_{ij}\delta\rho_{ji} } = f_1(E_i,E_j) \delta_{il}\delta_{jk} + f_2(E_i,E_k)\delta_{ij}\delta_{kl}.
\end{equation}
The first contraction corresponds to the `traditional' state-averaging ansatz, while the second correlates energy levels from different microcanonical windows. The function $f_2$ is small in the sense that  $f_2/f_1 \sim \e^{-S}$. This suppression guarantees that the modified ansatz does not change the on-shell wormhole amplitudes discussed in sections \ref{sec:wormholes} and \ref{sec:higher_d}. Information about $f_2$ can be extracted from the non-perturbative corrections to \eqref{thermsf}, and it is equivalent to the ramp in the spectral form factor. An intriguing feature about off-shell configurations is that, from the perspective of the boundary dual, the answer is not given by a saddle point approximation. The amplitude is instead determined by states that are very close to the edge of the integration domain. This edge marks the beginning of chaotic states (dual to black hole microstates) in the spectrum. 

It would be interesting to further explore the consequences of this ansatz to off-shell gravitational configurations.

\paragraph{Detecting semiclassical features of the black hole interior.}

We can consider maximally ignorant models of CFT states whose bulk semiclassical description includes portions of the black hole interior. One such case is the reduced density matrix $\rho_\Le$ to CFT$_\Le$ for the two-sided PETS, in the regime in which the entanglement wedge of CFT$_\Le$ contains the interior. In this case, all of the quantum information of the geometry and matter inside of the black hole is contained in $\rho_\Le$.

On the one hand, given our maximally ignorant model of a semiclassical state \eqref{eq:PETSensemblestates}, the coarse graining map $\rho_\Le \rightarrow \bar\rho_\Le$ erases all the semiclassical information of the black hole interior. As shown in section \ref{sec:higher_d}, when the black hole interior is large, the correlation functions of simple operators in the state $\overline{\rho_\Le}$ reduce to thermal correlation functions at the naive temperature of the black hole, and the entanglement entropy of the coarse-grained state $S(\overline{\rho_\Le})$ is given by the Bekenstein-Hawking entropy of the apparent horizon of the $L$ black hole. This agrees with ideas that the horizon of the black hole is a coarse-grained notion (see e.g. \cite{Balasubramanian:2005mg,Balasubramanian:2008da}) or more recent discussions in AdS/CFT \cite{Engelhardt:2017aux,Engelhardt:2018kcs,Chandra:2022fwi}.

On the other hand, the state-averaging ansatz \eqref{eq:PETSensemblestates} is a model which includes effective, and yet non-perturbative, information of the semiclassical state. The smooth amplitude of $\delta \rho = \rho - \overline{\rho}$ is semiclassically accessible, while the individual erratic matrix elements are not. Therefore, the smooth function $j_2(\bar{E}_{ij},\omega_{ij})$ in the state-averaging anstaz \eqref{eq:PETSensemblestates} must contain all of the semiclassical features of the black hole interior in the corresponding bulk state.

This amplitude can be detected in different ways. A direct way is to find exponentially small signals on correlations functions of simple operators. These signals will not be semiclassically accessible, will be erratic, and will depend on microscopic details of the particular microstate of the ensemble. An alternative possibility is to enhance this amplitude using an operator with semiclassical description which is correlated with the state. For instance, in the case of the PETS, one can consider an `anti-shell' operator which destroys the shell in the black hole interior. The correlation function of the anti-shell, in the state $\rho_\Le$, will detect the signal semiclassically. The signal can be made large with an appropriate choice of anti-shell operator which acts on the interior shell with $\order{1}$ probability. This way to act on the black hole interior has of course an inherent state-dependent flavor.

Moreover, one could consider finer non-linear measures that characterize the semiclassical state, such as the average von Neumann entropy $\overline{S(\rho_\Le)}$, given in the bulk by the Ryu-Takayanagi (RT) formula. The non-perturbatively small amplitude on each entry $\delta \rho_{ij}$ can have a relevant effect on these measures. Indeed, semiclassical configurations in which
\be\label{eq:conditionsmallentr}
\overline{S(\rho_\Le)} \ll  S(\bar{\rho}_\Le)\;,
\ee 
are easy to construct. In the semiclassical description, \eqref{eq:conditionsmallentr} simply means that the RT surface of the CFT$_L$ is not the apparent horizon of the $L$ black hole. Rather, the $L$ black hole contains an Einstein-Rosen bridge connected to a smaller black hole through its interior, which is in the entanglement wedge of CFT$_L$. Each member of the maximally ignorant microscopic ensemble \eqref{eq:state_averaging_ansatz} effectively knows this. One could even consider situations in which $\overline{S(\rho_\Le)}=O(G^0)$, for which the semiclassical interior contains a `bridge-to-nowhere' instead. As mentioned in the discussion after eq. \eqref{eq:coarse_graining_map}, in this case, the average relative entropy between the coarse-grained state and a state of the microscopic ensemble, $\overline{S(\rho_\Le\,||\,\bar\rho_\Le)}$, will actually be parametrically large. This suggests that distinguishing bridges-to-nowhere from Einstein-Rosen bridges from the perspective of an outside observer could be not that hard, with access to multiple copies of the state.\footnote{One such quantum information protocol is the SWAP test, which serves to distinguish density matrices with a few copies of the states if their second R\'{e}nyi entropy is very different.}

To sum up, distinguishing features of the semiclassical interior amounts to being able to detect the smooth envelope function $j_2(\bar{E}_{ij},\omega_{ij})$ of the state-averaging ansatz \eqref{eq:state_averaging_ansatz}, while it does not require the specification of the individual matrix elements of $\rho$, since the erratic phases on the matrix elements are washed out in the semiclassical description. 

\paragraph{The operator approach to 3D gravity.} As mentioned in the introduction, the principle of maximum entropy has a wide range of applicability and can be used to describe ensembles of operators or Hamiltonians. 

Recent proposals to model 3D gravity assert that this theory can be described using a tensor integral of the form 
\begin{equation}
    Z = \int \dd L_0\dd\bar L_0 \dd C \; \e^{- V(L_0,\bar L_0,C)},
\end{equation}
where $V$ is a potential to be determined. In \cite{Chandra:2022bqq}, the Gaussian part of the ensemble of operators was explored and matched to several wormhole amplitudes. In \cite{Belin:2023efa}, a quartic potential was chosen that favors configurations of OPE coefficients that are crossing symmetric. The priciple of maximum entropy makes yet another prediction for the potential $V(L_0,\bar L_0,C)$. 

To apply the principle of maximum ignorance we first need a set of observables that can be obtained from the semiclassical path integral. An evident choice that does not rely on additional external operators, like conical defects or matter fields, is to consider the torus and all higher genus partition functions. Extremizing the classical Shannon entropy with the inclusion of an infinite number of Lagrange multipliers that fix the partition functions, gives the following potential
\begin{equation}
\label{single-trace}
    V(L_0,\bar L_0,C) =  \Tr V_0(L_0)+\Tr \bar V_0(\bar L_0) + \sum_{ijk} V^{\text{sun.}}_2 {C_{ijk}^2 + V^{\text{dum.}}_2 C_{iij}C_{jkk}} + \sum  CCCC +\cdots.
\end{equation}
The configurations of OPE coefficients that are allowed to appear in this ansatz are single-trace configurations -- meaning that the functions $V_i$ are not allowed to include Kronecker deltas -- that are in a one-to-one correspondence between the different channels of the higher genus partition functions. For example, the genus-two partition function can be expressed in terms of two channels:
\begin{align}\notag
    \text{sunset:} \quad & C_{ijk}C_{ijk}\\ \notag
    \text{dumbbell:} \quad & C_{iij}C_{jkk}.
\end{align}
Meanwhile, there are five possible contractions for the genus-three partition function. 

The potential in \eqref{single-trace} gives a vanishing answer to contractions of the type 
\begin{equation}
   \opmean{ C_{1\clo2}C_{2\clo 3}C_{3\clo 1}}. 
\end{equation}
The reason is that the microcanonical average of these contractions is not universal and depends on the matter content of the theory. In a theory of pure gravity with only a few additional matter fields, these interactions should be incorporated by hand in both the boundary and the bulk description. We further comment and interpret these contractions in a follow-up paper \cite{deBoer:2024kat}.

\section*{Acknowledgements}

We would like to thank 
Igal Arav,
Vijay Balasubramanian,
Alexandre Belin,
Alejandra Castro,
Shira Chapman,
Lorenz Eberhardt,
Ben Freivogel,
Victor Gorbenko,
Albion Lawrence,
Javier Mag\'{a}n and
Julian Sonner
for stimulating discussions. MS is grateful to the University of Amsterdam and the Delta Institute for Theoretical Physics for their hospitality during the initial stages of the project.
JdB, DL and BP are supported by the European
Research Council under the European Unions Seventh Framework Programme (FP7/2007-2013),
ERC Grant agreement ADG 834878.
MS is supported in part by the U.S. Department of Energy through DE-SC0009986 and QuantISED DE-SC0020360. This preprint is assigned the code BRX-TH-6717.

\appendix

\section{The volume element $\dd\rho$}\label{app:volume_element}

In this paper, we have formalized mixed state-averaging using a measure $\mu(\rho)\dd\rho$ on the space of density matrices. 
In the main text, we have only described $\mu(\rho)$: maximum-entropy measures are typically of the form $\mu(\rho) \sim \exp(-V(\rho))$, so the precise form of $\dd\rho$ is a one-loop effect compared to the exponential suppression due to the potential $V$. Nonetheless, for completeness we will also describe the different choices one has for the volume element $\dd\rho$. 

A choice of metric on $\mathcal{D}$ defines a volume element $\dd\rho$. Natural choices include the Hilbert-Schmidt metric $ds^2 = \half\Tr d\rho^2$ or the quantum information metric $ds^2 = \half\Tr d\rho \,d\log\rho $ \cite{Hall1998,Zyczkowski_2001,Zyczkowski_2011}. Let us take a metric of the general form
\begin{equation}
    ds^2 = \half \sum_{i,j=1}^L g_{ij}[\Lambda]\,|d\rho_{ij}|^2
\end{equation}
where $g_{ij}$ depends on the eigenvalues $\Lambda  =\mathrm{diag}(\lambda_1,\dots,\lambda_L)$ of $\rho$.
To determine the volume element $\dd\rho$, we have to take the square root of the determinant of the metric. To evaluate the determinant, we have to diagonalize the metric. To do so, we first diagonalize the density matrix as $\rho = U^\dagger \Lambda U$, so that
\begin{equation}
    d\rho = d\Lambda + [d U,\Lambda]
\end{equation}
Evaluating in the basis in which $\Lambda$ is diagonal with eigenvalues $\lambda_i$, we can rewrite this as
\begin{equation}
    d\rho_{ij}  = \delta_{ij}d\lambda_i + (\lambda_i-\lambda_j)(dx_{ij} +i\,dy_{ij})
\end{equation}
where we have expressed $dU$ using real coordinates $x$ and $y$. Taking the absolute value squared to get $|d\rho_{ij}|^2$, the cross-terms vanish because $\delta_{ij}(\lambda_i-\lambda_j) = 0$, and we thus obtain
\begin{equation}
\begin{split}
    ds^2 &= \half \sum_{ij}g_{ij}[\Lambda]\Big(\delta_{ij}d\lambda_i^2 + (\lambda_i-\lambda_j)^2(dx_{ij}^2+dy_{ij}^2)\Big)\\[1em]
    &=\half \sum_i g_{ii}[\Lambda] \,d\lambda_i^2 + \sum_{i<j} g_{ij}[\Lambda](dx_{ij}^2+dy_{ij}^2)
\end{split}
\end{equation}
Since the metric is diagonal in the coordinates $(\lambda_i,x_{ij},y_{ij})$, we can easily compute the square root of the determinant, giving
\begin{equation}
    \dd V = 2^{-L/2} \prod_{i=1}^L \sqrt{g_{ii} [\Lambda]} \,d\lambda_i \,\prod_{k<l}(\lambda_k-\lambda_l)^2 g_{kl}[\Lambda]\,dx_{kl}dy_{kl}
\end{equation}
We recognize the Vandermonde determinant $\Delta(\Lambda) = \prod_{j>i}(\lambda_i-\lambda_j)$ as well as the Haar measure $\dd U \propto \prod_{k<l}dx_{kl}dy_{kl}$. Including the normalization condition and the positivity constraint, the volume element $\dd\rho$ becomes
\begin{equation}
    \dd\rho = \dd U \dd\Lambda \,\Delta(\Lambda)^2\, \delta(\Tr \Lambda -1) \theta(\Lambda)\,f(\Lambda) \eqqcolon [\dd U][\dd \Lambda]
\end{equation}
where we defined 
\begin{equation}
    f(\Lambda) = \prod_i \sqrt{g_{ii} [\Lambda]}\,\prod_{k<l}g_{kl}[\Lambda].
\end{equation}
Since the Haar measure is unitary invariant, we see that $\dd\rho$ is also fully $U(L)$ invariant for any choice of $g[\Lambda]$. As an example, if we take the Hilbert-Schmidt metric $ds^2 = \half \sum_{ij}|d\rho_{ij}|^2$, then $f(\Lambda)=1$. Another example comes from the quantum Fisher information metric,
\begin{equation}
    ds^2 = \half \Tr d\rho \,d\log \rho = \sum_{ij}\frac{\log\lambda_i-\log\lambda_j}{\lambda_i-\lambda_j} |d\rho_{ij}|^2
\end{equation}
Here it is understood that $g_{ii} = \lambda_i^{-1}$.
We encountered this quantity in the main text as the second-order expansion of the relative entropy. In this case the measure factor $f(\Lambda)$ depends non-trivially on the eigenvalues of $\rho$. Similar non-trivial measure factors can be derived from other information metrics, such as the Bures metric \cite{Hall1998} or Renyi divergences \cite{May_2018}.

\section{Fixed trace condition}\label{app:normalization}

In the main text we have assumed a Gaussian form for the ensemble $\int \dd A\,\mu(A)(\bullet)$, where
\begin{equation}
    \mu(A) 
    = \frac{1}{\mathcal{N}}\exp(-\frac{1}{2}\sum_{i,\alpha} \sigma(E_i,E_\alpha)^{-1}\, |A_{i\alpha}|^2).
\end{equation}
However, strictly speaking the integral is over the space of unit norm matrices, 
    $\Tr AA^\dagger = 1.$
We argued that this condition can be replaced by the average normalization condition $\overline{\Tr AA^\dagger} =1$ to a good approximation. In this appendix we will check that the corrections are indeed suppressed.

Specifically, we will compute the variance of $A$. We abbreviate $\sigma(E_i,E_\alpha) = \sigma_{i\alpha}$. We can represent the delta function as a Fourier transform, and then compute the Gaussian integrals:
\begin{align}
    \overline{|A_{i\alpha}|^2} &= \int \dd A\,\mu(A)\delta(1-\Tr AA^\dagger)\,|A_{i\alpha}|^2 \\[1.2em]
    &= \frac{1}{\mathcal{N}}\int\dd s\,\e^{-is}\int \dd A\exp(-\frac{1}{2}\sum_{n,\nu} (\sigma_{n\nu}^{-1}-2is)\, |A_{n\nu}|^2)\,|A_{i\alpha}|^2 \\[1.2em]
    &=\label{app:variance} \sigma_{i\alpha}\times \frac{1}{\mathcal{N}}\int\dd s\,\e^{-is}\Big( \prod_{n,\nu}\frac{1}{\sigma_{n\nu}^{-1}-2is}\Big) \frac{1}{1-2is\sigma_{i\alpha}}
\end{align}
We now expand the last fraction in \eqref{app:variance} for small $\sigma_{i\alpha}$, obtaining the result claimed in \eqref{eq:corrected_var},
\begin{equation}
    \overline{|A_{i\alpha}|^2} = \sigma(E_i,E_\alpha) \left(1+\sum_{n=1}^\infty c_n\, \sigma(E_i,E_\alpha)^n \right)
\end{equation}
where we defined the coefficients
\begin{equation}
c_n =  \frac{\int \dd s\,\e^{-is} d(s) (2is)^n}{\int \dd s\,\e^{-is} d(s)}, \qquad d(s) = \prod_{j\beta}(\sigma^{-1}_{j\beta}-2is)^{-1}.
\end{equation}

\section{Expansion of the relative entropy}\label{sec:stateperturbations}

In this appendix we expand the relative entropy $S(\rho||\bar\rho) = \Tr\rho\log\rho- \Tr\rho\log\bar\rho$ in a perturbative expansion around a given reference state $\bar{\rho}$. As we will make precise in this appendix, the perturbation theory is consistent as long as the variance $|\delta \rho_{ij}|^2$ is sufficiently small.

 We will make use of the following matrix identity:
\begin{equation}
\begin{split}
    \log(\bar{\rho} + \varepsilon\,\delta \rho) = \log \bar\rho + \varepsilon \int_0^\infty dx \frac{1}{\bar\rho + x \mathds{1}} \delta \rho \frac{1}{\bar\rho + x \mathds{1}} +\order{\varepsilon^2}.
\end{split}
\end{equation}
To zeroth order $S^{(0)}(\rho \,||\,\bar\rho) = S(\bar \rho \,||\,\bar\rho) = 0$, and to first order in $\varepsilon$ we find
\begin{equation}
    \frac{\dd}{\dd\varepsilon} S(\bar{\rho} + \varepsilon\,\delta \rho \,||\, \bar \rho)\,\Big|_{\varepsilon=0} = \Tr\left[ \bar\rho \int_0^\infty dx \frac{1}{\bar\rho + x \mathds{1}} \delta \rho \frac{1}{\bar\rho + x \mathds{1}}\right] = \Tr \delta \rho = 0,
\end{equation}
where the last equality holds thanks to the trace condition $\Tr \rho = \Tr \bar \rho = 1$. The first non-zero term comes in at second order:
\begin{equation}\label{eq:rel_entropy}
    S^{(2)}(\rho ||\bar\rho) = \Tr\left[ \delta \rho \frac{\dd}{\dd \varepsilon}\log(\bar \rho + \varepsilon\delta \rho)\,\Big |_{\varepsilon=0}\right] + \frac{1}{2}\Tr\left[\bar\rho \frac{\dd^2}{\dd\varepsilon^2}\log(\bar \rho + \varepsilon\delta\rho)\Big|_{\varepsilon=0}\right].
\end{equation}
We can simplify this expression by again using that $\Tr \delta \rho = 0$, which implies
\begin{equation}
    \Tr \rho(\varepsilon) \frac{\dd}{\dd\varepsilon}\log\rho(\varepsilon) \Big |_{\varepsilon=0} = \Tr \delta \rho = 0.
\end{equation}
Taking a $\varepsilon$-derivative of this identity gives 
\begin{equation}
    \Tr\left[ \delta \rho \frac{\dd}{\dd \varepsilon}\log(\bar \rho + \varepsilon\delta \rho)\Big |_{\varepsilon =0}\right] + \Tr\left[ \bar\rho \frac{\dd^2}{\dd\varepsilon^2}\log(\bar\rho+\varepsilon\delta\rho)\Big |_{\varepsilon=0}\right] = 0.
\end{equation}
Plugging this into \eqref{eq:rel_entropy} simplifies the expression for the quadratic part of the relative entropy:
\begin{equation}
    S^{(2)}(\rho ||\bar\rho) = \frac{1}{2} \Tr \left[\delta \rho \frac{\dd}{\dd\varepsilon}\log(\bar\rho + \varepsilon\delta\rho)\Big |_{\varepsilon=0}\right] = \frac{1}{2}\Tr \left[\delta \rho \int_0^\infty dx \frac{1}{\bar\rho + x \mathds{1}} \delta \rho \frac{1}{\bar\rho + x \mathds{1}}\right].
\end{equation}
We can compute the $x$-integral in the basis $|e_i\rangle$ in which $\bar\rho$ is diagonal with eigenvalues $\lambda_i$, $i=1,\dots,L$. Defining $\delta\rho_{ij}= \langle e_i |\delta \rho |e_j\rangle$, we find
\begin{equation}
    S(\rho\,||\,\bar\rho) = \frac{\varepsilon^2}{2}\sum_{ij}\frac{\log \lambda_i - \log \lambda_j}{\lambda_i-\lambda_j}\delta \rho_{ij}\delta\rho_{ji}+ \order{\varepsilon^3}.
\end{equation}
Using that $\delta \rho_{ji} = \delta \rho_{ij}^*$ by Hermiticity $\rho^\dagger = \rho$, and absorbing $\varepsilon$ into $\delta\rho$, the result \eqref{eq:rel_ent_expansion} follows.

\section{Operator averaging}\label{app:ETH}

In the introduction, we have shown that the eigenstate thermalization hypothesis follows from a principle of maximum ignorance. Although this is not a proof of ETH in chaotic systems, it gives an intuitive explanation of the ETH ansatz as the most entropic probability distribution for the matrix elements of $\mathcal{O}$ compatible with single-trace constraints. In this appendix we fill in some details about this distribution $\mu(\mathcal{O})$, most of which are well-known \cite{DAlessio:2015qtq}. 

Firstly, from the form of the Gaussian ETH stated in \eqref{eq:ETH} we can read off the relation between $f(E_i)$ and the thermal one-point function. From the constraint equation we get
\begin{equation}
\begin{split}
    F(\beta) \stackrel{!}{=} \opmean{\Tr \rho_\beta \mathcal{O} } &= \frac{1}{Z(\beta)}\sum_i \e^{-\beta E_i} f(E_i) \\
    &\approx \frac{1}{Z(\beta)}\int \dd E\,\e^{S(E)-\beta E}f(E)
    \end{split}
\end{equation}
where in the second line we made the continuum approximation. From this we see that $f(E)$ is the microcanonical one-point function. On the saddle-point $\beta = S'(E)$ we get $F(\beta)\approx f(E_\beta)$ in the thermodynamic limit.

Next, we relate $g(\bar E,\omega)$ to the connected two-point function. From the constraint equation we get
\begin{align}
    G(\beta,t)-F(\beta)^2 &\stackrel{!}{=}\opmean{\Tr \rho_\beta^{1/2} \mathcal{O}(0) \rho_\beta^{1/2} \mathcal{O}(t)} - \opmean{\Tr \rho_\beta \mathcal{O} }^2 \\[1.5em]
    &= \frac{1}{Z(\beta)}\sum_{ij}\e^{-\beta \bar E_{ij}+it\omega_{ij}}\e^{-S(\bar E_{ij})}g(\bar E_{ij},\omega_{ij})+ \Delta_f^2 \\[0.8em]
    \label{eq:app_C} &\approx \frac{1}{Z(\beta)}\int\dd \bar E\,\e^{S(\bar E)-\beta \bar E}\int \dd \omega \,\e^{it\omega}g(\bar E,\omega).
\end{align}
In the second line we introduced the thermal fluctuation $\Delta_f^2$, which vanishes in the thermodynamic limit
\begin{equation}\begin{split}
   \Delta_f^2 &\coloneqq \frac{1}{Z(\beta)} \sum_i \e^{-\beta E_i}f(E_i)^2 - \Big(\frac{1}{Z(\beta)}\sum_i \e^{-\beta E_i}f(E_i)\Big)^2 \approx f(E_\beta)^2 - f(E_\beta)^2 =0.
   \end{split}
\end{equation}
The approximation arises from taking the continuum limit $\sum_i \to \int \dd E\,\e^{S( E)}$ and evaluating the integrals on their saddle-point $E_\beta$. Equation \eqref{eq:app_C} shows that $g(\bar E,\omega)$ is the microcanonical version of the connected thermal two-point function $G(\beta,t)^{\mathrm{c}}$.

\section{Comments on off-shell configurations}
\label{sec:CJ}

In this appendix, we will consider the torus wormhole without any operator insertions:
\begin{equation}
    \eqimg{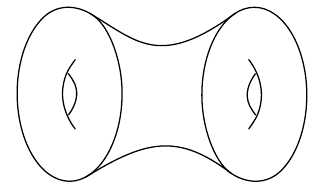} = \mean{\Tr \delta\rho_{\tt un} \Tr \delta\rho_{\tt un}}.
\end{equation}
To compute this amplitude, we need to consider unnormalized ensembles. Following the discussion of section \ref{sec:SAA}, we can also introduce an ansatz for unnormalized density matrices $\rho_{\tt un}$. The ansatz looks very similar to the normalized case, but without the extra condition that $\Tr \delta \rho_{\tt un} = 0$ and without the division by the partition function $Z(\beta)$,
\begin{equation}
    (\delta\rho_{\tt un})_{ij} = \e^{-\beta \bar E-\frac{S(\bar E)}{2}} j(\bar E,\omega)R_{ij},
\end{equation}
where $R_{ij}$ is a random variable with unit variance. 

Compared to previous examples, the gravitational amplitude of the torus wormhole depends more sensitively on the details of the spectrum of the CFT. To correctly reprodue this amplitude, for example, we have to to  use a resolution of the identity that takes into account the spin quantization:
\begin{equation}
\label{eq:idres}
    \bbi  =  \sum_{n\in \mathbb{Z}} \sum_{h-\bar h = n} \sum_{M, \bar M} \ket{h, M, \bar h, \bar M} \bra{h,M,\bar h,\bar M}.
\end{equation}
Using this resolution of the identity, the unnormalized version of the state-averaging ansatz in \eqref{eq:saa-conf} and the same arguments as in the previous sections, we find the following wormhole amplitude:
\begin{equation}
\label{eq:cotjensad}
     \eqimg{toruswormhole.pdf} \approx \sum_{n\in\mathbb{Z}}\int\dd h \dd \bar h \;\delta(h-\bar h -n) j'(\beta;h,\bar h)\chi(\beta;h)^2 \chi(\beta;\bar h)^2,
\end{equation}
with $\chi(\beta;\bar h)$ being the standard Virasoro character defined as 
\begin{equation}
    \chi(\beta;h) \coloneqq 
    \eta\left(\tau\right)^{-1}\e^{-\beta\left(h-\frac{c-1}{24}\right)}.
\end{equation}
The function $\eta$ is the Dedekind eta function and $\tau \coloneqq \frac{i\beta}{2\pi}$. 

 Unlike the torus wormhole with an operator insertion, this time there is no `effective' entropy in \eqref{eq:cotjensad} that can produce a saddle point. This is important because it means that the domain of integration is relevant. The most natural regime to integrate the amplitude is from $h,\bar h = \frac{c-1}{24}$ to $\infty$. This  condition has the interpretation that there should only be averaging over the states that are above the black hole threshold. Assuming the function $j'$ is a constant $j_0$ (as in the previous sections), we find the following expression for the wormhole amplitude\footnote{We used the Poisson resummation formula to write
\begin{equation}
\label{cot-jen-am}
    \sum_{n\in \mathbb{Z}} \delta(h-\bar h - n) = \sum_{n\in \mathbb{Z}} \e^{2\pi i (h-\bar h) n}. 
\end{equation}
}
\begin{equation}
\label{eq:wh-tor}
  \frac{j_0}{| \eta\left(\tau_1\right) \eta\left(\tau_2\right)|^2} \sum_{n=-\infty}^{\infty} \frac{1}{(\beta_1+\beta_2)^2 + 4\pi^2n^2}= \frac{j_0}{| \eta\left(\tau_1\right) \eta\left(\tau_2\right)|^2} \frac{\coth(\frac{\beta_1+\beta_2}{2})}{2(\beta_1+\beta_2)}.
\end{equation}
Here we have also slightly generalized the amplitude by allowing the left and right boundaries to have a different temperature $\beta_1$ and $\beta_2$ respectively. 

Comparing \eqref{eq:wh-tor} to the result one expects from RMT, and the Cotler and Jensen computation \cite{Cotler:2020ugk},  there is an important contribution that is missing: a factor of $\sqrt{\beta_1\beta_2}$ that multiplies the amplitude. This factor is important because it is needed to recover the ramp in the spectral form factor. 

One could argue that the factor of $\sqrt{\beta_1\beta_2}$ can be taken into account by the function $j'$ in the state-averaging ansatz. However, this is not the correct approach to recover the wormhole amplitude. The ramp in the spectral form factor is  a consequence of long-range eigenvalue repulsion, which is a measure of correlations between eigenvalues in different energy windows. 
In the context of RMT, these correlations are encoded in the measure $\langle \varrho(E_i)\varrho(E_j)\rangle$, and are given by the universal sine kernel formula (whose only input is the single sided density of states).  
In the context of state averaging, we can model these correlations by including an additional contraction in the ansatz of the two-point function:\footnote{Note that the second contraction can also be derived using single boundary input by studying the behaviour of the thermal correlator $\Tr\e^{-\beta_1 H}\clo\e^{-\beta_2 H}\clo$, as this contraction is equivalent to the ramp in the correlation function.}
\begin{equation}
    \mean{\delta\rho_{ij}\delta\rho_{kl}} = f_1(E_i, E_j) \delta_{il}\delta_{jk} + f_2(E_i, E_k) \delta_{ij}\delta_{kl}.
\end{equation}
The first contraction corresponds to the state-averaging ansatz, $f_1 \sim \e^{-S(\bar E)-2\beta \bar E}$. The second contraction can be used to measure correlations between energy eigenvalues at different energy windows. To reproduce the wormhole amplitude we will take the following ansatz for the non-trivial contraction:
\begin{equation}
\label{sec-con}
 \mean{\bra{h_1,\bar h_1} \delta\rho \ket{h_1,\bar h_1}\bra{h_2,\bar h_2}\delta\rho\ket{h_2,  \bar h_2}
 } = \delta_{J_1,J_2}\; f_2(\Delta_1,\Delta_2,J_1) \e^{-\beta_1\Delta_1-\beta_2 \Delta_2}\; \e^{-S(\Delta_1)-S(\Delta_2)}.
\end{equation}
Here we are using the notation 
\begin{equation}
    \Delta = h+\bar h \quad \text{and}\quad J= h-\bar h.
\end{equation}
One thing to note about \eqref{sec-con} is that the function $f_2$ is order $\e^{-2S}$ while $f_1$ is order $\e^{-S}$. The $f_2$ contraction is subleading in the on-shell computations of sections \ref{sec:punct-tor-worm} and \eqref{sec:genus-two-worm} because of this additional factor of the entropy. We would also like to point out that \eqref{sec-con} only correlates states that have the same spin $J_1 = J_2$.

Using the ansatz \eqref{sec-con} and ignoring for a moment descendant states, we arrive at the following expression for the $f_2$ contribution to the wormhole amplitude
\begin{multline}
\label{eq:inte-wh-am}
    \sum_n \int \dd h_1\dd h_2 \dd \bar h_1\dd \bar h_2 
    \delta(h_1-\bar h_1-n)\delta(h_2-\bar h_2-n) f_2(\Delta_1,\Delta_2,J_1)\e^{-\beta_1\Delta_1-\beta_2\Delta_2}\\ =\frac{1}{4} \sum_{n}\int \dd \Delta_1\dd\Delta_2\;  f_2(\Delta_1,\Delta_2,n)\e^{-\beta_1\Delta_1-\beta_2\Delta_2}.
\end{multline}
We would like to match this expression to the result of Cotler and Jensen, which after Poisson resummation reads 
\begin{equation}
    \sum_{n\in\mathbb{Z}} \frac{\sqrt{\beta_1\beta_2}}{(\beta_1+\beta_2)^2 + 4\pi^2 n^2} =  \frac{1}{2}\frac{\sqrt{\beta_1\beta_2}}{(\beta_1+\beta_2)} \sum_{n\in\mathbb{Z}} \e^{-|n|(\beta_1+\beta_2)}.
\end{equation}
The equation we would like to reproduce is 
\begin{equation}
    \frac{1}{2}\int \dd \Delta_1\dd\Delta_2 \;f_2(\Delta_1,\Delta_2,n)\;\e^{-\beta_1\Delta_1-\beta_2\Delta_2}=
    \frac{\sqrt{\beta_1\beta_2}}{(\beta_1+\beta_2)} \e^{-|n|(\beta_1+\beta_2)}.
\end{equation}
Which yields the following shape for the function $f_2$,
\begin{equation}
\label{eq:ei-rep}
 f_2(\Delta_1,\Delta_2,J) =-\frac{
 \Theta(\Delta_1-|J|)\Theta(\Delta_2-|J|)
 (\Delta_1+\Delta_2 -2 |J|)
 }{8\pi^2\sqrt{(\Delta_1-|J|)(\Delta_2-|J|)}}\frac{1}{(\Delta_1-\Delta_2)^2}.
\end{equation}
This is the familiar result one finds in the context of JT gravity. The function accurately captures long-range eigenvalue repulsion as it goes like $(\Delta_1-\Delta_2)^{-2}$. One might worry that the function is divergent when $E_i = E_j$, however this divergence disappears in RMT when taking into account non-pertubative effects.\footnote{The sine kernel is a regular function of the energy difference:
\begin{equation}
    -\frac{\sin^2(\pi \varrho(E)(E-E'))}{\pi^2(E-E')^2}.
\end{equation}
Using $2\sin^2(x) = 1 -\cos(2x)$, the sine kernel splits into a leading term and a rapidly oscillating term that scales like $\e^{i S(E)}$. The oscillating term gives a non-pertubative correction to $Z(\beta_1,\beta_2)$, and it is ultimately responsible for the plateau in the spectral form factor.}

The wormhole amplitude now has a term related to $f_1$ and a term related to $f_2$. These two factors added together lead to the following expression for the wormhole amplitude
\begin{equation}
  \frac{j_0 + \sqrt{\beta_1\beta_2}}{| \eta\left(\tau_1\right) \eta\left(\tau_2\right)|^2} \sum_{n=-\infty}^{\infty} \frac{1}{(\beta_1+\beta_2)^2 + 4\pi^2n^2}.
\end{equation}
To fully reproduce the wormhole amplitude, we need to cancel the piece proportional to $j_0$ that is associated to $f_1$. We can do this by including an additional piece in $f_2$ 
\begin{equation}
  f_2(\Delta_1,\Delta_2,J_1) =  f_2^{(1)}(\Delta_1,\Delta_2,J_1) +  f_2^{(2)}(h_1,h_2,J_1).
\end{equation}
In this notation, $f_2^{(1)}$ is given by \eqref{eq:ei-rep}. As discussed above, this expression for $f_2^{(1)}$ is valid 
when the difference $\Delta_1-\Delta_2$ is much less than one, but much larger than $\e^{-S}$. As $\Delta_1$ approaches $\Delta_2$, the regular part of $f_2$ (denoted by $f_2^{(2)}$) kicks in. Since $f_2^{(2)}$ is a function that is peaked at $h_1 = h_2$, we have define it in terms of these variables. To recover the gravitational amplitude, the function $f_2^{(2)}$ must be the negative of the contribution associated to $f_1$,
\begin{align}
\label{eq:f2-res}
\int \dd h_1\dd h_2 \;
    f^{(2)}_2(h_1,h_2,|n|)\e^{-\beta_1(h_1+|n|)-\beta_2(h_2+|n|)}
   =
   -j_0\int \dd h \;
    \e^{-\beta_1(h+|n|)-\beta_2(h+|n|)}.
\end{align}
Here the absolute value in $|n|$ appears depending on whether we first integrate the holomorphic or antiholomorpic part of the Dirac delta functions $\delta(h_1-\bar h_1 + n)$ in the full expression for the amplitude. Then the sum over the spin $n$, after doing Poisson resummation, matches the result  in \eqref{cot-jen-am} with the factors of $j_0$ and $-j_0$ canceling each other.

Note that the quantity $\mean{\rho_{ii}\rho_{jj}}$ is allowed to be negative, as it is not the average of a positive definite quantity. We can reproduce \eqref{eq:f2-res}, up to some tolerance, if we take $f^{(2)}_2(h_1,h_2,n)$ to be approximately given by a delta function $f^{(2)}_2\approx -j_0 \delta(h_1,h_2)$ in the conformal weights (this is similar to what we have seen in the previous examples.). The function $f_2^{(2)}$ is weighted by a factor of $\e^{-2S}$, so it is still a subleading correction to the state-averaging ansatz. However, due to the way indices contract in this particular example, it leads to an important contribution.

\bibliographystyle{utphys}
\bibliography{main}

\end{document}